\documentclass[twocolumn,showpacs,preprintnumbers,amsmath,amssymb]{revtex4-1} \usepackage{dcolumn}

\usepackage{graphicx} \usepackage{bm}

\newcommand{\hd}{\hat{d}}
\newcommand{\hf}{\hat{f}}
\newcommand{\hg}{\hat{g}}
\newcommand{\hk}{\hat{k}}
\newcommand{\hq}{\hat{q}}

\newcommand{\hDel}{\hat{\Delta}}
\newcommand{\hG}{\hat{G}}
\newcommand{\hLam}{\hat{\Lambda}}

\newcommand{\hI}{\hat{I}}
\newcommand{\hOm}{\hat{\Omega}}
\newcommand{\hS}{\hat{S}}
\newcommand{\hV}{\hat{V}}

\newcommand{\hbmg}{\hat{\bm g}}
\newcommand{\hbms}{\hat{\bm s}}
\newcommand{\hbmt}{\hat{\bm t}}

\newcommand{\hbmA}{\hat{\bm A}}
\newcommand{\hbmB}{\hat{\bm B}}
\newcommand{\hbmC}{\hat{\bm C}}
\newcommand{\hbmD}{\hat{\bm D}}
\newcommand{\hbmDel}{\hat{\bm \Delta}}
\newcommand{\hbmF}{\hat{\bm F}}
\newcommand{\hbmG}{\hat{\bm G}}

\newcommand{\hbmK}{\hat{\bm K}}

\newcommand{\hbmQ}{\hat{\bm Q}}
\newcommand{\hbmS}{\hat{\bm S}}
\newcommand{\hbmT}{\hat{\bm T}}
\newcommand{\hbmY}{\hat{\bm Y}}
\newcommand{\hbmZ}{\hat{\bm Z}}

\begin{document}

\title{Exact Green's function approach to RKKY interactions}
\date{\today}
\author{Tomasz M. Rusin } \email{email: tmr@vp.pl} \author{Wlodek Zawadzki}

\affiliation{Institute of Physics, Polish Academy of Sciences, Al. Lotnik\'ow 32/46, 02-688 Warsaw, Poland}

\begin{abstract}
The Green's function~(GF) of two localized magnetic moments embedded in the electron gas
is calculated exactly. The electrons are treated in the effective mass approximation
and the magnetic moments are coupled with electrons by a delta-like~$s-d$ interaction.
The resulting GF is obtained as a result of the exact summation of the Born series
using a generalization
of the method developed by Slater-Koster and Ziman to non-commuting spin operators with
the use of the Woodbury identities. For small~$s-d$ coupling~$J$ the exact GF reduces to
the RKKY case, for which the first two terms of the Born series are included.
In contrast to the standard RKKY, for the exact GF there is no symmetry between positive
and negative values of~$J$. The exact GF crucially depends on the value of the
one-electron Green's function at the origin, denoted as~$g_0$.
The Born series is convergent only if~$g_0$ is finite, which holds for electrons in
parabolic energy bands in~$1D$, but not in~$2D$ and~$3D$.
For this reason a simple model of RKKY interaction deserves to be reconsidered,
since the second term of the perturbation series is finite, and gives
the standard RKKY interaction, while the sum of remaining terms is divergent.
To ensure convergence of the Born series, a more realistic models of inter-spin interactions
have to be implemented. A finite value of~$g_0$ can be obtained once a cut-off for the
energy integration is introduced. In the general case, the exact GF includes nonlinear combination
of localized spins operators. A method of calculating matrix elements of these operators is given.
For spins~$S_a=S_b = 1/2$ the exact GF is expressed as a linear combination of
components of~$\hS_a, \hS_b$, and the exact range function~${\cal J}(r)$ is obtained
as a double integral over analytical expression.
For electron energy~$E=0$ and~$J g_0/2 \simeq 2$ or~$J g_0/2 \simeq-2/3$ the range function
and GF are singular. Poles of GF occur in the vicinities of singularity points
and the resulting energies of bound states
are calculated. The origin of asymmetry between positive and negative~$J$ values is
explained. The range function is analyzed within wide range of~$J$ values.
There are three regimes of~$J$. For~$|J| \ll |g_0|^{-1}$,
the range function~${\cal J}(r)$ resembles RKKY one: it has the same period~$\pi/k_F$,
the same decay character and a slightly different amplitude, usually within a few percent.
This regime occurs for nuclear spin ordering, magnetic interaction in
II-VI and IV-VI dilute magnetic semiconductors, III-V magnetic semiconductors,
some heavy fermion systems and bulk metal alloys.
For~$|J|$ comparable to~$|g_0|^{-1}$ the exact range function differs qualitatively
from RKKY one: it has much larger amplitude, non-oscillatory character and
it decays more slowly with inter-spin distance.
For~$|J| \gg |g_0|^{-1}$ the exact range function oscillates with the
same period and power-like decay as the usual RKKY function
but it has much lower amplitude decaying with growing~$|J|$.
In the limiting case of~$|J| \rightarrow \infty$ the range function vanishes.
This non-perturbative effect is explained.
A range of validity of the proposed model to real systems is discussed.
\end{abstract}

\maketitle

\section{Introduction \label{Sec_Intro}}

In 1954 Ruderman and Kittel described interaction between nuclear magnetic moments of
impurities in metals~\cite{Ruderman1954}.
The interaction was mediated by conduction electrons and had a long range character.
It was found that the second order correction to the energy of free electron gas due to the presence of
two nuclei is proportional to the product of the two spin operators and the range
function~${\cal J}_{RK}(r)$ depended on the distance between spins.
The range function oscillates in space with the period~$\pi/k_F$, where~$k_F$ is the Fermi vector,
and for large distances it decays as~$1/r^3$. Sometime later Kasuya~\cite{Kasuya1956}
and Yoshida~\cite{Yosida1957} pointed out
that exactly the same interaction appears between magnetic atom impurities in metals as a result
of~$s-d$ or~$s-f$ hybridization.

During last sixty years the RKKY interaction was investigated both theoretically and experimentally
in more realistic systems. The review works of RKKY can be found in Ref.~\cite{FreemanBook}
and many textbooks of solid state physics, see~\cite{KittelBook}.

In the present paper we propose a method of exact summation of the Born series for two
localized spins interacting with electron gas by the~$s-d$ interaction. Our calculations
generalize the RKKY theory by taking into account all terms of perturbation series instead
of retaining only terms of the second order in the~$s-d$ coupling constant~$J$.
We calculate the exact Green's function~(GF) of the system using a modification
of the method proposed by Slater-Koster-Ziman to potentials
including non-commuting spin operators~\cite{Koster1954}.
Having calculated the exact GF of the system we clarify the issues of convergence
of Born series and calculate the range function obtained from the exct GF. We also clarify
the issues related to behavior of GF and the range function for small and large values of~$|J|$
and discuss the possibility of existence of localized states. It appears that these
results have not been reported in literature.

Our intention is to compare the exact results with those obtained for standard
RKKY theory. For this reason we consider electrons in parabolic energy bands
described by the effective mass approximation. Within this approach we calculate the impact of
higher order terms of the Born series on the GF and range function of RKKY problem.
We mostly concentrate on~$3D$ case at~$T=0$.

The paper is organized as follows. Section~\ref{Sec_Prelim} outlines the derivation of RKKY
using the second order terms of the Born series and discusses some properties of singular potentials.
Section~\ref{Sec_GFDyson} introduces the Dyson equation of the problem and its solution
with use of Woodbury identities. In Section~\ref{Sec_GFArb} we express the exact GF for
arbitrary spins as nonlinear combination of localized spins operators. Section~\ref{Sec_GFME}
provides a method of calculating matrix elements of exact GF in Section~\ref{Sec_GFArb}.
Section~\ref{Sec_GFSpinOp} considers the case of spins~$\hS_a, \hS_b = 1/2$ and expresses the
exact GF as a linear function of products of spin operators.
Section~\ref{Sec_Grand} contains calculations of density of states
obtained from the exact GF, the grand canonical potential depending on localized
spins configuration and the corresponding range function.
Section~\ref{Sec_Appr} introduces a simplified model of exact GF, grand canonical potential
and the range function valid for fast decaying one-electron GF. This approximation allows us to
understand physical origin of several peculiarities existing in the exact results.
Section~\ref{Sec_OneE} discusses one-electron GF used in further calculations and introduces
an energy cut-off for one-electron GF at the origin. Section~\ref{Sec_Res} contains
numerical calculations of the exact range function for several values of key model parameters.
In section~\ref{Sec_Disc} we discuss our results.
The work is concluded by the Summary. Appendices and Supplemental material provide
auxiliary information related to the problems analyzed in this work.

\section{Preliminaries \label{Sec_Prelim}}

Let us consider the Dyson equation~$\hG = \hg + \hg \hV \hG$,
where~$\hV = \hV_a + \hV_b$,~$\hV_a$ and~$\hV_b$
are two non-overlapping potentials,~$\hg$ is the GF in absence of~$\hV$ and~$\hG$ is the GF in
the presence of~$\hV$. Iterating the Dyson equation one obtains the Born
series:~$\hG \simeq \hg + \hg \hV \hg + \hg \hV \hg \hV \hg + \ldots$.
The lowest order terms of this series depending on both~$\hV_a$ and~$\hV_b$ are
\begin{equation} \label{hGab_s2}
 \hG^{ab} \simeq \hg \hV_a \hg \hV_b \hg + \hg \hV_b \hg \hV_a \hg + \ldots.
\end{equation}
We consider the potentials~$\hV_c$ with~$c=a,b$ in the form of contact~$s-d$ interaction
\begin{equation} \label{Vrc}
 \hV_c({\bm r}) = J\delta({\bm r}-{\bm r}_c) \hbmS_c \hbms,
\end{equation}
where~$J$ is the~$s-d$ coupling constant measured in~$J \times m^D$ units,~$D$
is system dimensionality,~$\hbms ={\bm \sigma}/2$ is the electron spin operator,
and~${\bm \sigma}$ are the Pauli matrices in the standard notation. The operators~$\hbmS_c$
describe localized spins of atomic nuclei or magnetic impurities.
Taking the trace of~$\hG^{ab}$ one finds the density of
states~(DOS) of the system~$n(E)$ and the corresponding thermodynamic potential~$\Omega[n(E)]$.
For the one-electron Green's function~$\hg$ in the effective
mass approximation,~$D=3$ and~$T=0$ one obtains the well-known result~\cite{Ruderman1954}
\begin{equation} \label{Omega_RK}
 \Delta \Omega \simeq {\cal J}_{RK}\hbmS_a \hbmS_b,
\end{equation}
\begin{equation} \label{JRK}
 {\cal J}_{RK}(r) = \frac{J^2}{64\pi^3 r^4 \zeta} \left[ 2rk_F\cos(2k_Fr) -\sin(2k_Fr) \right],
\end{equation}
where~$\zeta = \hbar^2/(2m^*)$,~$k_F$ is the Fermi vector,~$m^*$ is electron effective mass,
and~$r$ is the distance between~$\hS_a$ and~$\hS_b$.
Equations~(\ref{Omega_RK}) and~(\ref{JRK}) describe the RKKY Hamiltonian and range function
of electrons interacting with localized spins.
The RKKY interaction in Eq.~(\ref{JRK}) is of second order effect in
terms of~$s-d$ coupling constant~$J$.

There appear questions about the validity of Eqs.~(\ref{Omega_RK}) and~(\ref{JRK}).
First, about the convergence of the Born series and the impact of remaining
infinite number of terms on the range function in Eq.~(\ref{JRK}).
Next, one may ask whether the Born series converges for arbitrary~$J$ or is there
a critical value of~$J$ above which the perturbation series diverges.
Finally, is it possible that for sufficiently large~$|J|$
there appear localized or resonant states.

Taking proper material band structure, reasonable physical parameters
and including other effects
appearing in solids (as e.g, phonons, disorder, many-body effects in electron gas
and in ion electrons), the RKKY theory correctly describes experimental results~\cite{FreemanBook}.
This implies that for RKKY problem the Born series converges and its higher
order terms do not alter significantly the results in Eqs.~(\ref{Omega_RK}) and~(\ref{JRK}).
Another implication is that even if there is a critical value of~$J$ leading to
divergence of the Born series, its magnitude is
much larger than~$|J|$ observed in real materials.

However, there are at least three hints indicating that the impact of higher order terms
in the Born series is more complicated and ambiguous.
First, as pointed in Refs.~\cite{Vertogen1966,Bowen1968},
the third order term of the perturbation series for RKKY energy is divergent.
However, there exists a suggestion of Kittel that,
possibly the whole Born series is convergent
irrespective of the fact that some of its terms diverge if calculated separately~\cite{Kittel1968}.
The second hint is that taking into account only spin parts of the potentials~$\hV_a$ and~$\hV_b$,
the higher order terms are more complicated functions of localized spins in Eq.~(\ref{Omega_RK}).
The last hint relates to analytical results obtained for the case of single scalar delta-like potential.
Let~$\hV_b=0$ and~$\hV_a = v_a\delta({\bm r}-{\bm r}_a)$, where~$v_a$ is potential strength.
Using the method proposed by Slater, Koster and Ziman and
others~\cite{Koster1954,Wolff1961,Clogston1962,ZimanBook}
one can sum the Born series to obtain
\begin{equation} \label{GF_Ziman}
 G({\bm r}_1, {\bm r}_2) = g({\bm r}_1,{\bm r}_2) +
 g({\bm r}_1,{\bm r}_a) \frac{v_a}{1 - g_0v_a} g({\bm r}_a,{\bm r}_2),
\end{equation}
where~$g_0 = g({\bm r}_a, {\bm r}_a)$ is one-electron GF at the origin.
The GF in Eq.~(\ref{GF_Ziman}) exists only when the quantity~$g_0$ is finite.
For~$|g_0v_a| \ll 1$ one may neglect~$|g_0v_a|$ in the denominator of Eq.~(\ref{GF_Ziman}) and
the GF is well approximated by its lowest order terms in~$v_a$.
By increasing~$|g_0v_a|$ the corrections due to the denominator in Eq.~(\ref{GF_Ziman}) are
more pronounced. For vanishing imaginary part of~$g_0$ and appropriate value of~$v_a$ there
appears a pole of GF, indicating an existence of localized states.
For~$|g_0v_a| \gg 1$ the second term in Eq.~(\ref{GF_Ziman}) gradually decreases
and for~$v_a \rightarrow \infty$ the GF does not depend on~$v_a$.
Finally, the GF in Eq.~(\ref{GF_Ziman}) is not symmetric for positive and negative values of~$v_a$.

The above hints suggest that the RKKY interaction obtained in a second order of
perturbation expansion, as given in Eqs.~(\ref{Omega_RK}) and~(\ref{JRK}), may overlook some important
properties of the system. The potentials~$\hV_a$ and~$\hV_b$ are products of
delta-like potentials and spin interactions between conduction electrons and localized moments.
Therefore the true GF of the system should include spin effects, e.g. its dependence
on relative spin orientations and effects related to
delta-like potentials, similar to those following from Eq.~(\ref{GF_Ziman}).

\section{The Green's Function of the system \label{Sec_GFDyson}}

We consider the electron gas perturbed by two localized
spins~$\hS_a$,~$\hS_b$ placed in~${\bm r}_a$,~${\bm r}_b$, respectively.
The potential of the~$s-d$ interaction between the spins and
the electron gas is
\begin{equation}\label{Vr}
 \hV({\bm r}) =\hbmZ_a \delta({\bm r}-{\bm r}_a) + \hbmZ_b\delta({\bm r}-{\bm r}_b) \equiv \hV_a + \hV_b,
\end{equation}
where we defined:~$\hbmZ_a= J \hbmS_a \hbms/2$
and~$\hbmZ_b= J \hbmS_b \hbms/2$, see Eq.~(\ref{Vrc}). Note the sign convention
in Eq.~(\ref{Vr}): positive sign of~$J$ corresponds to anti-ferromagnetic
coupling between impurity and electron spins. Then
\begin{equation} \label{hbmZc}
 \hbmZ_c
 = \frac{J}{2} \left(\begin{array}{cc} +\hS_c^z & \hS_c^- \\
 \hS_c^+ & -\hS_c^z \end{array} \right), \hspace*{1em} c=a,b.
\end{equation}
The main differences between the scalar potential in Eq.~(\ref{GF_Ziman}) and the spin dependent
potentials in Eqs.~(\ref{Vr}) and~(\ref{hbmZc})
are: i) the~$x, y,z$ components of~$\hV_a$ and~$\hV_b$ do not commute and ii)
the potentials~$\hV_a$ and~$\hV_b$ as given in Eq.~(\ref{Vrc}) do not commute, which can be
demonstrated by direct calculations. Then, in further calculation one has to ensure proper order of
spin operators and its components. Because of the nonzero commutator of~$\hV_a$ and~$\hV_b$
in our problem, we may not apply the results obtained for the Kondo
problem~\cite{Wiegmann1981,Andrei1983}.

We treat the electron gas in the single-particle approximation and assume
that the electron spin is a good quantum number, i.e. the periodic potential
of the lattice does not mix electron states of different spins.
The one electron states are then two-component
spinors~$|{\bm k}\nu\rangle = |{\bm k}\rangle \times| \nu\rangle$,
where~$\nu \in \{\uparrow, \downarrow \}$ is the~$s_z$ component of electron spin,
and~$|{\bm k}\rangle$ is the Bloch state of the conduction band.

The conduction band is filled by electrons up to the energy~$E_F$ and we neglect
interactions between electrons. The energy dispersion~$\epsilon({\bm k})$ may be
arbitrary, but spin-independent. Then the one-electron Green's function
is the a~$2 \times 2$ matrix diagonal in spin variables
\begin{equation} \label{hg}
 \hbmg({\bm r}_1,{\bm r}_2, \nu_1, \nu_2) = g({\bm r}_1,{\bm r}_2)
 \left(\begin{array}{cc} 1 & 0 \\ 0 & 1 \end{array} \right),
\end{equation}
where
\begin{equation} \label{hg1}
 g({\bm r}_1,{\bm r}_2, E)
 =\sum_{\bm k} \frac{|{\bm k} \rangle \langle {\bm k}|}{E-\epsilon_{\bm k}}.
\end{equation}
The only assumption for GF in Eq.~(\ref{hg1}) is that, for all energies~$E>0$,
the GF at the origin~$g_0$ is finite and nonzero
\begin{equation} \label{defg0}
 |g_0| = |\lim_{{\bm r}_1 \rightarrow {\bm r}_2} g({\bm r}_1,{\bm r}_2, E)| \in (0, \infty).
\end{equation}
In section~\ref{Sec_OneE} we consider the one-electron GF for parabolic energy band
in the effective mass approximation, which is a special case of GF in Eq.~(\ref{hg1}).

\subsection{The Dyson equation}

Within the model described above we solve the Dyson equation for the exact GF
of the system.
Let~$\hbmG$ be the Green's function of the electron gas in the presence
of external potential given in Eq.~(\ref{Vr}). The functions~$\hbmG$ and~$\hbmg$ are
related to each other by the Dyson equation:~$\hbmG= \hbmg+ \hbmg V\hbmG$.
In the position representation there is
\begin{equation} \label{hG_Dyson}
 \hbmG_{12}= \hbmg_{12}+ \int \hbmg_{13} V({\bm r}_3) \hbmG_{31} d^3{\bm r}_3.
\end{equation}
In Eq.~(\ref{hG_Dyson}) and below we use the notation:~$\hbmG_{12} = \hbmG({\bm r}_1, {\bm r}_2)$
and~$\hbmg_{12} = \hbmg({\bm r}_1, {\bm r}_2)$. Since the potential~$V({\bm r}_3)$
in Eq.~(\ref{Vr}) is the sum of delta functions multiplied by spin operators one obtains
\begin{equation} \label{hG_12_a}
 \hbmG_{12} = \hbmg_{12} + \hbmg_{1a}\hbmZ_a \hbmG_{a2} + \hbmg_{1b}\hbmZ_b \hbmG_{b2},
\end{equation}
where~$\hbmg^{\pm}_{12}$,~$\hbmg^{\pm}_{1a}$,~$\hbmg^{\pm}_{1b}$
are given in Eqs.~(\ref{hg}) and~(\ref{hg1}).
The function~$\hbmG_{12}$ is a~$2\times 2$ matrix and the main objective
of this paper is to obtain its four components in the analytical form.

To find~$\hbmG_{12}$ we generalize the method proposed by Slater-Koster and Ziman
to sum the Born-series for neutral delta-like impurity
embedded in the noninteracting electron gas~\cite{Koster1954,Wolff1961,Clogston1962,ZimanBook}.
By setting in Eq.~(\ref{hG_12_a}):~${\bm r}_1 \rightarrow {\bm r}_a$
and~${\bm r}_1 \rightarrow {\bm r}_b$ one
obtains two coupled equations for~$\hbmG_{a2}$ and~$\hbmG_{b2}$
\begin{eqnarray}
 \label{2Eqs_a}
 \hbmG_{a2} =\hbmg_{a2} + \hbmg_{aa}\hbmZ_a \hbmG_{a2} + \hbmg_{ab}\hbmZ_b \hbmG_{b2} \\
 \label{2Eqs_b}
 \hbmG_{b2} =\hbmg_{b2} + \hbmg_{ba}\hbmZ_a \hbmG_{a2} + \hbmg_{bb}\hbmZ_b \hbmG_{b2}.
\end{eqnarray}
We may rewrite Eqs.~(\ref{2Eqs_a}) and~(\ref{2Eqs_b}) in a matrix form
\begin{equation} \label{2Eqs_ab}
 \left(\begin{array}{cc} \hI- \hbmg_{aa}\hbmZ_a & -\hbmg_{ab}\hbmZ_b \\
 -\hbmg_{ba}\hbmZ_a & \hI-\hbmg_{bb}\hbmZ_b \end{array} \right)
 \left(\begin{array}{c} \hbmG_{a2} \\ \hbmG_{b2} \end{array} \right) =
 \left(\begin{array}{c} \hbmg_{a2} \\ \hbmg_{b2} \end{array} \right).
\end{equation}
In the above equation the matrix is a~$4 \times 4$ operator.
We write formally
\begin{equation} \label{2Eqs_Y}
 \left(\begin{array}{c} \hbmG_{a2} \\ \hbmG_{b2} \end{array} \right) =
 \left(\begin{array}{cc} \hI- \hbmg_{aa}\hbmZ_a & -\hbmg_{ab}\hbmZ_b \\
 -\hbmg_{ba}\hbmZ_a & \hI-\hbmg_{bb}\hbmZ_b \end{array} \right)^{-1}
 \left(\begin{array}{c} \hbmg_{a2} \\ \hbmg_{b2} \end{array} \right).
\end{equation}
To find the matrix in Eq.~(\ref{2Eqs_Y}) we consider two~$4 \times 4$
operators:~$\hbmY$ and~$\hbmt = \hbmY^{-1}$.
Let~$\hbmY$ be the matrix in Eq.~(\ref{2Eqs_ab})
\begin{equation} \label{DefY}
 \hbmY =
 \left(\begin{array}{cc} \hI- \hbmg_{aa}\hbmZ_a & -\hbmg_{ab}\hbmZ_b \\
 -\hbmg_{ba}\hbmZ_a & \hI-\hbmg_{bb}\hbmZ_b \end{array} \right)
 =\left[\begin{array}{cc} \hbmA & \hbmB \\ \hbmC & \hbmD \end{array} \right],
\end{equation}
and~$\hbmt$ be the matrix in Eq.~(\ref{2Eqs_Y}).
\begin{equation} \label{Deft}
 \hbmt =
 \left(\begin{array}{cc} \hI- \hbmg_{aa}\hbmZ_a & -\hbmg_{ab}\hbmZ_b \\
 -\hbmg_{ba}\hbmZ_a & \hI-\hbmg_{bb}\hbmZ_b \end{array} \right)^{-1}
 =\left(\begin{array}{cc}\hbmt^A &\hbmt^B \\ \hbmt^C & \hbmt^D\end{array} \right).
\end{equation}
In Eq.~(\ref{Deft}) the operators~$\hbmt^A, \hbmt^B, \hbmt^C, \hbmt^D$ are
undeterminate yet, and they are complicated
functions of~$\hbmA, \hbmB, \hbmC, \hbmD$, see below.
From Eq.~(\ref{2Eqs_Y}) we have
\begin{equation} \label{2Eqs_t}
 \left(\begin{array}{c} \hbmG_{a2} \\ \hbmG_{b2} \end{array} \right) =
 \left(\begin{array}{cc}\hbmt^A &\hbmt^B \\ \hbmt^C & \hbmt^D\end{array} \right)
 \left(\begin{array}{c} \hbmg_{a2} \\ \hbmg_{b2} \end{array} \right).
\end{equation}
The exact Green's function in Eq.~(\ref{hG_12_a}) is
\begin{eqnarray} \label{hbmG_12_b}
 \hbmG_{12} = \hbmg_{12} &+& \hbmg_{1a}[\hbmZ_a\ \hbmt^A]\hbmg_{a2} + \hbmg_{1a}[\hbmZ_a\ \hbmt^B]\hbmg_{b2} \ \ \nonumber \\
                         &+& \hbmg_{1b}[\hbmZ_b\ \hbmt^C]\hbmg_{a2} + \hbmg_{1b}[\hbmZ_b\ \hbmt^D]\hbmg_{b2}. \ \ \ \ \ \
\end{eqnarray}
Equation~(\ref{hbmG_12_b}) describes the Green's function of the two impurity problem
and it has a form of the Dyson equation for the~$\hbmT$-operator:~$\hbmG = \hbmg + \hbmg\hbmT\hbmg$~\cite{ZimanBook}.
In Eq.~(\ref{hbmG_12_b}),~$\hbmg_{1a}$,~$\hbmg_{1b}$ and~$\hbmg_{12}$ are scalars so below we
omit the matrix signs. The operators~$\hbmZ_a$,~$\hbmZ_b$,~$\hbmt^i$ with~$i\in\{A,B,C,D\}$ are~$2\times 2$ matrices.
The operators~$\hbmZ_a$,~$\hbmZ_b$ are given in Eq~(\ref{hbmZc}).
To determine~$\hbmt^A, \hbmt^B, \hbmt^C, \hbmt^D$ we use the Woodbury identities.

\subsection{Matrix inversion by Woodbury identities}

Let~$\hbmA$,~$\hbmB$,~$\hbmC$,~$\hbmD$ be noncommuting operators in Eq.~(\ref{DefY}).
Then the Woodbury formula states~\cite{Woodbury50}
\begin{equation} \label{Woodbury}
 \hbmY^{-1} = \left[\begin{array}{cc} +\hbmDel_1^{-1} & -\hbmDel_1^{-1} \hbmB \hbmD^{-1} \\
 -\hbmDel_2^{-1}\hbmC \hbmA^{-1} & +\hbmDel_2^{-1} \end{array} \right],
\end{equation}
where
\begin{eqnarray}
 \label{defDel1}
 \hbmDel_1 &=& \hbmA - \hbmB \hbmD^{-1} \hbmC,\\
\label{defDel2}
\hbmDel_2 &=& \hbmD - \hbmC \hbmA^{-1} \hbmB.
\end{eqnarray}
Turning to Eq.~(\ref{2Eqs_Y}) we note that in this case~$\hbmA$ commutes with~$\hbmC$,
and~$\hbmB$ commutes with~$\hbmD$. This gives:~$\hbmDel_1=\hbmD^{-1}\hbmF_1$
and~$\hbmDel_2=\hbmA^{-1}\hbmF_2$, where
\begin{eqnarray} \label{DefF1}
\hbmF_1 &=& \hbmD \hbmA - \hbmB \hbmC, \\
\label{DefF2}
\hbmF_2 &=& \hbmA \hbmD - \hbmC \hbmB,
\end{eqnarray}
respectively. Then we have from Eq.~(\ref{Woodbury})
\begin{equation} \label{Woodbury2}
 \hbmY^{-1} = \left[\begin{array}{cc} +\hbmF_1^{-1}\hbmD & -\hbmF_1^{-1}\hbmB \\
 -\hbmF_2^{-1}\hbmC & +\hbmF_2^{-1}\hbmA \end{array} \right],
\end{equation}
while from Eqs.~(\ref{2Eqs_Y}) and~(\ref{DefF1})--(\ref{Woodbury2}) there is
\begin{equation} \label{wyn_t}
 \hbmt = \left[\begin{array}{ll}
 \hbmF^{-1}_1 (\hI-g_0\hbmZ_b)& \hbmF^{-1}_1 g_{ab}\hbmZ_b,\\
 \hbmF^{-1}_2 g_{ba}\hbmZ_a & \hbmF^{-1}_2 (\hI-g_0\hbmZ_a) \end{array} \right],
\end{equation}
where~$g_{aa}=g_{bb} \equiv g_0$. We assume that~$g_0$ is finite, see Section~\ref{Sec_OneE}.
From Eqs.~(\ref{hbmG_12_b}) and~(\ref{wyn_t}) we have
\begin{eqnarray} \label{hbmG_12_c}
 \hbmG_{12} = g_{12}\hI + \hspace*{17em} \nonumber \\
            + g_{1a}[\hbmZ_a \hbmF_1^{-1} (\hI-g_0\hbmZ_b)]g_{a2}
            + g_{1a}[g_{ab}\hbmZ_a \hbmF_1^{-1} \hbmZ_b]g_{b2} \ \ \ \nonumber \\
            + g_{1b}[g_{ba}\hbmZ_b \hbmF_2^{-1} \hbmZ_a]g_{a2}
            + g_{1b}[\hbmZ_b \hbmF_2^{-1} (\hI-g_0\hbmZ_a)]g_{b2}. \ \ \
\end{eqnarray}
Introducing operators~$\hbmQ_1 = \hbmF_1^{-1}$ and~$\hbmQ_2 = \hbmF_2^{-1}$ we
rewrite Eq.~(\ref{hbmG_12_c}) as
\begin{eqnarray} \label{hbmG_12_d}
 \hbmG_{12} = g_{12}\hI + \hspace*{17em} \nonumber \\
            + g_{1a}[\hbmZ_a \hbmQ_1 (\hI-g_0\hbmZ_b)]g_{a2}
            + g_{1a}[g_{ab}\hbmZ_a \hbmQ_1 \hbmZ_b]g_{b2} \ \ \ \nonumber \\
            + g_{1b}[g_{ba}\hbmZ_b \hbmQ_2 \hbmZ_a]g_{a2}
            + g_{1b}[\hbmZ_b \hbmQ_2 (\hI-g_0\hbmZ_a)]g_{b2}. \ \ \
\end{eqnarray}
From Eqs.~(\ref{2Eqs_Y}) and~(\ref{DefF1})--(\ref{DefF2}) we find
\begin{eqnarray}
 \label{hbmF1}
 \hbmF_1 =\left[ (\hI-g_0\hbmZ_b)(\hI- g_0\hbmZ_a)- g_{ab} g_{ba} \hbmZ_b \hbmZ_a \right]=\hbmQ_1^{-1}, \ \ \ \\
 \label{hbmF2}
 \hbmF_2 = \left[ (\hI-g_0\hbmZ_a)(\hI- g_0\hbmZ_b)- g_{ab} g_{ba} \hbmZ_a \hbmZ_b \right]=\hbmQ_2^{-1}. \ \ \
\end{eqnarray}
Equations~(\ref{hbmG_12_d})--(\ref{hbmF2}) describe the exact GF of the considered system.
The operators~$\hbmQ_1$ and~$\hbmQ_2$ are~$2\times 2$ matrices defined as the
inversions of~$\hbmF_1$ and~$\hbmF_2$ matrices, which are combinations
of~$\hS_a$ and~$\hS_b$ operators.
In two limiting cases of small and large~$|J|$ the operators~$\hbmF_1$ and~$\hbmF_2$
can be inverted explicitly.
For arbitrary~$J$ we must invert~$\hbmF_1, \hbmF_2$ using the general form
of Woodbury identities in Eq.~(\ref{Woodbury}), see below.

For small~$s-d$ coupling
there is:~$g_0\hbmZ_a \ll \hI$,~$g_0\hbmZ_b \ll \hI$,~$g_{ab} g_{ba}\hbmZ_ b\hbmZ_a \ll \hI$,
so one can disregard these terms. Then one obtains in
Eqs.~(\ref{hbmF1})--(\ref{hbmF2}):~$\hbmF_1 \simeq \hI$,~$\hbmF_2 \simeq \hI$
and, consequently:~$\hbmQ_1, \hbmQ_2 \simeq \hI$. Then Eq.~(\ref{hbmG_12_d}) reduces to
\begin{eqnarray} \label{hbmG_12_lin}
 \hbmG_{12} \simeq \hbmg_{12}
 &+& g_{1a}\hbmZ_a g_{a2} + g_{1a} g_{ab} \hbmZ_a \hbmZ_b g_{b2} \nonumber \\
 &+&                        g_{1b} g_{ba} \hbmZ_b t\hbmZ_a g_{a2} + g_{1b}\hbmZ_b g_{b2}.
\end{eqnarray}
The equation~(\ref{hbmG_12_lin}) describes the second-order term of the
Born series for two-point spin-dependent potential
\begin{equation}
 \hG \simeq \hg + \hg(V_a+V_b)\hg + \hg(V_a+V_b)\hg(V_a+V_b)\hg + \ldots .
\end{equation}
Calculating the range function~${\cal J}(r)$ with use of GF in Eq.~(\ref{hbmG_12_lin}) one
obtains the standard result for RKKY interaction, see Appendix~\ref{App_RKKY}.

For the strong coupling there is~$g_0\hbmZ_a \gg \hI$,~$g_0\hbmZ_b \gg\hI$,
and~$g_{ab} g_{ba}\hbmZ_ b\hbmZ_a \gg \hI$, so that
one can disregard the identity operator~$\hI$ in Eqs.~(\ref{hbmF1}) and~(\ref{hbmF2}).
Then the expressions in Eqs.~(\ref{hbmF1}) and~(\ref{hbmF2}) reduce to products of two operators,
that can be inverted in the standard way. The GF in Eq.~(\ref{hbmG_12_d}) and the range function in
this limit are obtained and discussed in Appendix~\ref{App_Strong}.

\section{Exact Green's function for arbitrary spins \label{Sec_GFArb}}

Let~$\hbmF_1 = \left[\begin{array}{cc} \hf^{1A} & \hf^{1B} \\ \hf^{1C} & \hf^{1D} \end{array} \right]$,
in which
\begin{equation} \label{hF1A}
 \hf^{1A} = p_2(\hS_b^- \hS_a^+ + \hS_b^z \hS_a^z ) - p_1(\hS_b^z + \hS_a^z )+1,
\end{equation}
\begin{equation} \label{hF1B}
 \hf^{1B} = p_2(\hS_b^z \hS_a^- - \hS_b^- \hS_a^z) -p_1(\hS_b^- + \hS_a^-),
\end{equation}
\begin{equation} \label{hF1C}
 \hf^{1C} = p_2(\hS_b^+ \hS_a^z - \hS_b^z \hS_a^+ ) - p_1(\hS_b^+ +\hS_a^+),
\end{equation}
\begin{equation} \label{hF1D}
 \hf^{1D} = p_2(\hS_b^+ \hS_a^- + \hS_b^z \hS_a^z ) + p_1(\hS_b^z + \hS_a^z)+1,
\end{equation}
and
\begin{eqnarray} \label{defp1}
 p_1 &=& \frac{1}{2}Jg_0, \\
 \label{defp2}
 p_2 &=& \frac{1}{4} J^2(g_0^2-g_{ab}g_{ba}) \equiv J^2c_{ab}.
\end{eqnarray}
To obtain~$\hbmF_2$ one should exchange~$a$ and~$b$ indices in Eqs.~(\ref{hF1A})--(\ref{hF1D}). Let
\begin{equation}\label{hbmQ1a}
 \hbmQ_1 \equiv \hbmF_1^{-1} = \left[\begin{array}{cc} \hq^{1A} & \hq^{1B} \\ \hq^{1C} & \hq^{1D} \end{array} \right].
\end{equation}
Using Eq.~(\ref{Woodbury}) we find
\begin{eqnarray}
\label{hbmQ1b}
\hbmQ_1 &=& \left[\begin{array}{cc} \hDel_{1A}^{-1} & -\hDel_{1A}^{-1} \hf^{1B} \hf_{1D}^{-1} \\
 -\hDel_{1D}^{-1} \hf^{1C} (\hf^{1A})^{-1} & \hDel_{1D}^{-1} \end{array} \right],
 \end{eqnarray}
 in which
 \begin{eqnarray}
 \label{hDel1A}
 \hDel_{1A} &=& \hf^{1A} - \hf_{1B} (\hf^{1D})^{-1} \hf^{1C}, \\
 \label{hDel1D}
 \hDel_{1D} &=& \hf^{1D} - \hf^{1C} (\hf^{1A})^{-1} \hf^{1B}.
\end{eqnarray}
Similarly, let
\begin{equation} \label{hbmQ2a}
 \hbmQ_2 \equiv \hbmF_2^{-1} = \left[\begin{array}{cc} \hq^{2A} & \hq^{2B} \\ \hq^{2C} & \hq^{2D} \end{array} \right].
\end{equation}
Using Eq.~(\ref{Woodbury}) we find
\begin{eqnarray}
 \label{hbmQ2b}
\hbmQ_2 &=& \left[\begin{array}{cc} \hDel_{2A}^{-1} & -\hDel_{2A}^{-1} \hf^{2B} (\hf^{2D})^{-1} \\
 -\hDel_{2D}^{-1} \hf^{2C} (\hf^{2A})^{-1} & \hDel_{2D}^{-1} \end{array} \right],
 \end{eqnarray}
 in which
 \begin{eqnarray}
 \label{hDel2A}
 \hDel_{2A} &=& \hf^{2A} - \hf^{2B} (\hf^{2D})^{-1} \hf^{2C}, \\
 \label{hDel2D}
 \hDel_{2D} &=& \hf^{2D} - \hf^{2C} (\hf^{2A})^{-1} \hf^{2B}.
\end{eqnarray}
Then one obtains from Eqs.~(\ref{hbmG_12_d}),~(\ref{hbmQ1a}) and~(\ref{hbmQ2a})
\begin{equation}
 \hbmG=\hbmG^{aa} + \hbmG^{ab} + \hbmG^{ba}+\hbmG^{bb},
\end{equation}
which can be rewritten as a~$2 \times 2$ matrix equation
\begin{widetext}
\begin{equation}\label{hGgen}
 \left(\begin{array}{cc} (\hG)_{11} & (\hG)_{12} \\ (\hG)_{21}& (\hG)_{22} \end{array} \right) =
 \left(\begin{array}{cc} (\hG^{aa})_{11} & (\hG^{aa})_{12} \\ (\hG^{aa})_{21}& (\hG^{aa})_{22} \end{array} \right) +
 \left(\begin{array}{cc} (\hG^{ab})_{11} & (\hG^{ab})_{12} \\ (\hG^{ab})_{21}& (\hG^{ab})_{22} \end{array} \right) +
 \left(\begin{array}{cc} (\hG^{ba})_{11} & (\hG^{ba})_{12} \\ (\hG^{ba})_{21}& (\hG^{ba})_{22} \end{array} \right) +
 \left(\begin{array}{cc} (\hG^{bb})_{11} & (\hG^{bb})_{12} \\ (\hG^{bb})_{21}& (\hG^{bb})_{22} \end{array} \right),
\end{equation}
where
\begin{eqnarray} \label{hGaa11}
(\hG^{aa})_{11}=g_{1a} \left\{-\frac{J^2}{4} g_0 \left( \hS_a^- \hq^{1C} \hS_b^z - \hS_a^- \hq^{1D} \hS_b^+ - \hS_a^z \hq^{1A} \hS_b^z - \hS_a^z \hq^{1B} \hS_b^+ \right)+ \frac{J}{2} \left( \hS_a^- \hq^{1C} + \hS_a^z \hq^{1A} \right) \right\} g_{a2}, \\
(\hG^{aa})_{12}=g_{1a} \left\{-\frac{J^2}{4} g_0 \left( \hS_a^- \hq^{1C} \hS_b^- + \hS_a^- \hq^{1D} \hS_b^z - \hS_a^z \hq^{1A} \hS_b^- + \hS_a^z \hq^{1B} \hS_b^z \right) +\frac{J}{2} \left( \hS_a^- \hq^{1D} + \hS_a^z \hq^{1B} \right) \right\} g_{a2}, \\
(\hG^{aa})_{21}=g_{1a} \left\{-\frac{J^2}{4} g_0 \left( \hS_a^+ \hq^{1A} \hS_b^z - \hS_a^+ \hq^{1B} \hS_b^+ + \hS_a^z \hq^{1C} \hS_b^z + \hS_a^z \hq^{1D} \hS_b^+ \right) +\frac{J}{2} \left( \hS_a^+ \hq^{1A} - \hS_a^z \hq^{1C} \right) \right\} g_{a2}, \\
(\hG^{aa})_{22}=g_{1a} \left\{-\frac{J^2}{4} g_0 \left( \hS_a^+ \hq^{1A} \hS_b^- + \hS_a^+ \hq^{1B} \hS_b^z + \hS_a^z \hq^{1C} \hS_b^- - \hS_a^z \hq^{1D} \hS_b^z \right) +\frac{J}{2} \left( \hS_a^+ \hq^{1B} - \hS_a^z \hq^{1D} \right) \right\} g_{a2},
\end{eqnarray}
\begin{eqnarray}
(\hG^{ab})_{11}= g_{1a} \left\{ \frac{J^2}{4} g_{ab}\left( \hS_a^- \hq^{1C} \hS_b^z + \hS_a^- \hq^{1D} \hS_b^+ + \hS_a^z \hq^{1A} \hS_b^z + \hS_a^z \hq^{1B} \hS_b^+ \right) \right\} g_{b2}, \\
(\hG^{ab})_{12}= g_{1a} \left\{ \frac{J^2}{4} g_{ab}\left( \hS_a^- \hq^{1C} \hS_b^- - \hS_a^- \hq^{1D} \hS_b^z + \hS_a^z \hq^{1A} \hS_b^- - \hS_a^z \hq^{1B} \hS_b^z \right) \right\} g_{b2}, \\
(\hG^{ab})_{21}= g_{1a} \left\{ \frac{J^2}{4} g_{ab}\left( \hS_a^+ \hq^{1A} \hS_b^z + \hS_a^+ \hq^{1B} \hS_b^+ - \hS_a^z \hq^{1C} \hS_b^z - \hS_a^z \hq^{1D} \hS_b^+ \right) \right\} g_{b2}, \\
(\hG^{ab})_{22}= g_{1a} \left\{ \frac{J^2}{4} g_{ab}\left( \hS_a^+ \hq^{1A} \hS_b^- - \hS_a^+ \hq^{1B} \hS_b^z - \hS_a^z \hq^{1C} \hS_b^- + \hS_a^z \hq^{1D} \hS_b^z \right) \right\} g_{b2},
\end{eqnarray}
\begin{eqnarray}
(\hG^{ba})_{11}= g_{1b} \left\{ \frac{J^2}{4} g_{ba} \left( \hS_b^- \hq^{2C} \hS_a^z + \hS_b^- \hq^{2D} \hS_a^+ + \hS_b^z \hq^{2A} \hS_a^z + \hS_b^z \hq^{2B} \hS_a^+ \right) \right\} g_{a2}, \\
(\hG^{ba})_{12}= g_{1b} \left\{ \frac{J^2}{4} g_{ba} \left( \hS_b^- \hq^{2C} \hS_a^- - \hS_b^- \hq^{2D} \hS_a^z + \hS_b^z \hq^{2A} \hS_a^- - \hS_b^z \hq^{2B} \hS_a^z \right) \right\} g_{a2}, \\
(\hG^{ba})_{21}= g_{1b} \left\{ \frac{J^2}{4} g_{ba} \left( \hS_b^+ \hq^{2A} \hS_a^z + \hS_b^+ \hq^{2B} \hS_a^+ - \hS_b^z \hq^{2C} \hS_a^z - \hS_b^z \hq^{2D} \hS_a^+ \right) \right\} g_{a2}, \\
(\hG^{ba})_{22}= g_{1b} \left\{ \frac{J^2}{4} g_{ba} \left( \hS_b^+ \hq^{2A} \hS_a^- - \hS_b^+ \hq^{2B} \hS_a^z - \hS_b^z \hq^{2C} \hS_a^- + \hS_b^z \hq^{2D} \hS_a^z \right) \right\} g_{a2},
\end{eqnarray}
\begin{eqnarray}
(\hG^{bb})_{11}= g_{1b} \left\{-\frac{J^2}{4} g_0 \left( \hS_b^- \hq^{2C} \hS_a^z - \hS_b^- \hq^{2D} \hS_a^+ - \hS_b^z \hq^{2A} \hS_a^z - \hS_b^z \hq^{2B} \hS_a^+ \right) + \frac{J}{2} \left( \hS_b^- \hq^{2C} + \hS_b^z \hq^{2A} \right) \right\} g_{b2}, \\
(\hG^{bb})_{12}= g_{1b} \left\{-\frac{J^2}{4} g_0 \left( \hS_b^- \hq^{2C} \hS_a^- + \hS_b^- \hq^{2D} \hS_a^z - \hS_b^z \hq^{2A} \hS_a^- + \hS_b^z \hq^{2B} \hS_a^z \right) + \frac{J}{2} \left( \hS_b^- \hq^{2D} + \hS_b^z \hq^{2B} \right) \right\} g_{b2}, \\
(\hG^{bb})_{21}= g_{1b} \left\{-\frac{J^2}{4} g_0 \left( \hS_b^+ \hq^{2A} \hS_a^z - \hS_b^+ \hq^{2B} \hS_a^+ + \hS_b^z \hq^{2C} \hS_a^z + \hS_b^z \hq^{2D} \hS_a^+ \right) + \frac{J}{2} \left( \hS_b^+ \hq^{2A} - \hS_b^z \hq^{2C} \right) \right\} g_{b2}, \\
(\hG^{bb})_{22}= g_{1b} \left\{-\frac{J^2}{4} g_0 \left( \hS_b^+ \hq^{2A} \hS_a^- + \hS_b^+ \hq^{2B} \hS_a^z + \hS_b^z \hq^{2C} \hS_a^- - \hS_b^z \hq^{2D} \hS_a^z \right) + \frac{J}{2} \left( \hS_b^+ \hq^{2B} - \hS_b^z \hq^{2D} \right) \right\} g_{b2}.
 \label{hGbb22}
\end{eqnarray}
\end{widetext}

Equations~(\ref{hGgen})--(\ref{hGbb22}) describe the exact GF of electron gas in the presence of two point-like
impurities with arbitrary spins~$\hS_a$ and~$\hS_b$.
The operators~$\hq^{1\alpha},\hq^{2\alpha}$ with~$\alpha = A, B, C, D$ are defined
in Eqs.~(\ref{hbmQ1a}) and~(\ref{hbmQ2a}) respectively.
The terms~$\hG^{ab}$ and~$\hG^{ba}$ correspond, roughly, to interactions between spins,
while~$\hG^{aa}$ and~$\hG^{bb}$ describe one-site properties. By taking the limit~$J g_0 \rightarrow 0$
in Eqs.~(\ref{hF1A})--(\ref{defp2}) (corresponding to~$p_1, p_2 \rightarrow 0$)
we find:~$\hf^{1A}, \hf^{2A}, \hf^{1D}, \hf^{2D} \simeq 1$,
while the remaining terms vanish. There is also~$\hq^{1A}, \hq^{1D}, \hq^{2A}, \hq^{2D} \simeq 1$,
and the remaining terms vanish. Assuming~$g_{ab}=g_{ba}$ one obtains for the electron density~$n(E)$
\begin{eqnarray}
 n(E) &=& -\frac{1}{\pi}{\rm Im Tr}(\hbmG) \nonumber \\
 \label{nRKKY}
 &=& -\frac{J^2}{\pi} \left( {\rm Im} \int g_{1a}g_{b2}g_{ab} d^3 {\bm r} \right) \hbmS_a \hbmS_b,
\end{eqnarray}
which is the density of states obtained for the RKKY interaction, see Appendix~\ref{App_RKKY}.
In Eqs.~(\ref{hGaa11})--(\ref{hGbb22}) there is no symmetry between positive and
negative values of the coupling constant~$J$ because of the linear terms in~$J$.
The expressions in curly brackets in Eqs.~(\ref{hGaa11})--(\ref{hGbb22}) are the matrix elements
of the~$\hbmT$-operator.

For arbitrary spins~$\hS_a, \hS_b$ one can not find general expressions
for~$\hbmG$ in a closed form, because the operators~$\hq^{1\alpha}, \hq^{2\alpha}$ with~$\alpha = A, B, C, D$ in
Eqs.~(\ref{hGgen})--(\ref{hGbb22}) are nonlinear functions of~$\hS_a, \hS_b$, see Eqs.~(\ref{hbmQ1a})--(\ref{hDel2D}).
However, it is possible to obtain matrix elements of~$\hbmG$ using a method described in the next section.
Additionally, for~$\hS_a, \hS_b = 1/2$ it is possible to find analytical expressions
for~$\hq^{1\alpha}$ and~$\hq^{2\alpha}$.
This allows one to express the exact GF in Eqs.~(\ref{hGgen})--(\ref{hGbb22}) as a bilinear
combination of~$\hS_a, \hS_b$ components.

\section{Matrix elements of GF components \label{Sec_GFME}}

Here we present a general method of calculation of the matrix elements of~$\hbmG$ components,
as given in Eqs.~(\ref{hGgen})--(\ref{hGbb22}). This method may be applied for arbitrary
spins values~$\hS_a, \hS_b = 1/2, 1, 3/2, \ldots$ and we illustrate it for~$\hS_a, \hS_b=1/2$.

Consider the Zeeman basis for spins~$\hS_a, \hS_b$ in which each state~$|n\rangle$ is labeled by two~$z$-th
components the spins:~$|n\rangle =|S_a^z,S_b^z \rangle$.
For two~$S=1/2$ spins, the basis~${\cal B}_{1/2}$ consists of four vectors
\begin{eqnarray} \label{DefB12}
 {\cal B}_{1/2} &=&
 \{|\uparrow, \uparrow \rangle, |\uparrow, \downarrow \rangle, |\downarrow, \uparrow \rangle, |\downarrow, \downarrow \rangle \} \\
 & \equiv & \{|1\rangle, |2 \rangle, 3 \rangle, |4 \rangle \}, \nonumber
\end{eqnarray}
where the up and down arrows indicate states with~$S_z=+1/2$ and~$S_z=-1/2$, respectively.
For arbitrary spins such a basis consists of~$(2S_a+1)(2S_b+1)$ elements.
In the basis~${\cal B}_{1/2}$ the spin operators~$\hS_a^{\pm}$,~$\hS_b^{\pm}$,~$\hS_a^z$,~$\hS_b^z$
are~$4\times 4$ matrices
\begin{equation} \label{S05Sap}
 \hS_a^+ = \left(\begin{array}{cccc} 0 & 0 & 1 & 0 \\ 0 & 0 & 0 & 1 \\ 0 & 0 & 0 & 0 \\ 0 & 0 & 0 & 0 \end{array} \right),
\end{equation}
\begin{equation}\label{S05Sbp}
 \hS_b^+ = \left(\begin{array}{cccc} 0 & 1 & 0 & 0 \\ 0 & 0 & 0 & 0 \\ 0 & 0 & 0 & 1 \\ 0 & 0 & 0 & 0 \end{array} \right),
\end{equation}
and~$\hS_a^-=(\hS_a^+)^{\dagger}$,~$\hS_b^-=(\hS_b^+)^{\dagger}$. There is
also~$\hS_a^z = {\rm diag}(1/2,1/2,-1/2,-1/3)$ and~$\hS_b^z = {\rm diag} (1/2,-1/2,1/2,-1/2)$,
where~$'{\rm diag}'$ represents the diagonal matrix.
In this representation, each state~$|n\rangle$ with~$n=1,\ldots,4$ is a four-component column vector
with the~$n$-th element equal to unity and remaining elements equal to zero.
In the basis~${\cal B}_{1/2}$ the operators~$\hf_{1\alpha}, \hf_{2\alpha}$ with~$\alpha=A,B,C,D$
in Eqs.~(\ref{hF1A})--(\ref{hF1D}) are~$4 \times 4$ matrices, see Eqs.~(\ref{S05f1A})--(\ref{S05f2D})
in Supplemental material. Calculating appropriate products, sums and inverses
of these matrices, see Eqs.~(\ref{hq_m1})--(\ref{hq_m1}) and Eqs.~(\ref{q_1A_11})--(\ref{q_2C_34})
in Supplemental material, one obtains the~$4\times 4$ matrices
describing the~$\hq^{1\alpha}$,~$\hq^{2\alpha}$ operators. Inserting these matrices
to Eqs.~(\ref{hGgen})--(\ref{hGbb22}) one obtains~$\hbmG$, which is
also a~$4\times 4$ matrix in the representation~${\cal B}_{1/2}$. To find the matrix element of~$\hbmG$
between two states~$|n\rangle$ and~$|n'\rangle$, with~$n,n'=1,\ldots, 4$ one multiplies~$\hbmG$ by two appropriate
four-element vectors.

As an example of the above procedure we consider the third term of Eq.~(\ref{hGaa11})
\begin{equation} \label{Gab113x}
 (\hG^{ab})_{11|3}=\frac{1}{4}g_{1a} g_{ab} g_{b2} J^2 \hS_a^z \hq^{1A} \hS_b^z \equiv C_3 \hS_a^z \hq^{1A} \hS_b^z,
\end{equation}
where~$C_3=(J^2/4)g_{1a} g_{ab} g_{b2}$ is a~$c$-number. Using Eq.~(\ref{hq_m1}) from Supplemental material there is
\begin{widetext}
\begin{equation} \label{Gab113y}
 (\hG^{ab})_{11|3}= C
 \left(\begin{array}{rrrr} \frac{1}{2} & 0 & 0 & 0 \\ 0 & \frac{1}{2} & 0 & 0 \\ 0 & 0 & -\frac{1}{2} & 0 \\ 0 & 0 & 0 & -\frac{1}{2} \end{array} \right) \!
 \left(\begin{array}{cccc} q^{1A}_{11} & 0 & 0 & 0 \\ 0 & q^{1A}_{22} & q^{1A}_{23} & 0 \\ 0 & q^{1A}_{32} & q^{1A}_{33} & 0 \\ 0 & 0 & 0 & q^{1A}_{44}
 \end{array} \right)\!
 \left(\begin{array}{rrrr} \frac{1}{2} & 0 & 0 & 0 \\ 0 & -\frac{1}{2} & 0 & 0 \\ 0 & 0 & \frac{1}{2} & 0 \\ 0 & 0 & 0 & -\frac{1}{2} \end{array} \right) =
 \frac{C_3}{4}\left(\begin{array}{rrrr} q^{1A}_{11} & 0 & 0 & 0 \\ 0 & -q^{1A}_{22} & q^{1A}_{23} & 0 \\
 0 & q^{1A}_{32} & - q^{1A}_{33} & 0 \\ 0 & 0 & 0 & q^{1A}_{44} \end{array} \right),
\end{equation}
\end{widetext}
where~$q^{1A}_{22}, q^{1A}_{23}, q^{1A}_{32}, q^{1A}_{33}, q^{1A}_{44}$ are~$c$-numbers,
see Eqs.~(\ref{q_1A_11})--(\ref{q_2C_34}) in Supplemental material.
The matrix element of~$(\hG^{ab})_{11|3}$ between two states~$|1\rangle$ is then~$C_3q^{1A}_{11}/4$.

The procedure described above is convenient for calculation of the matrix elements of~$\hbmG$ for arbitrary spins.
Since the largest value of spin in stable isotopes is~$S=5$, corresponding to~$^{138}$La~\cite{Stone2005},
the largest number of basis states is~$(2S+1)^2=121$.

To find the matrix form of~$\hS_c^+, \hS_c^-, \hS_c^z$ operators~($c=a,b$) one uses the identities
\begin{eqnarray}
 \langle S,m'|\hS^z|S,m \rangle = m \delta_{m'm}, \\
 \langle S,m'|\hS^+|S,m \rangle = \delta_{m'm+1}\sqrt{S(S-1) -m'm}, \\
 \langle S,m'|\hS^-|S,m \rangle = \delta_{m+1'm}\sqrt{S(S-1) -m'm},
\end{eqnarray}
where~$S$ is an arbitrary spin whose~$S_z$ components are labeled by~$m=-S,-S+1, \ldots, S$.
Using the above identities one can construct operators~$\hS_a^{\pm}$,~$\hS_b^{\pm}$,~$\hS_a^z$,~$\hS_b^z$ analogous
to those in Eqs.~(\ref{S05Sap})--(\ref{S05Sbp}), which are now~$(2S_a+1)(2S_b+1) \times (2S_a+1)(2S_b+1)$ matrices.
Then the matrix elements of exact GF are obtained in the same way as those for~$S_a,S_b=1/2$ spins.

All numerical results obtained in Figures~\ref{Fig1}--\ref{Fig3}
can be derived using the method described above. We checked that they agree with
results obtained using expressions in Section~\ref{Sec_Grand}. However,
despite the fact that the described method is suitable for numerical calculation,
it gives little understanding of the physical nature of exact GF
and its dependence on the four physical parameters:~$m^*$,~$J$,~$r$ and~$E_F$.
For this reason, for the special case~$\hS_a, \hS_b=1/2$ we re-express exact GF in
terms of components of spins operators, which allows us to reduce the range
function~${\cal J}(r)$ to integrals of analytical expressions.

\section{Spin-operator form of GF components \label{Sec_GFSpinOp}}

Here we express the operators~$\hG_{11}$,~$\hG_{12}$,~$\hG_{21}$,~$\hG_{22}$
in Eqs.~(\ref{hGgen})--(\ref{hGbb22})
as linear combinations of spin operators~$\hS_a^{\pm}$,~$\hS_b^{\pm}$ and~$\hS_a^z$,~$\hS_b^z$.
This form of exact GF is more convenient for analysis the range function properties.

In the representation of Eq.~(\ref{DefB12}), both components of~$\hS_a, \hS_b$
spins and matrices~$\hq^{1\alpha}, \hq^{2\alpha}$
have at most fourteen non-zero elements. For all these matrices the elements~$(1,4)$ and~$(4,1)$ vanish.
Then, each term of RHS of Eqs.~(\ref{hGaa11})--(\ref{hGbb22}) can be expressed as
a linear combination of fourteen linearly independent~$4\times 4$ matrices~$\hLam_n$ having zero
elements~$(1,4)$ and~$(4,1)$. We define~$\hLam_n$ matrices as~$4 \times 4$ matrices having only one nonzero element
except elements~$(1,4)$ and~$(4,1)$. For~$\hLam_1$ we set the nonzero element to be~$(1,1)$, for~$\hLam_2$ the
element~$(1,2)$ etc., but we exclude elements~$(1,4)$ and~$(4,1)$. The last matrix in the set, i.e.,~$\hLam_{14}$
has nonzero element~$(4,3)$.

In the next step one expresses the matrices~$\hLam_n$ as combinations of
operators~$\hS_a^{\pm}$,~$\hS_b^{\pm}$,~$\hS_a^z$,~$\hS_b^z$ and their products, see Eqs.~(\ref{S05Sap})--(\ref{S05Sbp}).
This expansion is summarized below
\begin{widetext} \begin{equation} \label{S05M4x4}
 \begin{array}{|l|l|l|l|l|} \hline
 & |1\rangle=|\uparrow \uparrow \rangle & |2\rangle=|\uparrow \downarrow \rangle &
 |3\rangle=|\downarrow \uparrow \rangle & |4\rangle=|\downarrow \downarrow \rangle \\
 \hline
 |1\rangle=|\uparrow   \uparrow \rangle   & \hLam_{ 1}=\hd_{11}                  & \hLam_{ 2}=\hS_b^+ (\hI/2 + \hS_a^z) & \hLam_{ 3}=\hS_a^+ ( \hI/2 + \hS_b^z) & 0                     \\ \hline
 |2\rangle=|\uparrow   \downarrow \rangle & \hLam_{ 4}=\hS_b^- (\hI/2 + \hS_a^z) & \hLam_{ 5}=\hd_{22}                  & \hLam_{ 6}=\hS_a^+ \hS_b^-          & \hLam_{ 7}=\hS_a^+ (\hI/2 - \hS_b^z) \\ \hline
 |3\rangle=|\downarrow \uparrow \rangle   & \hLam_{ 8}=\hS_a^- (\hI/2 + \hS_b^z) & \hLam_{ 9}=\hS_a^- \hS_b^+           & \hLam_{10}=\hd_{33}                 & \hLam_{11}=\hS_b^+ (\hI/2 - \hS_a^z) \\ \hline
 |4\rangle=|\downarrow \downarrow \rangle & 0                     & \hLam_{12}=\hS_a^- (\hI/2 - \hS_b^z)  &  \hLam_{13}=\hS_b^- (\hI/2 - \hS_a^z) & \hLam_{14}=\hd_{44}                \\ \hline
 \end{array},
\end{equation} \end{widetext}
in which
\begin{eqnarray}
 \hd_{11}= \hI/4 + \hS_a^z/2 + \hS_b^z/2 + \hS_a^z\hS_b^z, \\
 \hd_{22}= \hI/4 + \hS_a^z/2 - \hS_b^z/2 - \hS_a^z\hS_b^z, \\
 \hd_{33}= \hI/4 - \hS_a^z/2 + \hS_b^z/2 - \hS_a^z\hS_b^z, \\
 \hd_{44}= \hI/4 - \hS_a^z/2 - \hS_b^z/2 + \hS_a^z\hS_b^z, \label{S05_hd44}
\end{eqnarray}
and~$\hI$ is the~$4 \times 4$ identity matrix. To explain notation used in Eq.~(\ref{S05M4x4})
let us consider the~$\hLam_{ 2}$ matrix. In the the Zeeman basis [upper line in Eq.~(\ref{S05M4x4})]
this matrix has one nonzero element~$(1,2)$. Direct calculation shows that matrix corresponding
to~$\hS_b^+ (\hI/2 + \hS_a^z)$ operator, see Eqs.~(\ref{S05Sap})--(\ref{S05Sbp}),
has also one non-vanishing element~$(1,2)$.
Then one assigns:~$\hLam_{ 2}=\hS_b^+ (\hI/2 + \hS_a^z)$, which is valid for~$\hS_a, \hS_b=1/2$.

Having defined operators~$\hLam_n$ we expand the
functions~$\hG^{\alpha\beta}_{ij}$ in Eqs.~(\ref{hGaa11})--(\ref{hGbb22}),
with~$c,d = a,b$ and~$i,j=1,2$ in linear combinations of~$\hLam_n$ operators
\begin{equation} \label{S05Gabij}
 \hG^{c,d}_{ij} = \sum_{n=1}^{14} C^{c,d}_{ij,n} \hLam_n,
\end{equation}
where~$C^{c,d}_{ij,n}$ are~$c$-numbers. Finally, using Eq.~(\ref{S05M4x4}), one
expresses each term of RHS of Eqs.~(\ref{hGaa11})--(\ref{hGbb22}) as a linear combination
of products of components of~$\hS_a, \hS_b$ operators. The formulas are shown in
in Eqs.~(\ref{hG_aa_11})--(\ref{hG_bb_22}) in Supplemental material.
These equations represent the exact GF
of a free electron gas interacting with two localized spin moments~$S=1/2$.
They are bilinear combinations of spin operators~$\{\hS_a^+, \hS_a^-, \hS_a^z, \hS_b^+, \hS_b^-, \hS_b^z \}$.
In contrast, the expressions in Eqs.~(\ref{hGaa11})--(\ref{hGbb22}) are nonlinear
combinations of spin operators because of the presence of~$\hq^{1\alpha}$,~$\hq^{2\alpha}$ operators.

Analytical expressions for elements of~$\hq^{1\alpha}$,~$\hq^{2\alpha}$ matrices are
shown in Eqs.~(\ref{q_1A_11})--(\ref{q_2C_34}) in Supplemental material.
The elements of this matrices, denoted as~$q^{1\alpha}_{ij}$ and~$q^{2\alpha}_{ij}$, are
complex numbers depending on~$p_1$ and~$p_2$ only,
see Eqs.~(\ref{defp1}) and~(\ref{defp2}). Both~$p_1$ and~$p_2$ depend
on the value of the one-electron GF at the origin~$g_0$,
which we assumed to be finite and nonzero, see Eq.~(\ref{defg0}).

To continue the example from Eqs.~(\ref{Gab113x}) we apply the above procedure
to~$(\hG^{ab})_{11|3}$ in Eq.~(\ref{Gab113y}) and obtain
\begin{eqnarray}
 (\hG^{ab})_{11|3} = \frac{C_3}{4} &&
 \left[ \hLam_{1} q^{1A}_{11} - \hLam_{5} q^{1A}_{22} - \hLam_{10} q^{1A}_{33} + \hLam_{14} q ^{1A}_{44} + \right. \nonumber \\
               && +\left. \hLam_{6} q^{1A}_{23} + \hLam_{9} q^{1A}_{32} \right].
\end{eqnarray}
Taking explicit forms of operators~$\hLam_1$,~$\hLam_5$,~$\hLam_6$,~$\hLam_9$,~$\hLam_{10}$,~$\hLam_{14}$,
see Eq.~(\ref{S05M4x4}), one finds
\begin{widetext}\begin{eqnarray}
 (\hG^{ab})_{11|3} &=& \frac{C_3}{4} \left( \frac{1}{4}\hI + \frac{1}{2}\hS_a^z + \frac{1}{2}\hS_b^z + \hS_a^z\hS_b^z \right) q^{1A}_{11}
 -\frac{C_3}{4} \left(\frac{1}{4}\hI + \frac{1}{2}\hS_a^z - \frac{1}{2}\hS_b^z - \hS_a^z\hS_b^z \right) q^{1A}_{22} \nonumber \\
 && -\frac{C_3}{4} \left(\frac{1}{4}\hI - \frac{1}{2}\hS_a^z + \frac{1}{2}\hS_b^z - \hS_a^z\hS_b^z \right) q^{1A}_{33}
 +\frac{C_3}{4} \left(\frac{1}{4}\hI - \frac{1}{2}\hS_a^z - \frac{1}{2}\hS_b^z + \hS_a^z\hS_b^z \right) q^{1A}_{44} \nonumber \\
 && + \frac{C_3}{4} \hS_a^+ \hS_b^- q^{1A}_{23} + \frac{C_3}{4} \hS_a^- \hS_b^+ q^{1A}_{32}. \label{Gab113z}
\end{eqnarray}
\end{widetext}
In Eq.~(\ref{Gab113z}) the quantity~$G^{ab}_{11|3}$ is a combination of products of localized spins components.
The remaining term of the exact GF are calculated in analogous way, and they are shown in
Eqs.~(\ref{hG_aa_11})--(\ref{hG_bb_22}).

\section{Grand canonical potential and range functions \label{Sec_Grand}}

Having obtained the exact GF one can calculate
observables measured experimentally. We calculate the density of
states~(DOS), the grand canonical potential, the range function and the energy of localized states.
All calculations are performed for~$T=0$ but they can be generalized to nonzero temperatures
using standard GF techniques, see Discussion.

\subsection{DOS and grand canonical potential}
The continuous energy spectrum of the system is determined by the discontinuity of the Green's function
along the cut of positive energy axis~\cite{EconomouBook}. Then the electron DOS is
\begin{equation} \label{Th_ne}
 n(E) = -\frac{1}{\pi} {\rm Im} \int{\rm Tr} \{\hbmG^+_{11} \} d^3 {\bm r},
\end{equation}
where~$\hbmG^+_{11} = \hbmG^+({\bm r}_1, {\bm r}_1)$ and
\begin{eqnarray}
 {\rm Tr} \{\hbmG_{11}^{+} \}&=& (\hbmG^{aa+})_{11} + (\hbmG^{aa+})_{22} + (\hbmG^{ab+})_{11} + (\hbmG^{ab+})_{22} + \nonumber \\
 \label{Th_Tr}               &+& (\hbmG^{ba+})_{11} + (\hbmG^{ba+})_{22} + (\hbmG^{bb+})_{11} + (\hbmG^{bb+})_{22}. \ \ \ \
\end{eqnarray}
Calculating the trace in Eq.~(\ref{Th_ne}) from Eqs.~(\ref{hGaa11})--(\ref{hGbb22})
or Eqs.~(\ref{hG_aa_11})--(\ref{hG_bb_22}) in Supplemental material we note
that~$\hbmG$ depends on spatial variables~${\bm r}_1$ and~${\bm r}_2$
by four products of one electron GFs,
namely:~$g_{1a}g_{a2}$,~$g_{1a}g_{b2}$,~$g_{1b}g_{a2}$,~$g_{1b}g_{b2}$, while the remaining
terms do not depend on~${\bm r}_1$ or~${\bm r}_2$. Taking the trace one obtains three integrals
\begin{eqnarray}
 \label{h_ab}
 h_{ab}^+ =h_{ba}^+ =& \int g_{1a}^+ g_{b1}^+ d^D{\bm r}_1 &= -\frac{\partial g_{ab}^+}{\partial E}, \\
 \label{h_0}
 h_0^+ = &\int g_{1a}^+ g_{a1}^+ d^D{\bm r}_1 &= \lim_{b\rightarrow a} h_{ab}^+,
\end{eqnarray}
where~$D=1,2,3$ is system's dimensionality, and~$g_{ab}^+=g_{ba}^+$.
In Eqs.~(\ref{h_ab}) and~(\ref{h_0}) we assumed the translational symmetry of one-electron GF.
To calculate quantities~$g_{ab}^+$,~$h_{ab}^+$ and~$h_0^+$ one needs to specify the
one electron GF. We address this point in Section~\ref{Sec_OneE}.

\subsection{Range function}
For non-interacting particles the generalized grand canonical potential is
\begin{equation} \label{Th_Omega}
 \hOm = -\int f(E) N(E) dE +\mu N,
\end{equation}
ant it satisfies the proper extremal properties of the total energy~\cite{Wildberger1995}.
Here~$\mu$ is the chemical
potential,~$N$ is the number of particles,~$f(E)$ is the Fermi-Dirac distribution function
and~$N(E)$ is the integrated density of states
\begin{equation} \label{Th_NE1}
 N(E) = \int_{-\infty}^{E} n(E') dE'.
\end{equation}
Our calculations are limited to~$T=0$, and below we approximate:~$f(E) = \Theta(E_F-E)$,
where~$\Theta(x)$ is the step function and~$E_F$ is the Fermi energy.

In Eq.~(\ref{Th_Omega}) the grand canonical potential~$\hOm$ depends on a configuration
of spins~$\hS_a$ and~$\hS_b$. For~$S_a= S_b = 1/2$ one defines the range function~${\cal J}(r)$
as a difference between~$\hOm$ for
parallel and anti-parallel configurations of~$\hS_a$ and~$\hS_b$ spins
\begin{equation} \label{defJ}
 {\cal J}(r) = \Omega_{\uparrow\uparrow} + \Omega_{\downarrow\downarrow}
           - \left(\Omega_{\uparrow\downarrow} + \Omega_{\downarrow\uparrow} \right),
\end{equation}
where
\begin{equation} \label{Omega_mn}
 \Omega_{\mu,\nu} = \langle \mu,\nu| \hOm |\mu,\nu \rangle,
\end{equation}
is the grand canonical potential for a given configuration~$\mu, \nu \in \{ \uparrow, \downarrow \}$
of~$\hS_a$ and~$\hS_b$. Then one can calculate~${\cal J}(r)$ numerically
with the use of Eqs.~(\ref{hGaa11})--(\ref{hGbb22}).

The range function~${\cal J}(r)$ in Eq.~(\ref{defJ}) can be conveniently calculated
for representation of GF given in Eqs.~(\ref{hG_aa_11})--(\ref{hG_bb_22}) in Supplemental material.
The derivation is based on the observation
that~${\cal J}(r)$ defined in Eq.~(\ref{defJ}) selects from Eqs.~(\ref{hG_aa_11})--(\ref{hG_bb_22})
only terms proportional to~$\hS_a^z\hS_b^z$. These terms we marked in Eqs.~(\ref{hG_aa_11})--(\ref{hG_bb_22})
by~$\maltese$ symbols. There are twelve such terms, and the trace in Eq.~(\ref{Th_Tr}) includes all of them.

Let~$\hbmG^{+S_a^z S_b^z}$ be the sum of terms proportional to~$\hS_a^z\hS_b^z$
and~$\Omega^{S_a^z S_b^z}$ be the part of the grand canonical
potential including~$\hbmG^{+S_a^z S_b^z}$. Then we have from Eqs.~(\ref{Th_Omega}) and~(\ref{Th_NE1})
\begin{equation} \label{Omega_SzSz1}
 \Omega^{S_a^z S_b^z} = \frac{1}{\pi} \int_0^{\infty} dE \int_{-\infty}^E dE'
 \left[ {\rm Im} \int{\rm Tr} \{\hbmG^{+S_a^z S_b^z} \} d^3 {\bm r} \right].
\end{equation}
Calculating the sum of twelve components of~$\hbmG^{+S_a^z S_b^z}$, and
taking the explicit form of elements~$\hq^{1\alpha}$ and~$\hq^{2\alpha}$ matrices, with~$\alpha=A,B,C,D$,
[see Eqs.~(\ref{q_1A_11})--(\ref{q_2C_34}) in Supplemental material], one obtains after some algebra
\begin{equation} \label{Omega_SzSz2}
 \Omega^{S_a^z S_b^z} = \Omega^{ab} + \Omega^{01} + \Omega^{02}.
\end{equation}
By~$\Omega^{ab}$ we denote the part of~$\Omega^{S_a^z S_b^z}$ depending on the inter-spin distance~$r$, and
by~$\Omega^{01} + \Omega^{02}$ we denote the part of~$\Omega^{S_a^z S_b^z}$ which does not depend on~$r$.
The indices~$1$ and~$2$ in~$\Omega^{01} + \Omega^{02}$ indicate powers of the coupling constant~$J$ entering into
these expressions. Then there is
\begin{equation} \label{Omega_ab}
 \Omega^{ab} = \frac{J^2}{\pi} \hS_a^z\hS_b^z\ {\rm Im} \int_0^{E_F}
 \left[ \int_{-\infty}^E g_{ab} h_{ab} w_{ab} dE' \right] dE,
\end{equation}
where~$g_{ab}$ is the one-electron GF at points~${\bm r}_a$
and~${\bm r}_b$, see Eq.~(\ref{hg1}), and~$h_{ab}$ is defined in Eq.~(\ref{h_ab}),
\begin{equation} \label{w_ab}
 w_{ab} = \frac{16(2p_1^2-4p_1-p_2+4)}{[8p_1(3p_2-4)-9p_2^2+8(p_2-2)](4p_1-p_2-4)},
\end{equation}
and~$p_1, p_2$ are given in Eqs.~(\ref{defp1}) and~(\ref{defp2}).
Similarly,
\begin{eqnarray} \label{Omega_01}
 \Omega^{01} &=& \frac{J}{\pi} \hS_a^z\hS_b^z\ {\rm Im} \int_0^{E_F} \left[ \int_{-\infty}^E h_0 w_{01} dE' \right] dE, \\
 \label{Omega_02}
 \Omega^{02} &=& \frac{J^2}{\pi} \hS_a^z\hS_b^z\ {\rm Im} \int_0^{E_F} \left[ \int_{-\infty}^E g_0 h_0 w_{02} dE' \right] dE,
\end{eqnarray}
in which
\begin{equation} \label{w_01}
 w_{01} = \frac{32(p_1(p_2+4)-4p_2)}{[8p_1(3p_2-4)-9p_2^2+8(p_2-2)](4p_1-p_2-4)},
 \end{equation}
\begin{equation} \label{w_02}
 w_{02} = \frac{-16(2p_1^2-4p_1-p_2+4)}{[8p_1(3p_2-4)-9p_2^2+8(p_2-2)](4p_1-p_2-4)},
\end{equation}
and~$h_0$ is given in Eq.~(\ref{h_0}).

First we analyze~$\Omega^{ab}$ term that gives the main contribution
to the range function~${\cal J}(r)$.
For small~$J$, we may expand~$w_{ab}$ in Taylor series. Assuming~$p_2=J^2c_{ab}$,
see Eq.~(\ref{defp2}), one obtains
\begin{eqnarray}
 w_{ab} &\simeq& 1 -J g_0 + \frac{9}{8}J^2 g_0^2 - \frac{17}{16} J^3 g_0^3 + \nonumber \\
 \label{w_ab_t}
 && +\frac{1}{32}J^4 (35g_0^4-11g_0^2c_2-14c_2^2) \ldots.
\end{eqnarray}
One observes from Eqs.~(\ref{Omega_ab})--(\ref{w_ab_t}):
i) By taking~$w_{ab}=1$ in Eq.~(\ref{Omega_ab}) one obtains the range function of the
RKKY interaction, see Appendix~\ref{App_RKKY}. ii) For arbitrary~$w_{ab}$, as given in Eq.~(\ref{w_ab}),
the double integral in Eq.~(\ref{Omega_ab}) may not be calculated analytically,
so calculations are performed numerically, see Section~\ref{Sec_Res}.
iii) Since~$g_0$ does not depend on the
distance~$r$ between localized spins, the second, third and fourth terms in
Eq.~(\ref{w_ab_t}) do not alter the
spatial oscillations of~$\Omega^{ab}$, they only affect its amplitude.
iv) For small~$J$ the difference between the exact and RKKY range functions is
on the order of~$\pm 2p_1 = \pm Jg_0$,
and usually it is on the order of a few percent.
v) The first modification of spatial dependence of~$w_{ab}$
appears in the fourth order of~$J$. This term includes~$c_{ab}$ which depends on~$r$, see Eq.~(\ref{defp2}).
vi) Since~$p_1\propto J$ and~$p_2 \propto J^2$, for~$J \rightarrow \infty$ there is~$w_{ab} \propto J^{-4}$
and~$\Omega^{ab} \propto J^{-2}$, which vanishes for large~$|J|$.
The last result is counter-intuitive since for large values of~$|J|$
one expects no difference between
configurations having parallel and antiparallel localized spins.
This issue can be clarified within our formalism,
see Section~\ref{Sec_Appr} and Appendix~\ref{App_Strong}.

Analyzing Eqs.~(\ref{Omega_01}) and~(\ref{Omega_02}) we consider first the case of small~$J$
and expand~$w_{01}$ and~$w_{02}$
in Eqs.~(\ref{w_01}) and~(\ref{w_02}) in power series of~$J$. One has
\begin{eqnarray}
 w_{01} &\simeq& Jg_0 -J^2(g_0^2+4c_2)/2 + \ldots, \\
 Jg_0 w_{02} &\simeq& - Jg_0 + J^2g_0^2 + \ldots,
\end{eqnarray}
i.e. the terms linear in~$J$ cancel out and one has
\begin{equation}
 \Omega^{01} + \Omega^{02} \simeq -\frac{2J^3}{\pi} \hS_a^z\hS_b^z\ {\rm Im} \int_0^{E_F} dE
 \int_{-\infty}^E h_0(c_2 + \ldots) dE'.
\end{equation}
We conclude: i) the terms~$\Omega^{01} + \Omega^{02}$ are of the third order
in the coupling constant~$J$, while the~$\Omega_{ab}$ term is of the second order in~$J$. ii)
Contrary to~$\Omega_{ab}$, the terms~$\Omega^{01} + \Omega^{02}$ include the product~$g_0 h_0$
which does not depend on~$r$, and for this reason these terms
weakly depend on the distance between spins.
iii) For large~$|J|$ the sun~$\Omega^{01} + \Omega^{02}$ vanishes as~$J^{-2}$, similarly to~$\Omega_{ab}$.
iv) Physically,~$\Omega^{01} + \Omega^{02}$ are generalization of the on-site energies
appearing in the second order of perturbation expansion.
Numerical calculations for~$3D$ range function show that, for reasonable~$r$, the contribution
of~$\Omega^{01} + \Omega^{02}$ to the range function is a few orders of magnitude smaller
than that of~$\Omega^{ab}$ term. Therefore, the impact of~$\Omega^{01} + \Omega^{02}$ terms
on the range function may be neglected.

\section{Approximate form of~$\Omega^{ab}$ in~$3D$ \label{Sec_Appr}}

Now we consider a simplified version of Eq.~(\ref{w_ab}) in which we assume that the one-electron GF
vanishes sufficiently fast with~$r$. This approximation works correctly
for electrons in parabolic energy bands in~$3D$ and~$2D$, see Section~\ref{Sec_OneE}. Let
\begin{equation} \label{defp2_a}
 p_2 = \frac{J^2}{4} \left( g_0^2 - g_{ab} g_{ba} \right) \simeq \frac{J^2 g_0^2}{4} = p_1^2,
\end{equation}
where~$p_1=Jg_0/2$, see Eq.~(\ref{defp1}).
Then from Eqs.~(\ref{Omega_ab}),~(\ref{w_ab}) and~(\ref{defp2_a}) we have
\begin{equation} \label{Omega_ab_p1}
 \Omega^{ab} \simeq \frac{16 J^2}{9\pi} \hS_a^z\hS_b^z\ {\rm Im} \int_0^{E_F} dE \int_{-\infty}^{E}
 \frac{g_{ab} h_{ab}\ dE'}{(p_1-2)^2(p_1+2/3)^2},
\end{equation}
and
\begin{equation} \label{Omega_012_p1}
 \Omega^{01} + \Omega^{02} \simeq 0.
\end{equation}
Equations~(\ref{Omega_ab_p1})--(\ref{Omega_012_p1}) give simple but complete
description of the spin-dependent part of the thermodynamical potential
and the range function~${\cal J}(r)$
in the whole range of model parameters.
First, taking~$p_1 \simeq 0$ one obtains
\begin{equation}
 \Omega^{ab} \simeq \frac{J^2}{\pi} \hS_a^z\hS_b^z\ {\rm Im} \int_0^{E_F} dE \int_{-\infty}^{E}
 g_{ab} h_{ab}\ dE,
\end{equation}
i.e. the thermodynamical potential and the range function for the RKKY interaction, see Appendix~\ref{App_RKKY}.
Next, for~$0 \le E \le E_F$ the quantity~$g_0$ entering~$p_1$
is a non-oscillating slowly varying function of energy.
Thus for~$p_1 \ll 2$ and~$-p_1 \ll 2/3$
the denominators in Eq.~(\ref{Omega_ab_p1}) are also slowly varying functions of energy.
These terms modify the amplitude of the range function
but not its oscillations. For large~$|p_1|$ and~$|J|$ the
denominators in Eq.~(\ref{Omega_ab_p1}) diminish the amplitude of range function and introduce an
additional phase shift to the oscillations. For very large~$|J|$ the range function vanishes as~$|J|^{-2}$,
as found previously. Finally, in the simplified model the one-site interactions do not give
any contribution to the range function in full analogy to the RKKY case.

The quantity~$g_0$ is a complex number:~$g_0= g_0^{R} + ig_0^{I}$. Usually,
the real part of~$g_0$ slowly varies with~$E$, while~$g_0^{I}$
is proportional to the density of states of the system.
For two values of~$J$ and appropriate energies there is:~$Jg_0^{R}/2 \simeq 2$
or~$Jg_0^{R}/2 \simeq -2/3$,
and the real part of~$(p_1-2)$ or~$(p_1+2/3)$ vanishes.
Then, one of the denominators in Eq.~(\ref{Omega_ab_p1}) becomes large,
especially for low energies. In this case one may expect a significant enhancement of~$\Omega^{ab}$
and consequently the range function~${\cal J}(r)$. This effect is
quite general, but its magnitude depends on one-electron GF
in the considered system.

The singular points of the integrand in Eq.~(\ref{Omega_ab_p1}) appear for~$p_1=2$ or~$(p_1=-2/3$ and the
vanishing imaginary part of~$g_0$. In~$3D$ this occurs for energies~$E\leq 0$,
since the density of states vanishes at or below the edge of the conduction band.
For a specific combination of parameters one may expect the presence
of localized states with discrete energies. This issue is discussed in Section~\ref{Sec_OneE}.
Note that for the general case of Eq.~(\ref{Omega_ab}) the singularities appear not
exactly at~$p_1=2$ or~$p_1=-2/3$, but in the vicinity of these points because of the
more complicated form of~$p_2$, see Eq.~(\ref{defp2}).

The above considerations suggest existence of three different regimes of parameters
in considered model.
For small coupling constants~$J$ the exact range function resembles the RKKY one,
with slightly altered amplitude but unchanged oscillation period.
For parameters meeting the conditions~$p_1 \simeq 2$ or~$p_1 \simeq -2/3$
the thermodynamical potential~$\Omega^{ab}$ and the range function~${\cal J}(r)$
are qualitatively different from the RKKY case
and discrete energy states appear. The third regime occurs for large values of~$|J|$
or~$|g_0|$. In this case the thermodynamic potential~$\Omega^{ab}$ and the range functions
resemble RKKY ones, but with additional phase shift in oscillations and much lower
amplitude vanish with increasing~$|J|$ or~$|g_0|$.
Numerical results in Section~\ref{Sec_Res} confirm the above predictions.

\subsection{Origin of model peculiarities}

The approximations in Eqs.~(\ref{defp2_a}) and~(\ref{Omega_ab_p1})
allow us to understand three peculiar features
of the exact GF, namely i) the asymmetry between positive (anti-ferromagnetic)
and negative (ferromagnetic) signs
of the coupling constant~$J$, ii) existence of two singularities for~$Jg_0/2 \in \{2, -2/3\}$,
and iii) disappearance of the range function for large~$|J|$ values.
Below we present the main steps in re-derivation of the density of states
entering the integrand of Eq.~(\ref{Omega_ab_p1}) in the approximate model
and explain the mathematical and physical origins of the peculiarities.

Consistently with the approximation given in Eq.~(\ref{defp2_a}) we neglect
in Eqs.~(\ref{hbmF1}) and~(\ref{hbmF2}) terms including products of~$g_{ab}g_{ba}$.
Then from Eqs.~(\ref{hbmF1}) and~(\ref{hbmF2}) one obtains
\begin{eqnarray}
 \label{DefKa}
 \hbmQ_1 \simeq \left(\hI- g_0\hbmZ_a \right)^{-1} \left(\hI-g_0\hbmZ_b\right)^{-1} \equiv \hbmK_a \hbmK_b, \\
 \label{DefKb}
 \hbmQ_2 \simeq \left(\hI- g_0\hbmZ_b \right)^{-1} \left(\hI-g_0\hbmZ_a\right)^{-1} \equiv \hbmK_b \hbmK_a.
\end{eqnarray}
In Eqs.~(\ref{DefKa})--(\ref{DefKb}) the quantities~$\hbmK_a, \hbmK_b$ are~$2\times 2$ matrices, whose
elements are combinations of~$\hS_a$ and~$\hS_b$ spin components, see below.
For finite and nonzero~$g_0$ we have
\begin{equation}
 \left(\hI- g_0\hbmZ_c \right)^{-1} \hbmZ_c = \frac{1}{g_0} (\hbmK_c -\hI),
\end{equation}
where~$c=a,b$. Note that~$\left(\hI- g_0\hbmZ_c \right)$ commutes with~$\hbmZ_c$.
From Eq.~(\ref{hbmG_12_d}) we have then
\begin{eqnarray} \label{hbmG_12_k}
 \hbmG_{12} &\simeq & g_{12}\hI + \frac{g_{1a}}{g_0}(\hbmK_a- \hI) g_{a2}
                + \frac{g_{1b}}{g_0}[\hbmK_b -\hI]g_{b2} \nonumber \\
            &+& \frac{g_{1a}}{g_0^2}[g_{ab}(\hbmK_a -\hI) (\hbmK_b - \hI)]g_{b2} \ \  \nonumber \\
            &+& \frac{g_{1b}}{g_0^2}[g_{ba}(\hbmK_b -\hI)(\hbmK_a-\hI)]g_{a2}. \ \ \
\end{eqnarray}
The first observation from Eqs.~(\ref{DefKa}),~(\ref{DefKa}), and~(\ref{hbmG_12_k}) is that, for large~$|J|$, the
operators~$\hbmK_a, \hbmK_b$ tend to zero and in this limit~$\hbmG_{12}$ in Eq.~(\ref{hbmG_12_k}) does
not depend on~$\hS_a$ and~$\hS_b$. In consequence, the thermodynamic potential does not depend
on spin configuration, so that the range function~${\cal J}(r)$ in Eq.~(\ref{defJ}) vanishes.
The derivation of this result for the general case is shown in Appendix~\ref{App_Strong}.

The next conclusion from Eq.~(\ref{hbmG_12_k}) is that, in the approximate model, the one-site parts of
the exact GF, given by the two last terms of first line in Eq.~(\ref{hbmG_12_k}), do not depend on
the inter-spin distance~$r$. This observation suggests, that
also in the general model discussed in the previous sections, these terms are negligible.

The density of states is proportional to the trace of~$\hbmG_{12}$. Let
\begin{equation} \label{DefKc}
 \hbmK_c = \left(\begin{array}{cc} \hk^{cA} & \hk^{cB} \\ \hk^{cC} & \hk^{cD} \end{array} \right),
\end{equation}
with~$c=a,b$.
Using the notation from Section~\ref{Sec_GFSpinOp}
find:~${\rm Tr} \{\hG \} = {\rm Tr} \{\hG_{ab} \} + {\rm Tr} \{\hG_{ba}\}$, where
\begin{widetext} \begin{eqnarray}
\label{TrK_ab}
 {\rm Tr} \{ \hG_{ab} \} = \frac{g_{ab} h_{ab}}{g_0^2}
 \left( \hk^{aA}\hk^{bA} +\hk^{aB}\hk^{bC} +\hk^{aC}\hk^{bB} +\hk^{aD}\hk^{bD} -\hk^{aA} -\hk^{aD} -\hk^{bA} -\hk^{bD} +2{\hat I} \right),\\
\label{TrK_ba}
 {\rm Tr} \{ \hG_{ba} \} = \frac{g_{ab} h_{ab}}{g_0^2}
 \left( \hk^{bA}\hk^{aA} +\hk^{bB}\hk^{aC} +\hk^{bC}\hk^{aB} +\hk^{bD}\hk^{aD} -\hk^{bA} -\hk^{bD} -\hk^{aA} -\hk^{aD} +2{\hat I} \right).
\end{eqnarray} \end{widetext}
Equation~(\ref{TrK_ab}) corresponds to the sum~$(\hG^{ab})_{11} + (\hG^{ab})_{22}$
in Eqs.~(\ref{hGaa11})--(\ref{hGbb22}), while Eq.~(\ref{TrK_ba}) corresponds
to the sum~$(\hG^{ba})_{11} + (\hG^{ba})_{22}$. The trace of GF obtained in Eqs.~(\ref{TrK_ab})--(\ref{TrK_ba})
is simpler than that in Eqs.~(\ref{hGaa11})--(\ref{hGbb22}).
Using the Woodbury identities in Eq.~(\ref{Woodbury}) and definition of~$\hbmK_c$
in Eqs.~(\ref{DefKa})--(\ref{DefKb}) one obtains
\begin{eqnarray} \label{hkaA}
 \hk^{aA} &=& [(\hI - p_1 \hS_a^z) - p_1^2 \hS_a^- (\hI + p_1 \hS_a^z)^{-1} \hS_a^+ ]^{-1}, \\
 \hk^{aD} &=& [(\hI + p_1 \hS_a^z) - p_1^2 \hS_a^+ (\hI - p_1 \hS_a^z)^{-1} \hS_a^- ]^{-1}, \\
 \hk^{aB} &=& p_1 \hk^{aD} \hS_a^- (\hI + p_1 \hS_a^z)^{-1}, \\
 \label{hkaC}
 \hk^{aC} &=& p_1 \hk^{aA} \hS_a^+ (\hI - p_1 \hS_a^z)^{-1},
\end{eqnarray}
and similarly for~$\hk^{b\alpha}$ with~$\alpha= A, B, C, D$.
For the spins~$\hS_a,\hS_b = 1/2$ the operators~$\hS_a^{\pm},\hS_a^z, \hS_b^{\pm},\hS_b^z$
are~$4\times 4$ matrices,
see Eqs.~(\ref{S05Sap})--(\ref{S05Sbp}). Then the operators~$\hk^{a\alpha}$ and~$\hk^{b\alpha}$
are also~$4\times 4$ matrices that can be calculated from Eqs.~(\ref{kaA11})--(\ref{kbB21}) in
Supplemental material. The matrix corresponding to~$\hk^{aA}$ operator is diagonal
\begin{equation} \label{hkaA_m}
 \hk^{aA} = {\rm diag}\left(t_{2},\ t_{2},\ t_{2} + r_{-2/3},\ t_{2} + t_{-2/3} \right),
\end{equation}
with~$t_{2}= 1/(2-p_1)$ and~$t_{-2/3}= 1/(3p_1+2)$.
The matrix in Eq.~(\ref{hkaA_m}) and the remaining matrices~$\hk^{a\alpha}$ and~$\hk^{b\alpha}$
have singularities for~$p_1 \in \{2, -2/3\}$, i.e., for the same~$p_1$ values as
the singularities of the thermodynamical potential in Eq.~(\ref{Omega_ab_p1}).
Thus, singularities of the exact GF appear when
the operators~$(\hI - g_0 \hbmZ_a)$ or~$(\hI - g_0 \hbmZ_b)$ may not be inverted.
For~$\hS_a,\hS_b = 1/2$ this occurs for two~$p_1$ values:~$p_1 =2$ or~$p_1= -2/3$.
Since~$p_1 = J g_0/2$, the non-reversibility of~$(\hI - g_0 \hbmZ_a)$ and~$(\hI - g_0 \hbmZ_b)$ operators
breaks the symmetry between positive (anti-ferromagnetic)
and negative (ferromagnetic) values of~$J$. This effect does not exist for the GF of the RKKY range function
since the latter depends on~$J^2$ and it is symmetric respect to positive or negative~$J$ values.

Having calculated matrices~$\hk^{a\alpha}$ and~$\hk^{b\alpha}$ the trace
in Eq.~(\ref{TrK_ab}) is
\begin{eqnarray} \label{TrK_ab_m}
{\rm Tr} \{\hG_{ab} \} = \left( \begin{array}{cccc}
 x_{11} & 0 & 0 & 0 \\ 0 & x_{22} & x_{23} & 0 \\
 0 & x_{32} & x_{33} & 0 \\ 0 & 0 & 0 & x_{44}\\ \end{array} \right) = \ \ \ \ \ \ \ \\
 \equiv e_1 \hI + e_2 \hS_a^z + e_3 \hS_b^z + e_4 \hS_a^z \hS_b^z + e_5 \hS_a^+ \hS_b^- + e_6 \hS_a^- \hS_b^+, \nonumber
\end{eqnarray}
where~$e_i$ are the coefficients to determinate and the~$x_{ij}$ are
listed in Eqs.~(\ref{xx11})--(\ref{xx23}) in Supplemental material.
The range function is defined as a coefficient~$e_4$
in front of~$\hS_a^z \hS_b^z$ see Eq.~(\ref{Omega_SzSz1}).
After some algebra we find~$e_4 = x_{11}-x_{22}-x_{33}+x_{44}$, which gives
\begin{equation} \label{TrK_SzSz}
 {\rm Tr} \{\hG_{ab}^{\hS_a^z \hS_b^z}\} = \left(\frac{8 J^2}{9\pi}\right) \frac{g_{ab}h_{ab}}{ (p_1-2)^2(p_1+2/3)^2}.
\end{equation}
Since~${\rm Tr} \{\hG_{ba}^{\hS_a^z \hS_b^z}\} = {\rm Tr} \{\hG_{ab}^{\hS_a^z \hS_b^z}\}$ one
finally obtains:~${\rm Tr} \{\hG^{\hS_a^z \hS_b^z}\} = 2{\rm Tr} \{\hG_{ab}^{\hS_a^z \hS_b^z}\}$,
i.e. the integrand in Eq.~(\ref{Omega_ab_p1}). On expanding it around~$p_1=0$ we find
\begin{equation}
 {\rm Tr} \{\hG^{\hS_a^z \hS_b^z}\} \simeq \frac{g_{ab}h_{ab}}{\pi} ( 1 - Jg_0 + \ldots ),
\end{equation}
i.e the same expansion as in Eq.~(\ref{w_ab_t}). This confirms the accuracy of the simplified form of
thermodynamical potential in Eq.~(\ref{Omega_ab_p1}).

\section{One-electron Green's function \label{Sec_OneE}}

The results for GF in Eqs.~(\ref{hGgen})--(\ref{hGbb22})
and~(\ref{hG_aa_11})--(\ref{hG_bb_22}) in Supplemental material
are valid for one-electron GF having arbitrary energy band dispersion
but a finite value of~$g_0$, see Eq.~(\ref{defg0}). We consider
electrons in the effective mass approximation in a parabolic energy band.
The use of such GF allows us to compare the range function obtained from
the exact GF with that obtained in the RKKY model.

\subsection{Parabolic energy bands}

Taking the Bloch states~$|\bm k\rangle$ in the form of plane waves
the one-electron GF in the effective mass approximation is
\begin{equation} \label{gpar}
 g_{ab} = \langle {\bm r}_a | \hg |{\bm r}_b \rangle =
 \frac{1}{(2\pi)^D} \int \frac{e^{i{\bm k}({\bm r}_a -{\bm r}_b)}}
 {E - \epsilon(k)} d^D {\bm k},
\end{equation}
\begin{equation} \label{epsilon_k}
 \epsilon(k) = \frac{\hbar^2 k^2}{2m^*} \equiv \zeta k^2,
\end{equation}
Here~$D$ is system's dimensionality,~$m^*$ is the electron effective mass and~$\zeta = \hbar^2/(2m^*)$.
For~$T=0$ the energy~$E$ is a real number with a small imaginary part.

For~$3D$ systems one has~\cite{EconomouBook}
\begin{equation} \label{hg_3D}
 g^{\pm}_{ab} = - \frac{ e^{\pm ik_0|{\bm r}_a-{\bm r}_b|}}{4\pi |{\bm r}_a-{\bm r}_b|\zeta}
 \equiv - \frac{e^{ik_0 r}}{4\pi r_{ab} \zeta},
\end{equation}
where~$k_0=\sqrt{|E|/\zeta}>0$,~$\Re(E) > 0$,~$r=|{\bm r}_a-{\bm r}_b|$,
and~$\pm$ signs correspond to the retarded and advanced
Green's function, respectively. From Eqs.~(\ref{h_ab}) and~(\ref{h_0}) one obtains
\begin{eqnarray}
 h_{ab}^+ &=&\frac{i e^{ik_0 r}}{8\pi k_0}, \\
 h_0^+ &=&\frac{i}{8\pi k_0}.
\end{eqnarray}

For~$2D$ systems~\cite{EconomouBook}
\begin{equation} \label{hg_2D}
 g^{\pm}_{ab} = -\frac{i}{4\pi \zeta} H_0(\pm k_0 r),
\end{equation}
where~$H_0(x)$ is the zeroth order Hankel function of the first kind.
For the~$1D$ systems~\cite{EconomouBook}
\begin{equation} \label{hg_1D}
 g^{\pm}_{ab} = \mp \frac{i}{2k_0\zeta} e^{\pm ik_0 x_{ab}}.
\end{equation}

As seen from Eqs.~(\ref{hg_3D})--(\ref{hg_1D}), the one-electron GF
at the origin~$g_0=g_{11}=g_{22}$
diverges in~$D=3$ and~$D=2$. In~$1D$ there is
\begin{equation} \label{hg0_1D}
 g^{\pm}_0 = \mp \frac{i}{2k_0\zeta},
\end{equation}
which is finite for~$k_0 \neq 0$. These results conclude
the issue of convergence of the perturbation series in the RKKY problem.
As follows from the above consideration, the latter stated in it's basic form
leads to divergent perturbation series for~$3D$ and~$2D$ systems.

\subsection{Cut-off energy}

There exist several effects in real materials which may eliminate divergence of~$g_0$.
Here we consider one of these effects,
i.e., a non-parabolicity of the energy band for large wave vectors. As seen in Eq.~(\ref{hg_3D}), the
singularity of one-electron GF at~$r=0$ arises from the divergence in the integral in Eq.~(\ref{gpar})
for large~$k$, while for real materials the parabolic band dispersion
is justified only for small~$k$. For~$k$ exceeding, roughly,
half of the first Brillouin zone, the curvatures
of energy bands change their signs and the parabolic model fails.

To overcome the problem of divergence of~$g_0$ for large~$k$ values
we follow method described in Refs.~\cite{Wolff1961,Clogston1962,ZimanBook}.
For~${\bm r} \neq 0$ we use the one-electron GF given
in Eq.~(\ref{gpar}), while for~${\bm r} = 0$ we take the GF in
the energy representation
\begin{eqnarray} \label{hg0_3Di}
 g_0^+ &=& \int_0^{\infty} \frac{n(E')}{E-E'+i\eta} dE' \nonumber \\
       &=& \int_0^{\infty}\! \!{\cal P} \frac{n(E')}{E-E'} dE' - i\pi\int_0^{\infty}\!\!\! n(E') \delta(E-E') dE', \ \ \ \ \
\end{eqnarray}
where~$n(E) \propto \sqrt{E} \Theta(E)$ is the density of states in~$3D$,~$\Theta(E)$
is the step function and~${\cal P}$ is the principal value of the integral.
For large energies the real part of~$g_0^+$ in Eq.~(\ref{hg0_3Di}) diverges.
To remove this divergence we introduce a cut-off energy~$E_m \gg E_F$ that ensures convergence
of the integrals in Eq.~(\ref{hg0_3Di}). We treat~$E_m$ as a model parameter. Similar
approach of dealing with divergence of the one-electron GF
was proposed in Ref.~\cite{Wiertz1976}. The density of states is then
\begin{equation}
 n(E) = \frac{1}{2\pi^2 \zeta^{3/2}}\sqrt{E}\ \Theta(E) \Theta(E_m-E).
\end{equation}
For~$E \geq 0$
\begin{eqnarray} \label{hg0_3Dp}
 g_0^+ &=& \frac{1}{2\pi^2 \zeta^{3/2}}
 \left[ -\sqrt{E}\ln\left(\frac{\sqrt{E_m}-\sqrt{E}}{\sqrt{E_m}+\sqrt{E}} \right)
 -2\sqrt{E_m} \right] + \nonumber \\
 && - \frac{i}{2\pi \zeta^{3/2}} \sqrt{E} \Theta(E_m-E),
\end{eqnarray}
while for~$E < 0$ there is
\begin{equation} \label{hg0_3Dm}
 g_0^+ = \frac{1}{2\pi^2 \zeta^{3/2}} \left[2 \sqrt{|E|} \arctan\left(\sqrt{\frac{E_m}{|E|}}\right) - 2\sqrt{E_m}\right],
\end{equation}
since~$n(E)$ is zero for~$E < 0$. For~$E \ll E_m$ the real part of~$g_0^+$ is
\begin{equation} \label{Rg0_3Dpa}
 {\rm Re}(g_0^+) \simeq \frac{1}{2\pi^2 \zeta^{3/2}} \left(-2\sqrt{E_m} + \frac{E}{2E_m} \right)
 \ \textrm {for}\ E \geq 0,
\end{equation}
\begin{equation} \label{Rg0_3Dma}
 {\rm Re}(g_0^+) \simeq \frac{1}{2\pi^2 \zeta^{3/2}} \left(-2\sqrt{E_m} + \pi \sqrt{|E|}-\frac{|E|}{2E_m}\right)
 \ \textrm {for}\ E < 0.
\end{equation}
For~$E >0$ the quantity~$g_0^+$ is complex while for~$E \leq 0$ it is real.
We choose~$E_m$ as the energy at~$k_m= \pi/a$, where~$a$ is the lattice constant.
For many lattices as, e.g. for the fcc lattice in the~$\Gamma X$ direction of~${\bm k}$,
the value of~$k_m$ corresponds to half of the Brillouin zone. Then
\begin{equation} \label{def_Em}
 E_m = \frac{\hbar^2 \pi^2}{2m^*a^2}.
\end{equation}
In~$2D$ systems the real part of~$g_0^+$ diverges as~$\ln(E_m)$ and the results
depend only weakly on~$E_m$.

For~$E=0$ it is possible to adjust~$J$,~$m^*$ and~$E_m$ in such a way
that~$Jg_0^+/2 \in \{2,-2/3\}$. In the vicinities of these two points the integral
in Eq.~(\ref{Omega_ab_p1}) has two singularities.
Using Eq.~(\ref{def_Em}) and~$\zeta = \hbar^2/(2m^*)$ we find that
the two singularities appear for~$p_1=p_1^s$, where
\begin{equation} \label{defp1_s}
 p_1^s = -\frac{1}{2}\frac{J\sqrt{E_m}}{\pi^2 \zeta^{3/2}} =
 -\frac{J m^*}{\pi \hbar^2 a} \in \{2, -2/3\}.
\end{equation}
The singularity~$p_1=2$ occurs for negative values of~$J$, i.e., for ferromagnetic
coupling between conduction electrons and atomic~$d$ states. The singularity~$p_1=-2/3$ occurs
for positive values of~$J$, i.e. for anti-ferromagnetic~$s-d$ coupling.
The two values of~$p_1^s$ indicate borders between three regimes of the model
parameters. Their positions depend on electron effective mass, elementary cell volume, lattice constant,
and~$s-d$ coupling constant. The two latter parameters do not change significantly between various compounds,
but the effective mass may vary more than two orders of the magnitude. For narrow gap
semiconductors such as InSb the effective mass can be below~$0.1 m_e$, while for some materials, e.g.
Sr$_{1-x}$La$_x$Ti$O_{3-y}$, it can exceed~$10 m_e$. In many compounds it possible to change~$m^*$
by changing electron concentration or by applying external pressure.
This may give a practical way of modifying~$p_1^s$ in Eq.~(\ref{defp1_s}).

\subsection{Discrete energy levels}

Discrete energy levels of a system are obtained from poles of~$\hbmG_{12}$ function~\cite{EconomouBook}.
For the exact GF given in Eqs.~(\ref{hG_aa_11})--(\ref{hG_bb_22}) in Supplemental material
the poles of GF are obtained from two alternative equations
\begin{eqnarray} \label{Zero_Gen1}
 4p_1-p_2-4 &=&0, \\
 32p_1^2(3p_2-4)-4p_1(15p_2^2+8p_2-16)+ \nonumber \\
 +(p_2+4)(9p_2^2-8p_2+16)&=&0. \label{Zero_Gen2}
\end{eqnarray}
These equations are difficult to analyze and they can be solved only numerically.
However, in~$3D$ and~$2D$ systems we may approximate~$p_2 \simeq p_1^2$ [see Eq.~(\ref{defp2_a})],
and obtain instead of Eqs.~(\ref{Zero_Gen1}) and~(\ref{Zero_Gen2})
the condition:~$(p_1-2)(p_1+2/3)=0$, which gives
\begin{equation} \label{Zero_Appr}
 \frac{Jg_0^+}{2}=2 \hspace{1em} {\rm or} \hspace{1em} \frac{Jg_0^+}{2}= -\frac{2}{3}.
\end{equation}
For~$E>0$ and~$E < E_m$ the conditions in Eq.~(\ref{Zero_Appr}) can not be satisfied.
However, for~$E \leq 0$ (i.e., below the conduction band edge)
the imaginary part of~$g_0^+$ vanishes and conditions in Eq.~(\ref{Zero_Appr}) may be
satisfied for some combination of parameters entering to the model.
Since we are interested in low-energy states,
we use the approximate form of~$g_0^+$ in Eq.~(\ref{Rg0_3Dma}).
From~(\ref{Zero_Appr}) we have
\begin{equation} \label{ZeroJ}
 \frac{J}{4\pi^2\zeta^{3/2}} \left(-2\sqrt{E_m} + \pi \sqrt{|E|} \right) = A,
\end{equation}
where~$A \in \{2, -2/3 \}$. It is convenient to introduce
\begin{equation} \label{ZeroJA}
 J^{\{A\}} = - \frac{2\pi^2 \zeta^{3/2} A}{\sqrt{E_m}}.
\end{equation}
and~$\delta J = J- J^{\{A\}}$. Assuming~$\delta J \ll J^{\{A\}}$
one obtains from Eq.~(\ref{ZeroJ})
\begin{equation} \label{ZeroE}
 \sqrt{|E|} =\frac{4\pi A \zeta^{3/2}(J^{\{A\}}-J)}{J^{\{A\}} J}
 \simeq -\frac{4\pi A \zeta^{3/2}(\delta J)}{(J^{\{A\}})^2}.
\end{equation}
The LHS of Eq.~(\ref{ZeroE}) is non-negative, which gives:~$-A(\delta J) \geq 0$.
For~$A=2$ one obtains:~$(\delta J) < 0$.
Since the singularity~$A=2$ corresponds to~$J < 0$, see the discussion after Eq.~(\ref{defp1_s}),
the bound states exist for~$J \leq J^{\{2\}}$. For~$A=-2/3$ there is:~$(\delta J) > 0$,
and the bound states exist for~$J \geq J^{\{-2/3\}}$. In both cases, the energies
of bound states appear for small values of~$\delta J$
in the vicinities of points~$p_1 \in \{2, -2/3 \}$.

\section{Numerical results \label{Sec_Res}}

Here we compare the range function~${\cal J}(r)$ of the standard RKKY
interaction with that obtained in Eq.~(\ref{defJ}) with use of the exact
GF and Eqs.~(\ref{Omega_SzSz1})--(\ref{w_02}).
We restrict the analysis to the~$3D$ case. The definite and indefinite
integrals in Eqs.~(\ref{Omega_ab})--(\ref{Omega_02})
are calculated by the Simpson method. To avoid singularities arising from~$E=0$
it is convenient to change the variable of integration~$E \rightarrow q^2$.
The model considered in this work depends on five parameters:~$s-d$ coupling constant~$J$,
values of localized spins~$\hS_a,~\hS_b$, electron effective mass~$m^*$,
the Fermi energy~$E_F = \hbar^2 k_F^2/(2m^*)$ and the
cut-off energy~$E_m$. In~$3D$ case the Fermi wave vector is
\begin{equation}
 k_F = \left( 3\pi^2 n_e \right)^{1 /3}.
\end{equation}

\begin{table}
\caption{Material parameters for Zn$_{1-x}$Mn$_x$Se used in calculations~\cite{Furdyna1998,Daniel2005}.
         Note the sign convention in Eq.~(\ref{Vr}) and~$\hS_a = \hS_b = 1/2$
         instead of~$\hS_a = \hS_b = 5/2$.}
\begin{tabular}{|l|c|c|c|}
\hline
parameter & symbol & value & unit\\
 \hline
  Localized spins value  & $\hS_a$, $\hS_b$ & $1/2$   & n.a.         \\
  $s-d$ coupling constant& $J$              & -11.85  & eV$\AA^3$    \\
  lattice constant       & $a$              & 5.67    & \AA          \\
  effective mass         & $m^*$            & 0.13    & $m_0$        \\
  electron concentration & $n_e$            & $6.0 \times 10^{1}9$ & cm$^{-3}$ \\
  cut-off energy         & $E_m$            & 8.99    & eV          \\
 \hline
  elementary cell volume & $\Omega_0$       & 45.57   & $\AA^3$     \\
  $s-d$ coupling energy  & $\alpha N_0=-J/\Omega_0$   & 0.26    & eV \\
  Fermi vector           & $k_F$            & 0.12    & $\AA^{-1}$  \\
  Fermi energy           & $E_F$            & 0.43    & eV          \\
  parameter $p_1^s$      & $p_1^s$          & 1.13\%  & n.a.        \\
\hline
\end{tabular} \label{Table1}
\end{table}

\begin{figure}
\includegraphics[width=8cm,height=8cm]{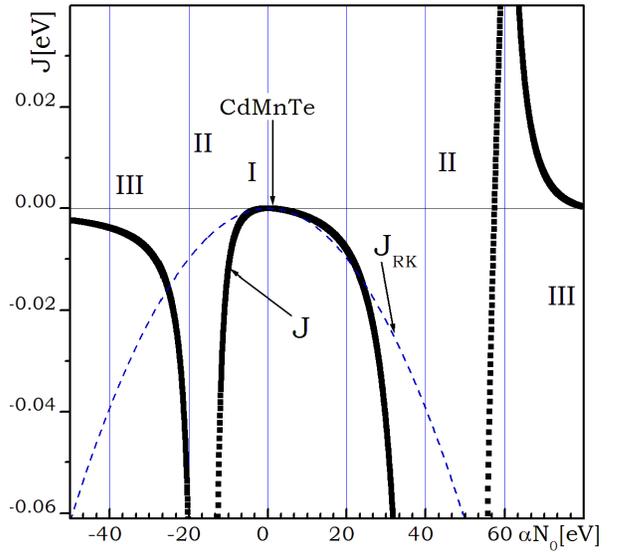}
\caption{Solid lines: Amplitude of the range function~${\cal J}(r)$ calculated
         from Eqs.~(\ref{defJ}) and~(\ref{Omega_SzSz1})--(\ref{w_02})
         versus~$\alpha N_0$ for NN cations distance~$r=4.01$\AA.
         The remaining model parameters are
         listed in Table~\ref{Table1}. Three regimes of the model are marked.
         Dashed line: Amplitude of the range function~${\cal J}_{RK}(r)$
         from Eq.~(\ref{JRK}) versus~$\alpha N_0$ for~$r=4.01$\AA.  } \label{Fig1}
\end{figure}

In Table~\ref{Table1} we list parameters corresponding to ZnMn$_x$Se$_{1-x}$,
but with~$\hS_a = \hS_b = 1/2$ instead of~$\hS_a = \hS_b = 5/2$~\cite{Furdyna1998,Daniel2005}.
These parameters are used in calculations shown in Figures~\ref{Fig1},~\ref{Fig3}
and in Table~\ref{Table2}.

In Figure~\ref{Fig1} we plot values of the range function~${\cal J}$ for NN cations
versus the coupling energy~$\alpha N_0 = -J/\Omega_0$, where~$\Omega_0$ is the elementary cell volume.
Note the sign convention in Eq.~(\ref{Vr}). The remaining material parameters
are taken from Table~\ref{Table1}. The range function~${\cal J}_{RK} \propto (\alpha N_0)^2$
is also indicated. This figure illustrates three regimes of model parameters discussed
qualitatively in Section~\ref{Sec_Grand}. The two extremes of the range function are located in the vicinities
of~$\alpha N_0 =-15.28$~eV, which corresponds to~$p_1^s = -2/3$ [see Eqs.~(\ref{defp1_s}) and~(\ref{Omega_ab_p1})]
and~$\alpha N_0 = 45.83$~eV corresponding to~$p_1^s = 2$. Both values of~$\alpha N_0$ are more than two
orders of magnitude larger than the experimental~$s-d$ coupling constant in Zn$_{1-x}$Mn$_x$Se, see Table~\ref{Table1}.

\begin{table}
\caption{Values of RKKY range function~${\cal J}_{RK}$ given in Eq.~(\ref{JRK}) and exact range function
         calculated using Eqs.~(\ref{defJ}) and~(\ref{Omega_SzSz1})--(\ref{w_02})
         for Zn$_{1-x}$Mn$_x$Se for several nearest neighbors distances.
         Material parameters are given in Table~\ref{Table1}.
         Magnitudes of range functions are in~$\mu eV$.
         The~$\pm$ signs indicate positive (anti-ferromagnetic) or negative (ferromagnetic)
         sign of the~$s-d$ coupling constant~$J$.
         In the last column:~$\Delta^{\pm} = [{\cal J}^{\pm} - {\cal J}_{RK}]/{\cal J}_{RK}$ is
         the relative change of the exact range function. From Eq.~(\ref{w_ab_t}) and Table~\ref{Table1}
         they should be~$|\Delta^{\pm}| = 2|p_1| \simeq 2.3\%$.}
\begin{tabular}{|c|c|c|c|c|c|}
\hline
 $r$ [\AA] & ${\cal J}(r)_{RK}$ & ${\cal J}(r)^-$ & $\Delta^-$ \\
 \hline
 4.0 & -2.599 & -2.647 & 1.8\% \\
 5.7 & -1.649 & -1.681 & 1.9\% \\
 8.0 & -0.960 & -0.979 & 2.0\% \\
11.3 & -0.431 & -0.439 & 1.9\% \\
11.4 & -0.420 & -0.428 & 1.9\% \\
\hline
 $r$ [\AA] & ${\cal J}(r)_{RK}$ & ${\cal J}(r)^+$ & $\Delta^+$ \\
 \hline
 4.0 & -2.599 & -2.551 & -1.8\% \\
 5.7 & -1.649 & -1.617 & -2.0\% \\
 8.0 & -0.960 & -0.940 & -2.0\% \\
11.3 & -0.431 & -0.422 & -2.1\% \\
11.4 & -0.420 & -0.411 & -2.1\% \\
\hline
\end{tabular} \label{Table2}
\end{table}

In Table~\ref{Table2} we compare the range functions calculated for several inter-spin distances~$r$
using the exact GF with that obtained within RKKY formalism [see Eq.~(\ref{JRK})]
for Zn$_{1-x}$Mn$_x$Se taking parameters from Table~\ref{Table1} with two signs of~$\alpha N_0$.
The parameters correspond to regime I of the model that is most common in nature.
The distance~$r=4.01\AA$ is the nearest neighbor distance of Mn cations in the lattice.
In our example~$p_1^s \simeq 1.13\%$. As follows from Eqs.~(\ref{Omega_ab}) and~(\ref{w_ab_t}),
for small~$p_1^s$ the difference between exact and RKKY range functions should be on
the order of~$|Jg_0| \simeq |2p_1| \simeq 2.3\%$.
Numbers shown in Table~\ref{Table2} confirm this expectation.
The exact and approximate functions oscillate with similar period~$\pi/k_F$ and similar amplitudes.
This result explains the efficiency and accuracy of the RKKY range function since for inter-spins
distances larger than~$r \ge 4\AA$ both models predict the same
ordering of localized spins.

It follows from Eq.~(\ref{defp1_s}) that the regime III of the model occurs for large values
of effective mass or large magnitude of the~$s-d$ coupling~$J$.
As an example of material in which the regime III may occur is thin film of
Sr$_{1-x}$La$_x$TiO$_{3-\delta}$ doped with magnetic ions.
This compound is one of perovskite-type transition-metal oxides in
which the dispersion of electrons is parabolic with a large effective mass~\cite{Benthema2001}.
As shown in Ref.~\cite{Ravichandran2011}, by varying concentration of La atoms it is possible
to change simultaneously the electron effective mass and carrier concentration.
In our example it is assumed that a thin film of Sr$_{1-x}$La$_x$TiO$_{3-\delta}$
is doped with magnetic atoms having spin~$\hS=1/2$. We take the ferromagnetic
coupling constant between conduction electrons and
that of the magnetic impurity~$J=-15.48$eV$\AA^3$.
This corresponds to~$\alpha N_0=0.26$eV, i.e. to the experimental value for Zn$_{1-x}$Mn$_x$Se.
Since the conduction band in Sr$_{1-x}$La$_x$TiO$_{3-\delta}$
is created mostly from the Ti~$3d_{t2g}$ states, the parameter~$J$ may not be
interpreted as the~$s-d$ coupling constant but as~$3d-f$ or~$3d-nd$ couplings.
As follows from Ref.~\cite{Brooks1991,Brooks1997},
for rare-earth atoms the exchange integrals are ferromagnetic with
magnitudes of~$J_{4f-5d} = -J/(2 \Omega_0)$~\cite{OtherDef} on the order of~$180-140$~meV
depending on the number of electrons in the~$4f$ shell,
but other hybridization mechanisms lead to larger values of~$J$.

\begin{table}
\caption{Values of RKKY range function~${\cal J}_{RK}$ given in Eq.~(\ref{JRK}) and exact range function
         calculated using Eq.~(\ref{defJ}) and Eqs.~(\ref{Omega_SzSz1})--(\ref{w_02})
         for nearest neighbor magnetic impurities in Sr$_{1-x}$La$_x$TiO$_{3-\delta}$.
         The inter-spin distance is~$r=3.905\AA$.
         Concentrations and effective masses are taken from Ref.~\cite{Ravichandran2011},
         localized spins are~$\hS_a =\hS_b = 1/2$ and the coupling constant
         between conduction and magnetic impurity electrons is ferromagnetic~$J=-15.48$eV$\AA^3$.}
\begin{tabular}{|c|c|c|c|c|c|}
\hline
 X & $n_e$ [cm$^{-3}$] & $m^*/m_0$ & $p_1^s$ & ${\cal J}_{RK}$ [eV] & ${\cal J}$ [eV] \\
 \hline
 A & 1.9$\times 10^{20}$ & 5.6  & 0.93	& -5.60$\times 10^{-4}$ & -4.29 $\times 10^{-4}$ \\
 B & 3.1$\times 10^{21}$ & 6.0  & 0.99	& -2.39$\times 10^{-3}$ & -1.84 $\times 10^{-3}$ \\
 C & 2.7$\times 10^{21}$ & 6.1  & 1.01	& -2.56$\times 10^{-3}$ & -1.84 $\times 10^{-3}$ \\
 D & 1.1$\times 10^{21}$ & 7.1  & 1.18	& -2.52$\times 10^{-3}$ & -1.59 $\times 10^{-3}$ \\
 E & 2.1$\times 10^{21}$ & 7.1  & 1.18	& -3.05$\times 10^{-3}$ & -1.95 $\times 10^{-3}$ \\
 F & 1.2$\times 10^{20}$ & 7.2  & 1.19	& -4.40$\times 10^{-4}$ & -4.74 $\times 10^{-4}$ \\
 G & 6.0$\times 10^{20}$ & 8.3  & 1.37	& -2.05$\times 10^{-3}$ & -1.40 $\times 10^{-3}$ \\
 H & 1.8$\times 10^{19}$ & 9.2  & 1.52	& -1.02$\times 10^{-4}$ & -3.02 $\times 10^{-4}$ \\
 I & 5.0$\times 10^{18}$ & 13.5 & 2.24	& -4.23$\times 10^{-5}$ & -4.18 $\times 10^{-5}$ \\
 J & 3.1$\times 10^{18}$ & 18.6 & 3.08	& -3.66$\times 10^{-5}$ & -1.76 $\times 10^{-6}$ \\
\hline
\end{tabular} \label{Table3}
\end{table}

\begin{figure}
\includegraphics[width=8.5cm,height=8cm]{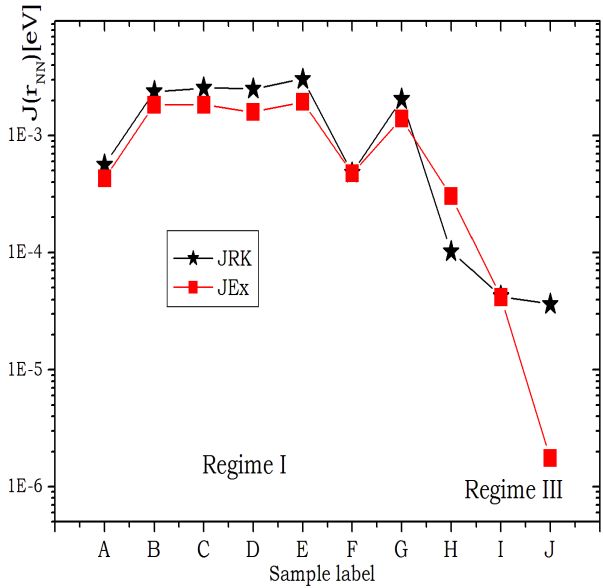}
                 \caption{Logarithms of~$|{\cal J}_{RK}|$ given in Eq.~(\ref{JRK})
                 and exact range function ~$|{\cal J}|$ calculated using
                 Eqs.~(\ref{defJ}) and~(\ref{Omega_SzSz1})--(\ref{w_02})
                 for nearest neighbor magnetic impurities in Sr$_{1-x}$La$_x$TiO$_{3-\delta}$.
                 Points are labeled according to Table~\ref{Table3}.
                 The coupling constant between conduction and magnetic
                 impurity electrons is ferromagnetic~$J=-15.48$eV$\AA^3$.
                 Results correspond to regimes I and III of model.} \label{Fig2}
\end{figure}

In Table~\ref{Table3} and Figure~\ref{Fig2} we compare the exact and RKKY range functions
for this films of Sr$_{1-x}$La$_x$TiO$_{3-\delta}$ doped with magnetic ions taking the
effective mass and concentration from Ref.~\cite{Ravichandran2011}.
Parameter~$p_0^s$ is calculated
from Eq.~(\ref{defp1_s}). Both range functions are calculated for~$r=a=3.905\AA$, i.e.
for the nearest neighbors atoms. In this example the parameter~$p_1^s$ varies from~$0.93$ to~$3.08$,
which corresponds to regimes I ($p_1^s < 2$) and III ($p_1^s > 2$) of the model, see Figure~\ref{Fig1}.
For~$p_1^s$ on the order of unity the values of exact range function are a few times
larger that those for the RKKY one. For larger~$p_1^s$ the exact range
function is much smaller than the RKKY counterpart.
For large~$p_1^s$ the ratio of exact range function to RKKY
one is~$(p_0^s)^{-4}$, see Eq.~(\ref{Omega_ab_p1}), and a similar ratio is
obtained for~$p_0^s=3.08$. The results of Figure~\ref{Fig2} suggest a possible
method to observe experimental deviation of the exact function~${\cal J}(r)$ from
the RKKY one, since by changing concentration of La atoms both models predict significantly
different values of coupling between neighboring magnetic impurities and, consequently, different
Curie temperatures.

\begin{figure}
\includegraphics[width=8cm,height=8cm]{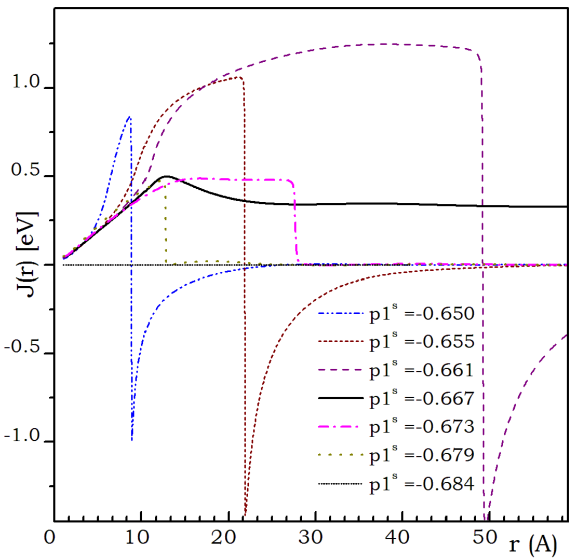}
\caption{ Range function~${\cal J}(r)$ calculated exactly using
         Eqs.~(\ref{defJ}) and~(\ref{Omega_SzSz1})--(\ref{w_02}) in regime II of the model,
         see Figure~1. Curves are labeled by values of~$p_1^s$, see Eq.~(\ref{defp1_s}).
         The remaining parameters (except~$J$ and~$\alpha N_0$) are listed
         in Table~\ref{Table1}. R} \label{Fig3}
\end{figure}

In Figure~\ref{Fig3} we plot the range function in the vicinity of~$p_1^s = -2/3$,
corresponding to~$\alpha N_0 \simeq -15.28$~eV. In this regime the range function does not oscillate,
and it has a very large amplitude.
We present these results without detailed discussion  because for parabolic energy bands
the one-electron GF  diverges at the origin and~$g_0$ in Eq.~(\ref{defg0}) is infinite.
The approximation of~$g_0$ by a finite value gives reasonable results in two other regimes of parameters,
but in the vicinities of singularities a more accurate one-electron GF is required.

Tables~\ref{Table2},~\ref{Table3} and Figures~\ref{Fig2},~\ref{Fig3} provide
three representative examples of the range function~${\cal J}(r)$ obtained
from the exact GF. The behavior of~${\cal J}(r)$ confirms the qualitative description presented
previously and, in particular, the predictions of the
simplified model in Eq.~(\ref{Omega_ab_p1}).

\section{Discussion \label{Sec_Disc}}

In the previous sections we described four main results for the exact GF of the system
and the range function~${\cal J}(r)$. In Eqs.~(\ref{hGgen})--(\ref{hGbb22}) the exact GF is
expressed as a non-linear combination of~$\hS_a, \hS_b$ components and we provided a method
of calculating the matrix elements of consecutive terms. These results are valid for
arbitrary spin values but in practice such calculations can be done only numerically.
For the spins~$\hS_a, \hS_b = 1/2$ we re-expressed the exact GF in terms of linear
combinations of localized spins components, [see Eqs.~(\ref{hG_aa_11})--(\ref{hG_bb_22})
in Supplemental material] and calculated the exact range function,
see Eqs.~(\ref{defJ}) and~(\ref{Omega_SzSz1})--(\ref{w_02}). The exact GF
is obtained analytically, and the range function is found as integrals of analytical
expressions, see Eqs.~(\ref{Omega_ab}),~(\ref{Omega_01}) and~(\ref{Omega_02}).
Both quantities depend on two dimensionless parameters~$p_1$ and~$p_2$, see
Eqs.~(\ref{defp1}) and~(\ref{defp2}). This form of GF and range function is still exact
and suitable for numerical calculations but it also does not explain the physical nature
of the problem.

The third form of results is approximate and assumes
that~$p_2 \simeq p_1^2$ [see Eq.~(\ref{defp2_a})]. This holds for one-electron
GF vanishing sufficiently fast with the increasing distance~$r=|{\bm r}_a-{\bm r}_b|$.
In practice, this is quite a good approximation in~$3D$ systems and possibly in~$2D$ systems.
This approximation allows one to understand the three main physical features of the model:
the existence of three regimes for small, large and intermediate values of~$|J|$,
the asymmetry between ferromagnetic and anti-ferromagnetic values of~$s-d$ coupling constant,
and possible existence of bound states corresponding
to the poles of exact GF in the vicinities of points~$p_1 \in \{2,-2/3\}$.

The fourth result is that the Born series is convergent if and only if the
one-electron GF is finite at the origin. As a consequence,
for the parabolic energy band dispersion in~$3D$ and~$2D$ systems
the Born series diverges, while in~$1D$ it converges.
Then, formally, the second order GF in Eq.~(\ref{hGab_s2})
and the range function in Eq.~(\ref{JRK}) are not sufficiently precise,
since one approximates the
divergent series by a finite result. However, in real solids the parabolic energy
approximation works roughly to half of the Brillouin zone and for larger wave vectors
the band energies tend to a finite value. By taking a realistic band structure one
introduces an energy cut-off related to a finite size of the Brillouin zone. Then
the one-electron GF at the origin is finite and the Born series converges.
This reasoning restores the validity of RKKY results in Eq.~(\ref{JRK}),
since after introducing the cut-off energy one approximates the convergent
Born series by its second order term given in Eq.~(\ref{hGab_s2}).
Calculating the range function using GF approximated by this term
one makes second approximation extending some energy integrals
to the infinity, instead to the cut-off energy. Then one finally obtains
the analytical result for the range function in Eq.~(\ref{JRK}).

Since many issues related to the main results have been already discussed,
here we only comment on the points related to other physical aspects of
the considered problem.
Calculations of the thermodynamic potential~$\Omega$ in Eq.~(\ref{Th_Omega})
and the range function in Eq.~(\ref{defJ})
can be also performed for finite temperatures. In this case one should use the
standard form of the Fermi-Dirac distribution function for finite~$T$.
Such calculations were reported in the literature for RKKY case~\cite{Kim1995}
and it turns out that at nonzero
temperatures the oscillations have a similar period as for~$T=0$ case,
but the amplitude decreasing with temperature.

Calculating the one-electron GF in Eq.~(\ref{gpar}) with band energy in Eq.~(\ref{epsilon_k})
one should take the velocity (or momentum) effective mass
\begin{equation} \label{m_eff}
 \frac{1}{m^*} = \frac{1}{\hbar^2 k^2} \frac{d\epsilon(k)}{dk}.
\end{equation}
This mass is well defined both for parabolic and non-parabolic energy bands.
As discussed in Ref.~\cite{Zawadzki1974},
this effective mass can be obtained from cyclotron resonance experiments, dc transport phenomena
or free carrier optics.
In many systems there exists an anisotropy of the effective masses.
In this case one may not use an "average" or "density" effective mass,
but one should calculate the one-electron GF in Eq.~(\ref{hg_3D}) taking into account this anisotropy.

In our approach we assumed that the potential of the crystal lattice does not mix electrons
states with different spins. Thus, in our considerations we neglect the spin-orbit interaction.
This approximation is valid for electrons in conduction bands of metals or wide-gap semiconductors,
but not for the holes, since usually the band structure of holes is strongly affected by
the spin-orbit coupling. On the other hand, our model is valid for an arbitrary shape of electron bands.
As an example, by taking the non-parabolic energy dispersion
\begin{equation}
 \epsilon({\bm k}) = \frac{\hbar^2 k^2}{2m^*} \left( 1 \pm A k^2 \right),
\end{equation}
where~$ A$ is parameter of non-parabolicity, one obtains from Eqs.~(\ref{gpar})
and~(\ref{defg0}) a finite value of~$g_0$. The same occurs for the tight-binding dispersion as, e.g.
\begin{equation}
 \epsilon({\bm k}) = E_0 - t\cos(k_xa)\cos(k_ya)\cos(k_za),
\end{equation}
where~$E_0$ and~$t$ are parameters of the tight-binding Hamiltonian. Then
the integration over~${\bm k}$ in Eq.~(\ref{gpar}) is restricted to the first Billowing zone
and one also obtains a finite value of~$g_0$.
The two above examples show hat the existence of~$g_0$ is a separate problem,
independent of the derivation of the exact GF. In this work we considered
parabolic energy bands because our intention was to compare the results obtained from the summation of
the infinite series (exact GF) with results obtained for the lowest order terms (RKKY model)
in the parabolic approximation.

Our approach can be generalized to many energy bands and include the spin-orbit interaction.
Assume for simplicity that one considers~$2^j$ energy bands, where~$j$ is a positive integer.
Then in order to invert the operators~$\hbmF_1$ and~$\hbmF_2$ in Eqs.~(\ref{hbmF1})--(\ref{hbmF2})
one should apply the Woodbury identities~$j+1$ times, see Eqs.~(\ref{Woodbury})--(\ref{defDel2}).
In practice, it can be done only numerically.
We expect that such a procedure gives similar results to those obtained in this paper.

The divergence of the perturbation series in~$2D$ and~$3D$
resembles difficulties arising for delta-like potentials for~$2D$ and~$3D$ systems.
As discussed in~\cite{Calkin1987}, the presence of delta potential is inconsistent with the assumption
that the electron wave function is finite at the origin.
Such a problem does not exists in~$1D$ or for systems with non-parabolic energy dispersion.
Other peculiarities of singular potentials are discussed in Ref.~\cite{Case1950}.

Crucial assumption in our work is the zero-range potential in Eq.~(\ref{Vr}), since
only for delta-like potentials the Dyson equation in Eq.~(\ref{hG_Dyson}) can be converted into
algebraic equations. In practice this potential
is realized by two kinds of physical objects: atom nuclei or magnetic impurity atoms. The diameter of
nucleus varies from~$1,8$~fm for hydrogen to c.a.~$11.7$~fm for uranium. Both diameters are more than
five orders of magnitude smaller than the lattice constant of metals, semiconductors or heavy fermion
compounds. Therefore the assumption of the zero-range potential
is justified for all nuclear systems interacting with electrons in a crystal lattice~\cite{Frisken1986}.
The approximation of zero-range potential is less evident for magnetic moments
occurring from the hybridization between~$d$ or~$f$ electrons of a magnetic impurity atom and band
electrons~\cite{Furdyna1998}.
The radius of an impurity atom is on the order of a half lattice constant, which is typically around~$3\AA$.
The period of oscillations of the range function is~$\pi/k_F$, where~$k_F \propto n_e^{1/3}$.
The approximation of the~$s-d$ interaction
by the~$\delta$ like potential is justified if~$\pi/k_F \gg a/2$, which determines the maximum concentration
of electrons in the sample.

The described model assumes presence of only two localized spins in the lattice.
This assumption is valid for sufficiently diluted systems, as e.g.
diluted magnetic or ferromagnetic semiconductors, in which one can disregard interactions
between three or mores spins. But there are systems like the Kondo-lattice~\cite{Coleman2015},
in which all atoms (or cations) are coupled by the RKKY interaction, whose spatial
decay is described by the standard formula for the RKKY range function.
In these systems the assumption of low impurity concentration is not fulfilled both for the exact and
the RKKY range functions. However, because of the fast decay of range functions with inter-spin
distance the presence of more distant magnetic atoms may be neglected. Nevertheless,
some caution is needed when applying the results
given in Figures~\ref{Fig2} and~\ref{Fig3} to such systems.

An exponential decay of the RKKY interaction was proposed in literature to fit experimental
values of the Curie temperature in some systems~\cite{de Gennes1962}.
However, as explained in
Ref.~\cite{Bulaevski1986}, the exponential decay of RKKY
interaction results not from exponential form of the range function, but rather from averaging over
random distribution of magnetic impurities in the lattice. The same arguments can be applied to
the exact range function regimes I and III of the model, because in these regimes the exact range
function resembles the RKKY one.

In Ref.~\cite{Rusin2018} we successfully removed the divergence of~$g_0$
for the Friedel oscillations using
the regularization procedure. This approach may not be applied in the present case
because in the exact GF there exist several divergent terms. In consequence,
each term of GF should be regularized using different
regulators, i.e., different values of~$J^{eff}$.
In the present work we used a different approach and introduced only one
effective parameter, namely the cut-off energy~$E_m$, see Eq.~(\ref{gpar}).
Therefore all terms of the exact GF are calculated using the same approximation.

The exact GF calculated in this work relates to the problem of two magnetic impurities
interacting via~$s-d$ interaction. However, this is not a a problem of two-impurity
Anderson Hamiltonian. The reason is that the RKKY interaction, obtained
in the second order of perturbation in terms of~$s-d$ coupling constant, differs
from the interaction obtained in the fourth order of the~$V_{sd}$ hybridization parameter
of the Anderson models, since the latter includes some extra terms that are not
present in RKKY~\cite{Proetto1982}. The same terms are omitted in the calculation of the exact GF.

The results given in Eqs.~(\ref{hGgen})--(\ref{hGbb22})
and~(\ref{hG_aa_11})--(\ref{hG_bb_22}) in Supplemental material,
are valid for any system dimension~$D$. The case of~$D=3$ was analyzed in previous
sections, so here we briefly discuss the exact range function in one and two dimensions.
In~$2D$ systems the exact range function oscillates with the period~$T= \pi/k_F$
and for large~$r$ it vanishes as~$1/r^{2}$. We expect the existence of similar
three regimes for small, intermediate and large~$s-d$ coupling,
analogous to those shown in Figure~\ref{Fig1}.
For parabolic energy bands in~$2D$ the real part of~$g_0$ diverges as~$\ln(E)$
and in order to eliminate this divergence one also should add the cut-off energy~$E_m$,
see Eq.~(\ref{hg0_3Dp}). But because of the logarithmic divergence of~${\rm Re}(g_0)$
in~$2D$, the quantity~$g_0$ is less is sensitive to the cut-off energy than
its counterpart in~$3D$. Finally, for large~$r$ in~$2D$ the one-electron GF
in Eq.~(\ref{hg_2D}) decays as~$1/r^{0.5}$ and the approximate form of thermodynamical
potential~$\Omega_{ab}$ in Eq.~(\ref{Omega_ab_p1}) is less justified than in~$3D$.

In one dimension the exact GF and the exact range function differ significantly
from those in~$3D$ and~$2D$. First, in~$1D$ the quantity~$g_0$ in Eq.~(\ref{hg0_1D})
for a parabolic energy band is finite and imaginary. Next, the one-electron GF
diverges for~$E=0$, and this singularity gives a nonzero contribution to the
range function~${\cal J}(r)$~\cite{Yafet1987,Litvinov1998,Rusin2017}.
Because of the presence of the singularity one may not decide about
the existence of localized states. Finally, in~$1D$ the one-electron GF
in Eq.~(\ref{hg_1D}) oscillates in space
with a constant amplitude, so the contributions of~$w_{01}$ and~$w_{02}$
terms in Eqs.~(\ref{w_01})--(\ref{w_02}) become comparable to that of~$w_{ab}$,
while in~$3D$ the contributions of~$w_{01}$ and~$w_{02}$ to the range function
are negligible. However it seems that there are no real~$1D$ systems
with electrons described by the effective mass approximation
with spin-independent parabolic energy dispersion.
For this reason we did not investigate the~$1D$ case in more detail.

The method of calculating GF proposed in this work applies only
to delta-like~$s-d$ interactions, and it can not be directly extended to
models including exchange, correlations,
screening, the presence of phonons, strain etc. Nevertheless, it is possible to include
these effects indirectly in a way similar to the RKKY
interaction, see~\cite{Kittel1968,FreemanBook}.
This method is based on the observation that the RKKY range
function~${\cal J}_{RK}(r)$ is the Fourier transform of the
susceptibility~$\chi_0({\bm q})$ of a free electron gas
\begin{equation} \label{DefChi0}
 {\cal J}_{RK}(r) = A_0 \sum_{\bm q} e^{i{\bm q} {\bm r}} \chi_0({\bm q}),
\end{equation}
where~$A_0$ is a constant. Then one may replace in Eq.~(\ref{DefChi0}) the
susceptibility~$\chi_0({\bm q})$ by the susceptibility~$\chi({\bm q})$ of electron gas
calculated including many body effects, non-local character of~$J$, or screening.
The same procedure can be applied to the exact range function~${\cal J}(r)$ in
regimes~I and~III of the model, since in these regimes the exact and the RKKY range functions
differ by the scaling factor and the phase shift, see Table~\ref{Table2} and Figure~\ref{Fig2}.
In the regime~II the exact range function does not resemble the RKKY one, see Figure~\ref{Fig3},
and there is no simple method of incorporating many body effects to
the range function.

In rare-earth materials the Coulomb exchange interaction between conduction electrons
and~$4f$-shell electrons is
\begin{equation} \label{HLiu}
 \hV = \sum_{{\bm k}, {\bm k}'} -2J_{sf}({\bm k}, {\bm k}')(\acute{g}-1)
 \left( \hat{\bm J}_{a} + \hat{\bm J}_{b} \right) \hbms,
\end{equation}
where~$\hat{\bm J}$ is the operator of the total angular momentum of~$4f$ electrons
and~$\acute{g}$ is the Lande factor~\cite{Liu1961}.
This approximation is valid if the wavelength of the conduction electron is large
compared with the size of the~$4f$ shell and if one neglects the
dependence of the electron wave function on the direction in space.
Our approach can be directly used to systems with the exchange potential given
in Eq.~(\ref{HLiu}) if the integral~$J_{sf}({\bm k}, {\bm k}')$
may be approximated by the delta function.
This could be valid for low electron concentrations resulting in large periods~$\pi/k_F$
of RKKY oscillations. When the exchange parameter~$J({\bm k}, {\bm k}')$
can be approximated by~$J_{sf}({\bm q})$ with~${\bm q}= {\bm k}-{\bm k}'$, we may
apply the spin susceptibility formalism from Eq.~(\ref{DefChi0}) and make a substitution
\begin{equation} \label{DefChiLiu}
 \chi_0({\bm q}) \rightarrow (\acute{g}-1)^2|J_{sf}({\bm q})|^2 \chi({\bm q}).
\end{equation}
This method may be used for~${\cal J}(r)$ in regimes~I and~III of parameters
shown in Figure~\ref{Fig1}.

In modern approaches, the RKKY range function are obtained with use of Lloyd's formula~\cite{Gorman2014},
which gives the difference between integrated densities of states~$N(E)$ [see Eq.~(\ref{Th_NE1})]
obtained from~$\hg(E)$ and~$\hG(E)$
\begin{equation} \label{LLoyd}
 \Delta N(E) = -\frac{1}{\pi} {\rm Im Tr} \ln \left(1 - \hg(E)\hV_a -\hg(E)\hV_b \right),
\end{equation}
where~$\hV_a, \hV_b$ are given in Eq.~(\ref{Vr})~\cite{LLoyd1967}.
The identity~(\ref{LLoyd}) is exact for
arbitrary~$\hg(E)$ and external potentials. The problem with Eq.~(\ref{Vr}) is how to
evaluate of the logarithm for operators~$\hV_a, \hV_b$ having non-commuting components.
In Eqs.~(\ref{hGaa11})--(\ref{hGbb22}) and~(\ref{hG_aa_11})--(\ref{hG_bb_22})
we calculated the exact GF of the system, and
we may obtain~$N(E)$ in Eq.~(\ref{Th_NE1}) by taking the trace over the GF and performing
the indefinite integration of~$n(E)$ over the energy. Then the results in Eq.~(\ref{Th_NE1})
should be equal to the expression of the RHS of Eq.~(\ref{LLoyd}).

However, there are two differences between our approach and LLoyd's formula. First, the
exact GF in Eqs.~(\ref{hGaa11})--(\ref{hGbb22}) and~(\ref{hG_aa_11})--(\ref{hG_bb_22}) is more general
than the intergraded electron density in Eq.~(\ref{LLoyd}).
For the calculation of thermodynamic properties of the system,
which depend on electron densities~$n(E)$ or~$N(E)$, the LLoyd's formula may be more convenient
than our approach. However, if one calculates quantities
depending on the GF of the system e.g., discrete energy states (as in Section~\ref{Sec_OneE}) or
the conductivity tensor, the knowledge of GF is necessary.
Second, our approach is limited to delta-like potentials, while the Lloyd's formula is valid
for arbitrary potentials and within this formalism one can include more
physical effects (screening, phonons etc.) than by our approach. However, Lloyd's
approach requires calculation of the logarithm of non-commuting operators in Eq.~(\ref{LLoyd})
which in practice can be done only by the perturbation expansion.

The~$s-d$ coupling constant~$J$ in Eq.~(\ref{Vr}) is expressed in~$eV$\AA$^D$,
where~$D$ is the system dimensionality. Experimentally one measures the coupling
constants~$J_{sd}$,~$J_{sf}$,~$J_{df}$ etc. expressed in~$eV$.
They are related to~$J$ in Eq.~(\ref{Vr}):~$J=-J_{sd}\Omega_0$, where~$\Omega_0$
is the elementary cell volume and the minus sign follows from sign convention in Eq.~(\ref{Vr}).
In the theory of diluted magnetic semiconductors one uses
notation~$J_{sd}=\alpha N_0$ and~$N_0=1/\Omega_0$~\cite{Furdyna1998}.

To observe experimentally a deviation of~${\cal J}(r)$ in Eq.~(\ref{Omega_SzSz2})--(\ref{w_02})
from the RKKY range function in Eq.~(\ref{JRK}) one should
meet the following conditions. First, both the~$s-d$ coupling~$J$ and the range function should be
measured independently with sufficient accuracy. Second, both the exchange, correlation
and screening terms in Eqs.~(\ref{DefChi0}), and~(\ref{DefChiLiu}) should be small.
Finally, proper value of~$g_0$ in the material should be known.

Is seems difficult to observe difference between two range functions
in systems belonging to the regime~I of parameters, (see Table~\ref{Table2}),
since in this case the difference between the exact and approximate range functions is
on the order of~$\pm 2p_1$, which is typically a few percent. In practice such a small difference
makes it impossible to distinguish experimentally between the two cases.
A more promising way of experimental verification of the results given
in Section~\ref{Sec_Res} is the regime~III in Figure~\ref{Fig2}.
In the latter, characterized by large~$s-d$ coupling~$|J|$ or large effective mass,
see Eq.~(\ref{defp1_s}), there is significant difference between magnitudes
of the exact and RKKY range functions. In consequence,
by measuring independently the coupling constant~$J$ and
the range function~${\cal J}(r)$ it should be possible to distinguish between the exact
and approximate range functions even in the presence of additional terms in the generalized
susceptibility of Eq.~(\ref{DefChiLiu}).
Another promising way to confirm the results obtained in this work is to observe the bound
states predicted in Section~\ref{Sec_OneE}. Experimental difficulty in such measurements is
the narrow range of parameters for which there should exist bound states.

\section{Summary \label{Sec_Sum}}

The Green's function and the range function of two localized spins in electron gas is calculated exactly by
summing the Born series using a generalization of the method of Slater-Koster and Ziman
to non-commuting spin operators. Our calculations generalize the RKKY results that are obtained from the
second order terms of the Born series. We obtained four specific results. First, the
exact GF is expressed as a nonlinear combination of localized spins components.
This form of exact GF is valid for
arbitrary spins. Second, for spins~$1/2$ we re-expressed the exact GF as a linear combination of
localized spin components. Third, an approximation is proposed for the exact GF that clearly
explains the physical nature of the problem. Fourth, it is shown that the Born series converges if and
only if the one-electron GF at the origin~$g_0$ is finite. This occurs for electrons in parabolic energy bands
in~$1D$ but not in~$2D$ or~$3D$. However, by introducing a proper cut-off energy in the calculation of
one-electron GF one obtains finite value of~$g_0$ and the convergent Born series.

For spins~$S_a=S_b = 1/2$ there are three regimes of the model. For~$|J| \ll |g_0|^{-1}$,
the range function~${\cal J}_{3D}(r)$ resembles the RKKY one: it has the same period~$\pi/k_F$,
the same decay character and a slightly different amplitude, usually differing by a few percent.
This regime occurs most frequently in nature.
For~$|J|$ comparable to~$|g_0|^{-1}$, the exact range function differs qualitatively
from the RKKY one: it has a much larger amplitude, non-oscillatory character and
it decays more slowly with inter-spin distance.
For~$|J| \gg |g_0|^{-1}$ the exact range function oscillates with the
same period and power-like decay as the RKKY one,
but it has much lower amplitude decreaing with growing~$|J|$.
In the limiting case~$|J| \rightarrow \infty$ the range function vanishes.

For the electron energy~$E=0$ and~$p_1 \simeq 2$ or~$p_1 \simeq -2/3$, [see Eq.~(\ref{Omega_ab_p1})],
the range function and GF are singular, the
poles of GF occur in the vicinities of the singularity points. The energies of bound states
are calculated. In contrast to the standard RKKY approach, for the exact GF and the range function there is
no symmetry between ferromagnetic and anti-ferromagnetic values of~$s-d$ coupling constant~$J$.
The asymmetry follows from the singularities of the operators~$(\hI - p_1\hbmZ_c)^{-1}$
for~$p_1 \in \{ 2,-2/3\}$. We calculated the exact range function for one representative material
using realistic model parameters. We also report results for the exact range function~${\cal J}(r)$
in the wide range of values of~$s-d$ coupling constants~$J$. We compared our results with other
theoretical approaches existing in the literature. Promising ways to confirm
experientially the results of this work are: i) independent measurement of the~$s-d$
coupling constant~$J$ and the range function~${\cal J}(r)$ in the regime~$|J| \gg |g_0|^{-1}$
because there the amplitude of exact range function significantly differs from its RKKY counterpart.
ii) detection of bound states in the vicinities of points~$p_1 \in \{ 2,-2/3\}$.
We hope that the exact results reported in this
paper will be useful in analyzes of similar problems.

\appendix
\section{Woodbury identities \label{App_Wood}}
In this section we prove the Woodbury identities used in Section~\ref{Sec_Prelim}.
They differ slightly from those given in Ref.~\cite{Woodbury50}.
First we prove Eq.~(\ref{Woodbury}), i.e. show that
\begin{equation}
 \left[\begin{array}{cc} \hbmDel_1^{-1} & -\hbmDel_1^{-1} \hbmB \hbmD^{-1} \\
 -\hbmDel_2^{-1}\hbmC \hbmA^{-1} & \hbmDel_2^{-1} \end{array} \right] \cdot
 \left[\begin{array}{cc} \hbmA & \hbmB \\ \hbmC & \hbmD \end{array} \right]
 = \left[\begin{array}{cc} \hat{\bm 1} & \hat{\bm 0} \\ \hat{\bm 0} & \hat{\bm 1} \end{array} \right],
\end{equation}
with~$\hbmDel_1$ and~$\hbmDel_2$ defined in Eqs.~(\ref{defDel1}) and Eqs.~(\ref{defDel2}), respectively.
We have then
\begin{eqnarray}
 \hbmDel_1^{-1} \hbmA - \hbmDel_1^{-1} \hbmB \hbmD^{-1} \hbmC = \nonumber \\
 \hbmDel_1^{-1}(\hbmA - \hbmB \hbmD^{-1} \hbmC) = \hbmDel_1^{-1} \hbmDel_1 &=& \hat{\bm 1}.
\end{eqnarray}
Similarly
\begin{eqnarray}
 -\hbmDel_2^{-1}\hbmC \hbmA^{-1} \hbmB + \hbmDel_2^{-1} \hbmD = \nonumber \\
 \hbmDel_2^{-1}(\hbmD - \hbmC \hbmA^{-1} \hbmB) = \hbmDel_2^{-1} \hbmDel_2 &=& \hat{\bm 1}.
\end{eqnarray}
Finally
\begin{eqnarray}
 \hbmDel_1^{-1}\hbmB - \hbmDel_1^{-1} \hbmB \hbmD^{-1} \hbmD = \hbmDel_1^{-1}\hbmB - \hbmDel_1^{-1} \hbmB = \hat{\bm 0},\ \ \\
 -\hbmDel_2^{-1}\hbmC \hbmA^{-1} \hbmA + \hbmDel_2^{-1} \hbmC = -\hbmDel_2^{-1}\hbmC + \hbmDel_2^{-1}\hbmC = \hat{\bm 0}.\ \
\end{eqnarray}
This proves Eq.~(\ref{Woodbury}).
Now we prove Eq.~(\ref{Woodbury2}) for~$[\hbmA, \hbmC] =0$ and~$[\hbmB, \hbmD]=0$.
There is
\begin{eqnarray}
\hbmDel_1^{-1} &=& (\hbmA - \hbmB \hbmD^{-1} \hbmC)^{-1} = (\hbmD^{-1} \hbmD \hbmA - \hbmD^{-1}\hbmB \hbmC)^{-1} = \nonumber \\
 &=&\left[\hbmD^{-1}( \hbmD \hbmA - \hbmB \hbmC) \right]^{-1} = \hbmF_1^{-1} \hbmD,
\end{eqnarray}
\begin{eqnarray}
\hbmDel_2^{-1} &=& (\hbmD - \hbmC \hbmA^{-1} \hbmB)^{-1} = (\hbmA^{-1} \hbmA \hbmD - \hbmA^{-1}\hbmC \hbmB)^{-1} = \nonumber \\
 &=&\left[\hbmA^{-1}( \hbmA \hbmD - \hbmC \hbmB) \right]^{-1} = \hbmF_2^{-1} \hbmA,
\end{eqnarray}
\begin{eqnarray}
 -\hbmDel_1^{-1} \hbmB \hbmD^{-1} = \hbmF_1^{-1} \hbmD \hbmB \hbmD^{-1} = \hbmF_1^{-1} \hbmB, \\
 -\hbmDel_2^{-1} \hbmC \hbmA^{-1} = \hbmF_2^{-1} \hbmA \hbmC \hbmA^{-1} = \hbmF_1^{-1} \hbmC.
\end{eqnarray}
This completes the proof.

\section{RKKY range function:~$w_{ab} \simeq 1$ \label{App_RKKY}}

Here we calculate the range function~${\cal J}_{RK}(r)$ for the grand canonical
potential~$\Omega^{ab}$ in Eq.~(\ref{Omega_ab}) in the limit~$w_{ab}=1$, i.e. by truncating the
Born series to the terms of the second order in the~$s-d$ coupling constant~$J$.
We begin with Eq.~(\ref{hGab_s2}), i.e. from the lowest order terms of the
Born series including both~$\hV_a$ and~$\hV_b$ potentials. Using the notation
introduced in Section~\ref{Sec_GFArb} one obtains from Eq.~(\ref{hGab_s2})
\begin{equation} \label{hGab_s2x}
 \hbmG_{1,2}^{ab} \simeq g_{1a}\hbmZ_a g_{ab} \hbmZ_b g_{b2} + g_{1b}\hbmZ_b g_{ba} \hbmZ_a g_{a2}.
\end{equation}
Since~$g_{ab} = g_{ba}$ one gets for the trace of~$\hbmG_{1,2}^{ab}$
\begin{eqnarray}
 {\rm Tr} \{\hbmG_{1,2}^{ab} \} &=& \left(\int d^3 {\bm r}_1 g_{1a} g_{b1} g_{ab}\right)
        {\rm Tr} \left\{ \hbmZ_a \hbmZ_b + \hbmZ_b \hbmZ_b  \right \} \nonumber \\
       &=& J^2 g_{ab} h_{ab} \hS_a \hS_b.
\end{eqnarray}
Then the~$S_a^zS_b^z$ part of the thermodynamic potential is
\begin{equation} \label{Omega_ab_A}
 \Omega^{ab} = \frac{J^2}{\pi} \hS_a^z \hS_b^z \int_0^{E_F} \left[ \int g_{ab} h_{ab} dE' \right] dE,
\end{equation}
which is the limit given in Eq.~(\ref{Omega_ab}) for~$w_{ab}=1$.
Using the retarded one-electron GF
\begin{equation} \label{g_ab_rk}
 g_{ab}^+ \equiv g_{ab}= -\frac{\exp(ir\sqrt{E/\zeta})}{4\pi \zeta r},
\end{equation}
with~$\zeta = \hbar^2/(2m^*)$, one obtains from Eq.~(\ref{h_ab})
\begin{equation} \label{h_ab_rk}
 h_{ab} = -\frac{dg_{ab}}{dE} = -\frac{\exp(ir\sqrt{E/\zeta})}{8\pi \zeta^{3/2} \sqrt{E}}.
\end{equation}
The one-electron density of states~$n(E)$ in Eq.~(\ref{Th_ne}) is then
\begin{equation}
 n(E) = \frac{\cos(2r\sqrt{E/\zeta})}{32\pi^3 \zeta^{5/2} \sqrt{E} r}.
\end{equation}
Calculating the double integral in Eq.~(\ref{Omega_ab_A})
with~$g_{ab}^+$ and~$h_{ab}$ given in Eqs.~(\ref{g_ab_rk})--(\ref{h_ab_rk})
and taking~$k_F= \sqrt{E_F/\zeta}$ we find
\begin{equation} \label{JRK_A}
 {\cal J}_{RK}(r) = \frac{J^2}{64\pi^3 r^4 \zeta} \left[ 2rk_F\cos(2k_Fr) -\sin(2k_Fr) \right],
\end{equation}
which is the RKKY range function for electrons in a parabolic energy band in~$3D$.

\section{GF and range function for strong coupling \label{App_Strong}}

Consider the exact GF for large~$s-d$ coupling~$J$. In this limit we approximate in
Eqs.~(\ref{hbmF1})--(\ref{hbmF2})
\begin{equation}
 \hI-g_0\hbmZ_c \simeq -g_0\hbmZ_c,
\end{equation}
where~$c=a,b$. Then we have
\begin{eqnarray}
 \label{hbmF1x}
 \hbmF_1 \simeq (g_0^2 - g_{ab} g_{ba}) \hbmZ_b \hbmZ_a = c_{ab}\hbmZ_b \hbmZ_a, \\
 \label{hbmF2x}
 \hbmF_2 \simeq (g_0^2 - g_{ab} g_{ba}) \hbmZ_a \hbmZ_b = c_{ab}\hbmZ_b \hbmZ_a,
\end{eqnarray}
where~$c_{ab} = g_0^2- g_{ab}g_{ba}$, see Eq.~(\ref{defp2}). In consequence there is
\begin{eqnarray}
 \label{hbmQ1x}
 \hbmQ_1 =& \hbmF_1^{-1} \simeq & \hbmZ_a^{-1} \hbmZ_b^{-1}/c_{ab}, \\
 \label{hbmQ2x}
 \hbmQ_2 =& \hbmF_2^{-1} \simeq & \hbmZ_b^{-1} \hbmZ_a^{-1}/c_{ab}.
\end{eqnarray}
From Eq.~(\ref{hbmG_12_d}) one obtains
\begin{eqnarray} \label{hbmG_12_dx}
 \hbmG_{12} \simeq g_{12}\hI + \hspace*{17em} \nonumber \\
            + g_{1a}[\hbmZ_a \hbmQ_1 (-g_0\hbmZ_b)]g_{a2}
            + g_{1a}[g_{ab}\hbmZ_a \hbmQ_1 \hbmZ_b]g_{b2} \ \ \ \nonumber \\
            + g_{1b}[g_{ba}\hbmZ_b \hbmQ_2 \hbmZ_a]g_{a2}
            + g_{1b}[\hbmZ_b \hbmQ_2 (-g_0\hbmZ_a)]g_{b2}, \ \ \ \
\end{eqnarray}
Inserting the approximate forms of~$\hbmQ_1$,~$\hbmQ_2$ into Eq.~(\ref{hbmG_12_dx})
one finally obtains
\begin{eqnarray} \label{hbmG_12_dJ}
 &&\hbmG_{12} \simeq g_{12}\hI + \nonumber \\
 && \frac{1}{c_{ab}}
 \left(- g_0 g_{1a} g_{a2} + g_{ab} g_{1a} g_{b2} +g_{ba}g_{1b}g_{a2} - g_0 g_{1b} g_{b2} \right). \ \ \ \ \
\end{eqnarray}
As seen from Eq.~(\ref{hbmG_12_dJ}), for large~$J$ the GF does not
depend on~${\bm S}_1$ and~${\bm S}_2$, and it has an universal character.
Such behavior of GF for large perturbing potentials is known in the literature~\cite{Rusin2018}
and it appears even in simple models of one spinless impurity, see Eq.~(\ref{GF_Ziman}).

The range function of the RKKY interaction is defined as a difference of the grand canonical potential
for parallel and antiparallel spins, see Eq.~(\ref{defJ}). However, since the electron
density~$n_e \propto {\rm Im} \{ {\rm Tr}(\hbmG_{12}) \}$, as given in Eq.~(\ref{hbmG_12_dJ}),
does not depend on~${\bm S}_1$ and~${\bm S}_2$, the grand canonical
potential~$\Omega_{\mu,\nu}$ in Eq.~(\ref{Omega_mn}) also does not depend on spin configuration.
The range function in Eq.~(\ref{defJ}) is a sum of two positive and two negative terms.
For large~$|J|$ all the four terms tend to a common value
not depending on spin configurations. Thus for large~$|J|$ the range function~${\cal J}(r)$
vanishes, which explains the disappearance of~$w_{ab}$ term in Eq.~(\ref{w_ab}) for large~$|J|$.


\clearpage
\widetext
\begin{center}
\textbf{\large Supplemental Material}
\end{center}
\setcounter{equation}{0}
\setcounter{figure}{0}
\setcounter{table}{0}
\setcounter{page}{1}
\makeatletter
\renewcommand{\theequation}{S\arabic{equation}}
\renewcommand{\thefigure}{S\arabic{figure}}
\renewcommand{\bibnumfmt}[1]{[S#1]}
\renewcommand{\citenumfont}[1]{S#1}

\subsection{Green's function for two~$S=1/2$ spins}

Here we show the final formulas for the exact GF for~$\hS_a, \hS_b=1/2$ obtained
from Eqs.~(\ref{hGaa11})--(\ref{hGbb22}) of the main text using a method described in Section~\ref{Sec_GFSpinOp}.
Some of these formulas were derived explicitly as an example of the
calculations in Eqs.~(\ref{Gab113x}),~(\ref{Gab113y}) and~(\ref{Gab113z}) of the main text.
The terms proportional to~$\hS_a^z \hS_b^z$ are marked by~$\maltese$ symbol.
The coefficients~$\hq^{1\alpha}_{ij}$ and~$\hq^{2\alpha}_{ij}$ with~$\alpha=A,B,C,D$ and~$i,j=1,2,3,4$
are~$c$-numbers and they are shown in Eqs.~(\ref{q_1A_11})--(\ref{q_2C_34}).
They depend only on~$p_1$ and~$p_2$, see Eqs.~(\ref{defp1}) and~(\ref{defp2}) of the main text.

\begin{widetext}
\begin{eqnarray} \label{hG_aa_11}
&&(\hG^{aa})_{11}=\\ &+& g_{1a} g_{a2} J (+q^{1A}_{11}/16+q^{1A}_{22}/16-q^{1D}_{33}/16-q^{1D}_{44}/16+q^{1C}_{13}/8+q^{1C}_{24}/8)\\ &+&\hS_b^z g_{1a} g_{a2} J (+q^{1A}_{11} /8-q^{1D}_{33} /8+q^{1D}_{44} /8+q^{1C}_{13} /4-q^{1C}_{24} /4)\\ &+&\hS_a^z g_{1a} g_{a2} J (+q^{1A}_{11} /8+q^{1A}_{22} /8+q^{1D}_{33} /8+q^{1D}_{44} /8-q^{1C}_{13} /4-q^{1C}_{24} /4)\\ \maltese &+& \hS_a^z \hS_b^z g_{1a} g_{a2} J (+q^{1A}_{11} /4-q^{1A}_{22} /4+q^{1D}_{33} /4-q^{1D}_{44} /4-q^{1C}_{13} /2+q^{1C}_{24} /2)\\ &+&\hS_a^- \hS_b^+ g_{1a} g_{a2} J (-q^{1D}_{32} /4+q^{1B}_{21} /2)\\ &+&\hS_a^+ \hS_b^- g_{1a} g_{a2} J (+q^{1A}_{23} /4)\\ &+& g_{1a} g_{a2} g_0 J^2 (-q^{1A}_{11}/64+q^{1A}_{22}/64+q^{1D}_{33}/64-q^{1D}_{44}/64-q^{1B}_{21}/32+q^{1B}_{43}/32-q^{1C}_{13}/32+q^{1C}_{24}/32-q^{1D}_{23}/16)\\ &+&\hS_b^z g_{1a} g_{a2} g_0 J^2 (-q^{1A}_{11} /32+q^{1D}_{33} /32+q^{1D}_{44} /32-q^{1B}_{43} /16-q^{1C}_{13} /16-q^{1C}_{24} /16+q^{1D}_{23} /8)\\ &+&\hS_a^z g_{1a} g_{a2} g_0 J^2 (-q^{1A}_{11} /32+q^{1A}_{22} /32-q^{1D}_{33} /32+q^{1D}_{44} /32-q^{1B}_{21} /16-q^{1B}_{43} /16+q^{1C}_{13} /16-q^{1C}_{24} /16+q^{1D}_{23} /8)\\ \maltese &+&\hS_a^z \hS_b^z g_{1a} g_{a2} g_0 J^2 (-q^{1A}_{11} /16-q^{1A}_{22} /16-q^{1D}_{33} /16-q^{1D}_{44} /16+q^{1B}_{21} /8+q^{1B}_{43} /8+q^{1C}_{13} /8+q^{1C}_{24} /8-q^{1D}_{23} /4)\\ &+&\hS_a^- \hS_b^+ g_{1a} g_{a2} g_0 J^2 (-q^{1D}_{32} /16+q^{1B}_{31} /8+q^{1B}_{21} /8-q^{1D}_{11} /4)\\ &+&\hS_a^+ \hS_b^- g_{1a} g_{a2} g_0 J^2 (-q^{1A}_{23} /16)
\end{eqnarray}
\begin{eqnarray}
&&(\hG^{aa})_{12}=\\ &+&\hS_b^- g_{1a} g_{a2} J (+ q^{1B}_{21} /8- q^{1B}_{43} /8+ q^{1D}_{23} /4)\\ &+&\hS_a^- g_{1a} g_{a2} J (- q^{1B}_{31} /8- q^{1B}_{42} /8+ q^{1D}_{11} /4+ q^{1D}_{22} /4)\\ &+&\hS_a^z \hS_b^- g_{1a} g_{a2} J (+q^{1B}_{21} /4+q^{1B}_{43} /4-q^{1D}_{23} /2)\\ &+&\hS_a^- \hS_b^z g_{1a} g_{a2} J (-q^{1B}_{31} /4+q^{1B}_{42} /4+q^{1D}_{11} /2-q^{1D}_{22} /2)\\ &+&\hS_b^- g_{1a} g_{a2} g_0 J^2 (- q^{1A}_{22} /16+ q^{1D}_{44} /16+ q^{1B}_{21} /32- q^{1B}_{43} /32- q^{1C}_{24} /8+ q^{1D}_{23} /16)\\ &+&\hS_a^- g_{1a} g_{a2} g_0 J^2 (+ q^{1D}_{32} /16- q^{1B}_{31} /32+ q^{1B}_{42} /32- q^{1B}_{21} /8+ q^{1D}_{11} /16- q^{1D}_{22} /16)\\ &+&\hS_a^z \hS_b^- g_{1a} g_{a2} g_0 J^2 (-q^{1A}_{22} /8-q^{1D}_{44} /8+q^{1B}_{21} /16+q^{1B}_{43} /16+q^{1C}_{24} /4-q^{1D}_{23} /8)\\ &+&\hS_a^- \hS_b^z g_{1a} g_{a2} g_0 J^2 (+q^{1D}_{32} /8-q^{1B}_{31} /16-q^{1B}_{42} /16-q^{1B}_{21} /4+q^{1D}_{11} /8+q^{1D}_{22} /8)
\end{eqnarray}
\begin{eqnarray}
&&(\hG^{aa})_{21}=\\ &+&\hS_b^+ g_{1a} g_{a2} J (+ q^{1D}_{32} /4- q^{1B}_{21} /8+ q^{1C}_{34} /8)\\ &+&\hS_a^+ g_{1a} g_{a2} J (+ q^{1D}_{33} /4+ q^{1D}_{44} /4- q^{1C}_{13} /8- q^{1C}_{24} /8)\\ &+&\hS_a^z \hS_b^+ g_{1a} g_{a2} J (+q^{1D}_{32} /2-q^{1B}_{21} /4-q^{1C}_{34} /4)\\ &+&\hS_a^+ \hS_b^z g_{1a} g_{a2} J (+q^{1D}_{33} /2-q^{1D}_{44} /2-q^{1C}_{13} /4+q^{1C}_{24} /4)\\ &+&\hS_b^+ g_{1a} g_{a2} g_0 J^2 (+ q^{1D}_{32} /16- q^{1B}_{31} /8- q^{1B}_{21} /32+ q^{1C}_{34} /32+ q^{1D}_{11} /16- q^{1D}_{33} /16)\\ &+&\hS_a^+ g_{1a} g_{a2} g_0 J^2 (- q^{1D}_{33} /16+ q^{1D}_{44} /16- q^{1B}_{43} /8+ q^{1C}_{13} /32- q^{1C}_{24} /32+ q^{1D}_{23} /16)\\ &+&\hS_a^z \hS_b^+ g_{1a} g_{a2} g_0 J^2 (+q^{1D}_{32} /8-q^{1B}_{31} /4-q^{1B}_{21} /16-q^{1C}_{34} /16+q^{1D}_{11} /8+q^{1D}_{33} /8)\\ &+&\hS_a^+ \hS_b^z g_{1a} g_{a2} g_0 J^2 (-q^{1D}_{33} /8-q^{1D}_{44} /8+q^{1B}_{43} /4+q^{1C}_{13} /16+q^{1C}_{24} /16-q^{1D}_{23} /8)
\end{eqnarray}
\begin{eqnarray}
&&(\hG^{aa})_{22}=\\ &+& g_{1a} g_{a2} J (+q^{1B}_{31}/8+q^{1B}_{42}/8-q^{1D}_{11}/16-q^{1D}_{22}/16+q^{1D}_{33}/16+q^{1D}_{44}/16)\\ &+&\hS_b^z g_{1a} g_{a2} J (+q^{1B}_{31} /4-q^{1D}_{11} /8+q^{1D}_{33} /8-q^{1D}_{44} /8)\\ &+&\hS_a^z g_{1a} g_{a2} J (+q^{1B}_{31} /4+q^{1B}_{42} /4-q^{1D}_{11} /8-q^{1D}_{22} /8-q^{1D}_{33} /8-q^{1D}_{44} /8)\\ \maltese &+&\hS_a^z \hS_b^z g_{1a} g_{a2} J (+q^{1B}_{31} /2-q^{1B}_{42} /2-q^{1D}_{11} /4+q^{1D}_{22} /4-q^{1D}_{33} /4+q^{1D}_{44} /4)\\ &+&\hS_a^- \hS_b^+ g_{1a} g_{a2} J (+q^{1D}_{32} /4)\\ &+&\hS_a^+ \hS_b^- g_{1a} g_{a2} J (+q^{1B}_{43} /2-q^{1D}_{23} /4)\\ &+& g_{1a} g_{a2} g_0 J^2 (-q^{1D}_{32}/16+q^{1B}_{31}/32-q^{1B}_{42}/32+q^{1B}_{21}/32-q^{1C}_{34}/32-q^{1D}_{11}/64+q^{1D}_{22}/64+q^{1D}_{33}/64-q^{1D}_{44}/64)\\ &+&\hS_b^z g_{1a} g_{a2} g_0 J^2 (-q^{1D}_{32} /8+q^{1B}_{31} /16+q^{1B}_{21} /16-q^{1C}_{34} /16-q^{1D}_{11} /32+q^{1D}_{33} /32+q^{1D}_{44} /32)\\ &+&\hS_a^z g_{1a} g_{a2} g_0 J^2 (-q^{1D}_{32} /8+q^{1B}_{31} /16-q^{1B}_{42} /16+q^{1B}_{21} /16+q^{1C}_{34} /16-q^{1D}_{11} /32+q^{1D}_{22} /32-q^{1D}_{33} /32+q^{1D}_{44} /32)\\ \maltese &+&\hS_a^z \hS_b^z g_{1a} g_{a2} g_0 J^2 (-q^{1D}_{32} /4+q^{1B}_{31} /8+q^{1B}_{42} /8+q^{1B}_{21} /8+q^{1C}_{34} /8-q^{1D}_{11} /16-q^{1D}_{22} /16-q^{1D}_{33} /16-q^{1D}_{44} /16)\\ &+&\hS_a^- \hS_b^+ g_{1a} g_{a2} g_0 J^2 (-q^{1D}_{32} /16)\\ &+&\hS_a^+ \hS_b^- g_{1a} g_{a2} g_0 J^2 (-q^{1D}_{44} /4+q^{1B}_{43} /8+q^{1C}_{24} /8-q^{1D}_{23} /16)
\end{eqnarray}

\begin{eqnarray}
&&(\hG^{ab})_{11}=\\ &+& g_{1a} g_{b2} g_{ab} J^2 (+q^{1A}_{11}/64-q^{1A}_{22}/64-q^{1D}_{33}/64+q^{1D}_{44}/64+q^{1B}_{21}/32-q^{1B}_{43}/32+q^{1C}_{13}/32-q^{1C}_{24}/32+q^{1D}_{23}/16)\\ &+&\hS_b^z g_{1a} g_{b2} g_{ab} J^2 (+q^{1A}_{11} /32-q^{1D}_{33} /32-q^{1D}_{44} /32+q^{1B}_{43} /16+q^{1C}_{13} /16+q^{1C}_{24} /16-q^{1D}_{23} /8)\\ &+&\hS_a^z g_{1a} g_{b2} g_{ab} J^2 (+q^{1A}_{11} /32-q^{1A}_{22} /32+q^{1D}_{33} /32-q^{1D}_{44} /32+q^{1B}_{21} /16+q^{1B}_{43} /16-q^{1C}_{13} /16+q^{1C}_{24} /16-q^{1D}_{23} /8)\\ \maltese &+&\hS_a^z \hS_b^z g_{1a} g_{b2} g_{ab} J^2 (+q^{1A}_{11} /16+q^{1A}_{22} /16+q^{1D}_{33} /16+q^{1D}_{44} /16-q^{1B}_{21} /8-q^{1B}_{43} /8-q^{1C}_{13} /8-q^{1C}_{24} /8+q^{1D}_{23} /4)\\ &+&\hS_a^- \hS_b^+ g_{1a} g_{b2} g_{ab} J^2 (+q^{1D}_{32} /16-q^{1B}_{31} /8-q^{1B}_{21} /8+q^{1D}_{11} /4)\\ &+&\hS_a^+ \hS_b^- g_{1a} g_{b2} g_{ab} J^2 (+q^{1A}_{23} /16)
\end{eqnarray}
\begin{eqnarray}
&&(\hG^{ab})_{12}=\\ &+&\hS_b^- g_{1a} g_{b2} g_{ab} J^2 (+ q^{1A}_{22} /16- q^{1D}_{44} /16- q^{1B}_{21} /32+ q^{1B}_{43} /32+ q^{1C}_{24} /8- q^{1D}_{23} /16)\\ &+&\hS_a^- g_{1a} g_{b2} g_{ab} J^2 (- q^{1D}_{32} /16+ q^{1B}_{31} /32- q^{1B}_{42} /32+ q^{1B}_{21} /8- q^{1D}_{11} /16+ q^{1D}_{22} /16)\\ &+&\hS_a^z \hS_b^- g_{1a} g_{b2} g_{ab} J^2 (+q^{1A}_{22} /8+q^{1D}_{44} /8-q^{1B}_{21} /16-q^{1B}_{43} /16-q^{1C}_{24} /4+q^{1D}_{23} /8)\\ &+&\hS_a^- \hS_b^z g_{1a} g_{b2} g_{ab} J^2 (-q^{1D}_{32} /8+q^{1B}_{31} /16+q^{1B}_{42} /16+q^{1B}_{21} /4-q^{1D}_{11} /8-q^{1D}_{22} /8)
\end{eqnarray}
\begin{eqnarray}
&&(\hG^{ab})_{21}=\\ &+&\hS_b^+ g_{1a} g_{b2} g_{ab} J^2 (- q^{1D}_{32} /16+ q^{1B}_{31} /8+ q^{1B}_{21} /32- q^{1C}_{34} /32- q^{1D}_{11} /16+ q^{1D}_{33} /16)\\ &+&\hS_a^+ g_{1a} g_{b2} g_{ab} J^2 (+ q^{1D}_{33} /16- q^{1D}_{44} /16+ q^{1B}_{43} /8- q^{1C}_{13} /32+ q^{1C}_{24} /32- q^{1D}_{23} /16)\\ &+&\hS_a^z \hS_b^+ g_{1a} g_{b2} g_{ab} J^2 (-q^{1D}_{32} /8+q^{1B}_{31} /4+q^{1B}_{21} /16+q^{1C}_{34} /16-q^{1D}_{11} /8-q^{1D}_{33} /8)\\ &+&\hS_a^+ \hS_b^z g_{1a} g_{b2} g_{ab} J^2 (+q^{1D}_{33} /8+q^{1D}_{44} /8-q^{1B}_{43} /4-q^{1C}_{13} /16-q^{1C}_{24} /16+q^{1D}_{23} /8)
\end{eqnarray}
\begin{eqnarray}
&&(\hG^{ab})_{22}=\\ &+& g_{1a} g_{b2} g_{ab} J^2 (+q^{1D}_{32}/16-q^{1B}_{31}/32+q^{1B}_{42}/32-q^{1B}_{21}/32+q^{1C}_{34}/32+q^{1D}_{11}/64-q^{1D}_{22}/64-q^{1D}_{33}/64+q^{1D}_{44}/64)\\ &+&\hS_b^z g_{1a} g_{b2} g_{ab} J^2 (+q^{1D}_{32} /8-q^{1B}_{31} /16-q^{1B}_{21} /16+q^{1C}_{34} /16+q^{1D}_{11} /32-q^{1D}_{33} /32-q^{1D}_{44} /32)\\ &+&\hS_a^z g_{1a} g_{b2} g_{ab} J^2 (+q^{1D}_{32} /8-q^{1B}_{31} /16+q^{1B}_{42} /16-q^{1B}_{21} /16-q^{1C}_{34} /16+q^{1D}_{11} /32-q^{1D}_{22} /32+q^{1D}_{33} /32-q^{1D}_{44} /32)\\ \maltese &+&\hS_a^z \hS_b^z g_{1a} g_{b2} g_{ab} J^2 (+q^{1D}_{32} /4-q^{1B}_{31} /8-q^{1B}_{42} /8-q^{1B}_{21} /8-q^{1C}_{34} /8+q^{1D}_{11} /16+q^{1D}_{22} /16+q^{1D}_{33} /16+q^{1D}_{44} /16)\\ &+&\hS_a^- \hS_b^+ g_{1a} g_{b2} g_{ab} J^2 (+q^{1D}_{32} /16)\\ &+&\hS_a^+ \hS_b^- g_{1a} g_{b2} g_{ab} J^2 (+q^{1D}_{44} /4-q^{1B}_{43} /8-q^{1C}_{24} /8+q^{1D}_{23} /16)
\end{eqnarray}

\begin{eqnarray}
&&(\hG^{ba})_{11}=\\ &+& g_{1b} g_{a2} g_{ba} J^2 (+q^{2D}_{11}/64-q^{2D}_{22}/64-q^{2D}_{33}/64+q^{2D}_{44}/64+q^{2B}_{31}/32-q^{2B}_{42}/32+q^{2C}_{12}/32-q^{2C}_{34}/32+q^{2D}_{32}/16)\\ &+&\hS_b^z g_{1b} g_{a2} g_{ba} J^2 (+q^{2D}_{11} /32-q^{2D}_{33} /32-q^{2D}_{44} /32+q^{2B}_{31} /16+q^{2B}_{42} /16+q^{2C}_{34} /16-q^{2D}_{32} /8)\\ &+&\hS_a^z g_{1b} g_{a2} g_{ba} J^2 (+q^{2D}_{11} /32-q^{2D}_{22} /32+q^{2D}_{33} /32-q^{2D}_{44} /32-q^{2B}_{31} /16+q^{2B}_{42} /16+q^{2C}_{12} /16+q^{2C}_{34} /16-q^{2D}_{32} /8)\\ \maltese &+&\hS_a^z \hS_b^z g_{1b} g_{a2} g_{ba} J^2 (+q^{2D}_{11} /16+q^{2D}_{22} /16+q^{2D}_{33} /16+q^{2D}_{44} /16-q^{2B}_{31} /8-q^{2B}_{42} /8-q^{2C}_{12} /8-q^{2C}_{34} /8+q^{2D}_{32} /4)\\ &+&\hS_a^- \hS_b^+ g_{1b} g_{a2} g_{ba} J^2 (+q^{2D}_{32} /16)\\ &+&\hS_a^+ \hS_b^- g_{1b} g_{a2} g_{ba} J^2 (+q^{2D}_{23} /16-q^{2B}_{21} /8-q^{2C}_{13} /8+q^{2D}_{11} /4)
\end{eqnarray}
\begin{eqnarray}
&&(\hG^{ba})_{12}=\\ &+&\hS_b^- g_{1b} g_{a2} g_{ba} J^2 (- q^{2D}_{23} /16+ q^{2B}_{21} /32- q^{2B}_{43} /32+ q^{2C}_{13} /8- q^{2D}_{11} /16+ q^{2D}_{33} /16)\\ &+&\hS_a^- g_{1b} g_{a2} g_{ba} J^2 (+ q^{2D}_{33} /16- q^{2D}_{44} /16- q^{2B}_{31} /32+ q^{2B}_{42} /32+ q^{2C}_{34} /8- q^{2D}_{32} /16)\\ &+&\hS_a^z \hS_b^- g_{1b} g_{a2} g_{ba} J^2 (-q^{2D}_{23} /8+q^{2B}_{21} /16+q^{2B}_{43} /16+q^{2C}_{13} /4-q^{2D}_{11} /8-q^{2D}_{33} /8)\\ &+&\hS_a^- \hS_b^z g_{1b} g_{a2} g_{ba} J^2 (+q^{2D}_{33} /8+q^{2D}_{44} /8-q^{2B}_{31} /16-q^{2B}_{42} /16-q^{2C}_{34} /4+q^{2D}_{32} /8)
\end{eqnarray}
\begin{eqnarray}
&&(\hG^{ba})_{21}=\\ &+&\hS_b^+ g_{1b} g_{a2} g_{ba} J^2 (+ q^{2D}_{22} /16- q^{2D}_{44} /16+ q^{2B}_{42} /8- q^{2C}_{12} /32+ q^{2C}_{34} /32- q^{2D}_{32} /16)\\ &+&\hS_a^+ g_{1b} g_{a2} g_{ba} J^2 (- q^{2D}_{23} /16+ q^{2B}_{21} /8+ q^{2C}_{13} /32- q^{2C}_{24} /32- q^{2D}_{11} /16+ q^{2D}_{22} /16)\\ &+&\hS_a^z \hS_b^+ g_{1b} g_{a2} g_{ba} J^2 (+q^{2D}_{22} /8+q^{2D}_{44} /8-q^{2B}_{42} /4-q^{2C}_{12} /16-q^{2C}_{34} /16+q^{2D}_{32} /8)\\ &+&\hS_a^+ \hS_b^z g_{1b} g_{a2} g_{ba} J^2 (-q^{2D}_{23} /8+q^{2B}_{21} /4+q^{2C}_{13} /16+q^{2C}_{24} /16-q^{2D}_{11} /8-q^{2D}_{22} /8)
\end{eqnarray}
\begin{eqnarray}
&&(\hG^{ba})_{22}=\\ &+& g_{1b} g_{a2} g_{ba} J^2 (+q^{2D}_{23}/16-q^{2B}_{21}/32+q^{2B}_{43}/32-q^{2C}_{13}/32+q^{2C}_{24}/32+q^{2D}_{11}/64-q^{2D}_{22}/64-q^{2D}_{33}/64+q^{2D}_{44}/64)\\ &+&\hS_b^z g_{1b} g_{a2} g_{ba} J^2 (+q^{2D}_{23} /8-q^{2B}_{21} /16+q^{2B}_{43} /16-q^{2C}_{13} /16+q^{2D}_{11} /32-q^{2D}_{33} /32-q^{2D}_{44} /32)\\ &+&\hS_a^z g_{1b} g_{a2} g_{ba} J^2 (+q^{2D}_{23} /8-q^{2B}_{21} /16-q^{2B}_{43} /16-q^{2C}_{13} /16+q^{2C}_{24} /16+q^{2D}_{11} /32-q^{2D}_{22} /32+q^{2D}_{33} /32-q^{2D}_{44} /32)\\ \maltese &+&\hS_a^z \hS_b^z g_{1b} g_{a2} g_{ba} J^2 (+q^{2D}_{23} /4-q^{2B}_{21} /8-q^{2B}_{43} /8-q^{2C}_{13} /8-q^{2C}_{24} /8+q^{2D}_{11} /16+q^{2D}_{22} /16+q^{2D}_{33} /16+q^{2D}_{44} /16)\\ &+&\hS_a^- \hS_b^+ g_{1b} g_{a2} g_{ba} J^2 (+q^{2D}_{44} /4-q^{2B}_{42} /8-q^{2C}_{34} /8+q^{2D}_{32} /16)\\ &+&\hS_a^+ \hS_b^- g_{1b} g_{a2} g_{ba} J^2 (+q^{2D}_{23} /16)
\end{eqnarray}

\begin{eqnarray}
&&(\hG^{bb})_{11}=\\ &+& g_{1b} g_{b2} J (+q^{2D}_{11}/16-q^{2D}_{22}/16+q^{2D}_{33}/16-q^{2D}_{44}/16+q^{2C}_{12}/8+q^{2C}_{34}/8)\\ &+&\hS_b^z g_{1b} g_{b2} J (+q^{2D}_{11} /8+q^{2D}_{33} /8+q^{2D}_{44} /8-q^{2C}_{34} /4)\\ &+&\hS_a^z g_{1b} g_{b2} J (+q^{2D}_{11} /8-q^{2D}_{22} /8-q^{2D}_{33} /8+q^{2D}_{44} /8+q^{2C}_{12} /4-q^{2C}_{34} /4)\\ \maltese &+&\hS_a^z \hS_b^z g_{1b} g_{b2} J (+q^{2D}_{11} /4+q^{2D}_{22} /4-q^{2D}_{33} /4-q^{2D}_{44} /4-q^{2C}_{12} /2+q^{2C}_{34} /2)\\ &+&\hS_a^- \hS_b^+ g_{1b} g_{b2} J (+q^{2D}_{32} /4)\\ &+&\hS_a^+ \hS_b^- g_{1b} g_{b2} J (-q^{2D}_{23} /4+q^{2C}_{13} /2)\\ &+& g_{1b} g_{b2} g_0 J^2 (-q^{2D}_{11}/64+q^{2D}_{22}/64+q^{2D}_{33}/64-q^{2D}_{44}/64-q^{2B}_{31}/32+q^{2B}_{42}/32-q^{2C}_{12}/32+q^{2C}_{34}/32-q^{2D}_{32}/16)\\ &+&\hS_b^z g_{1b} g_{b2} g_0 J^2 (-q^{2D}_{11} /32+q^{2D}_{33} /32+q^{2D}_{44} /32-q^{2B}_{31} /16-q^{2B}_{42} /16-q^{2C}_{34} /16+q^{2D}_{32} /8)\\ &+&\hS_a^z g_{1b} g_{b2} g_0 J^2 (-q^{2D}_{11} /32+q^{2D}_{22} /32-q^{2D}_{33} /32+q^{2D}_{44} /32+q^{2B}_{31} /16-q^{2B}_{42} /16-q^{2C}_{12} /16-q^{2C}_{34} /16+q^{2D}_{32} /8)\\ &+&\hS_a^- g_{1b} g_{b2} g_0 J^2 (-q^{2D}_{32} \hS_b^+/16)\\ \maltese &+&\hS_a^z \hS_b^z g_{1b} g_{b2} g_0 J^2 (-q^{2D}_{11} /16-q^{2D}_{22} /16-q^{2D}_{33} /16-q^{2D}_{44} /16+q^{2B}_{31} /8+q^{2B}_{42} /8+q^{2C}_{12} /8+q^{2C}_{34} /8-q^{2D}_{32} /4)\\ &+&\hS_a^+ \hS_b^- g_{1b} g_{b2} g_0 J^2 (-q^{2D}_{23} /16+q^{2B}_{21} /8+q^{2C}_{13} /8-q^{2D}_{11} /4)
\end{eqnarray}
\begin{eqnarray}
&&(\hG^{bb})_{12}=\\ &+&\hS_b^- g_{1b} g_{b2} J (- q^{2B}_{21} /8- q^{2B}_{43} /8+ q^{2D}_{11} /4+ q^{2D}_{33} /4)\\ &+&\hS_a^- g_{1b} g_{b2} J (+ q^{2B}_{31} /8- q^{2B}_{42} /8+ q^{2D}_{32} /4)\\ &+&\hS_a^z \hS_b^- g_{1b} g_{b2} J (-q^{2B}_{21} /4+q^{2B}_{43} /4+q^{2D}_{11} /2-q^{2D}_{33} /2)\\ &+&\hS_a^- \hS_b^z g_{1b} g_{b2} J (+q^{2B}_{31} /4+q^{2B}_{42} /4-q^{2D}_{32} /2)\\ &+&\hS_b^- g_{1b} g_{b2} g_0 J^2 (+ q^{2D}_{23} /16- q^{2B}_{21} /32+ q^{2B}_{43} /32- q^{2C}_{13} /8+ q^{2D}_{11} /16- q^{2D}_{33} /16)\\ &+&\hS_a^- g_{1b} g_{b2} g_0 J^2 (- q^{2D}_{33} /16+ q^{2D}_{44} /16+ q^{2B}_{31} /32- q^{2B}_{42} /32- q^{2C}_{34} /8+ q^{2D}_{32} /16)\\ &+&\hS_a^z \hS_b^- g_{1b} g_{b2} g_0 J^2 (+q^{2D}_{23} /8-q^{2B}_{21} /16-q^{2B}_{43} /16-q^{2C}_{13} /4+q^{2D}_{11} /8+q^{2D}_{33} /8)\\ &+&\hS_a^- \hS_b^z g_{1b} g_{b2} g_0 J^2 (-q^{2D}_{33} /8-q^{2D}_{44} /8+q^{2B}_{31} /16+q^{2B}_{42} /16+q^{2C}_{34} /4-q^{2D}_{32} /8)
\end{eqnarray}
\begin{eqnarray}
&&(\hG^{bb})_{21}=\\ &+&\hS_b^+ g_{1b} g_{b2} J (+ q^{2D}_{22} /4+ q^{2D}_{44} /4- q^{2C}_{12} /8- q^{2C}_{34} /8)\\ &+&\hS_a^+ g_{1b} g_{b2} J (+ q^{2D}_{23} /4- q^{2C}_{13} /8+ q^{2C}_{24} /8)\\ &+&\hS_a^z \hS_b^+ g_{1b} g_{b2} J (+q^{2D}_{22} /2-q^{2D}_{44} /2-q^{2C}_{12} /4+q^{2C}_{34} /4)\\ &+&\hS_a^+ \hS_b^z g_{1b} g_{b2} J (+q^{2D}_{23} /2-q^{2C}_{13} /4-q^{2C}_{24} /4)\\ &+&\hS_b^+ g_{1b} g_{b2} g_0 J^2 (- q^{2D}_{22} /16+ q^{2D}_{44} /16- q^{2B}_{42} /8+ q^{2C}_{12} /32- q^{2C}_{34} /32+ q^{2D}_{32} /16)\\ &+&\hS_a^+ g_{1b} g_{b2} g_0 J^2 (+ q^{2D}_{23} /16- q^{2B}_{21} /8- q^{2C}_{13} /32+ q^{2C}_{24} /32+ q^{2D}_{11} /16- q^{2D}_{22} /16)\\ &+&\hS_a^z \hS_b^+ g_{1b} g_{b2} g_0 J^2 (-q^{2D}_{22} /8-q^{2D}_{44} /8+q^{2B}_{42} /4+q^{2C}_{12} /16+q^{2C}_{34} /16-q^{2D}_{32} /8)\\ &+&\hS_a^+ \hS_b^z g_{1b} g_{b2} g_0 J^2 (+q^{2D}_{23} /8-q^{2B}_{21} /4-q^{2C}_{13} /16-q^{2C}_{24} /16+q^{2D}_{11} /8+q^{2D}_{22} /8)
\end{eqnarray}
\begin{eqnarray} \label{hG_bb_22}
&&(\hG^{bb})_{22}=\\ &+& g_{1b} g_{b2} J (+q^{2B}_{21}/8+q^{2B}_{43}/8-q^{2D}_{11}/16+q^{2D}_{22}/16-q^{2D}_{33}/16+q^{2D}_{44}/16)\\ &+&\hS_b^z g_{1b} g_{b2} J (+q^{2B}_{21} /4+q^{2B}_{43} /4-q^{2D}_{11} /8-q^{2D}_{33} /8-q^{2D}_{44} /8)\\ &+&\hS_a^z g_{1b} g_{b2} J (+q^{2B}_{21} /4-q^{2B}_{43} /4-q^{2D}_{11} /8+q^{2D}_{22} /8+q^{2D}_{33} /8-q^{2D}_{44} /8)\\ \maltese &+&\hS_a^z \hS_b^z g_{1b} g_{b2} J (+q^{2B}_{21} /2-q^{2B}_{43} /2-q^{2D}_{11} /4-q^{2D}_{22} /4+q^{2D}_{33} /4+q^{2D}_{44} /4)\\ &+&\hS_a^- \hS_b^+ g_{1b} g_{b2} J (+q^{2B}_{42} /2-q^{2D}_{32} /4)\\ &+&\hS_a^+ \hS_b^- g_{1b} g_{b2} J (+q^{2D}_{23} /4)\\ &+& g_{1b} g_{b2} g_0 J^2 (-q^{2D}_{23}/16+q^{2B}_{21}/32-q^{2B}_{43}/32+q^{2C}_{13}/32-q^{2C}_{24}/32-q^{2D}_{11}/64+q^{2D}_{22}/64+q^{2D}_{33}/64-q^{2D}_{44}/64)\\ &+&\hS_b^z g_{1b} g_{b2} g_0 J^2 (-q^{2D}_{23} /8+q^{2B}_{21} /16-q^{2B}_{43} /16+q^{2C}_{13} /16-q^{2D}_{11} /32+q^{2D}_{33} /32+q^{2D}_{44} /32)\\ &+&\hS_a^z g_{1b} g_{b2} g_0 J^2 (-q^{2D}_{23} /8+q^{2B}_{21} /16+q^{2B}_{43} /16+q^{2C}_{13} /16-q^{2C}_{24} /16-q^{2D}_{11} /32+q^{2D}_{22} /32-q^{2D}_{33} /32+q^{2D}_{44} /32)\\ &+&\hS_a^- g_{1b} g_{b2} g_0 J^2 (-q^{2D}_{44} \hS_b^+/4+q^{2B}_{42} \hS_b^+/8+q^{2C}_{34} \hS_b^+/8-q^{2D}_{32} \hS_b^+/16)\\ \maltese &+&\hS_a^z \hS_b^z g_{1b} g_{b2} g_0 J^2 (-q^{2D}_{23} /4+q^{2B}_{21} /8+q^{2B}_{43} /8+q^{2C}_{13} /8+q^{2C}_{24} /8-q^{2D}_{11} /16-q^{2D}_{22} /16-q^{2D}_{33} /16-q^{2D}_{44} /16)\\ &+&\hS_a^+ \hS_b^- g_{1b} g_{b2} g_0 J^2 (-q^{2D}_{23} /16)
\end{eqnarray}
\end{widetext}

\subsection{Explicit form of~$\hf_{1\alpha}$ and~$\hf^{2\alpha}$ operators}

Here we list matrices corresponding to~$\hf_{1\alpha}$ and~$\hf^{2\alpha}$ operators ($\alpha=A,B,C,D$)
defined in Eqs.~(\ref{hF1A})--(\ref{hF1D}) in the main text.
Using Eqs.~(\ref{S05Sap})--(\ref{S05Sbp}) in the main text one obtains for~$\hf_{1\alpha}$

\begin{equation} \label{S05f1A}
 \hf_{1A} = \left(\begin{array}{cccc} 1 - p_1 + \frac{p_2}{4} & 0 & 0 & 0 \\ 0 & 1 - \frac{p_2}{4} & p_2 & 0 \\ 0 & 0 & 1 - \frac{p_2}{4} & 0 \\ 0 & 0 & 0 & 1 + p_1 + \frac{p_2}{4} \end{array} \right), \ \
\end{equation}
\begin{equation} \label{S05f1B}
 \hf_{1B} = \left(\begin{array}{cccc} 0 & 0 & 0 & 0 \\ - p_1 - \frac{p_2}{2} & 0 & 0 & 0 \\ - p_1 + \frac{p_2}{2} & 0 & 0 & 0 \\ 0 & - p_1 - \frac{p_2}{2} & -p_1 + \frac{p_2}{2} & 0 \end{array} \right),
\end{equation}
\begin{equation} \label{S05f1C}
 \hf_{1C} = \left(\begin{array}{cccc} 0 & - p_1 + \frac{p_2}{2} & -p_1 - \frac{p_2}{2} & 0 \\ 0 & 0 & 0 & - p_1 + \frac{p_2}{2} \\ 0 & 0 & 0 & - p_1 - \frac{p_2}{2} \\ 0 & 0 & 0 & 0 \end{array} \right),
\end{equation}
\begin{equation} \label{S05f1D}
 \hf_{1D} = \left(\begin{array}{cccc} 1 + p_1 + \frac{p_2}{4} & 0 & 0 & 0 \\ 0 & 1 - \frac{p_2}{4} & 0 & 0 \\ 0 & p_2 & 1 - \frac{p_2}{4} & 0 \\ 0 & 0 & 0 & 1 - p_1 + \frac{p_2}{4} \end{array} \right). \ \
\end{equation}
Similarly, for~$\hf_{2\alpha}$ there is
\begin{equation} \label{S05f2A}
\hf_{2A} = \left(\begin{array}{cccc} 1 - p_1 + \frac{p2}{4} & 0 & 0 & 0 \\ 0 & 1 - \frac{p_2}{4} & 0 & 0 \\ 0 & p_2 & 1 - \frac{p_2}{4} & 0 \\ 0 & 0 & 0 & 1 + p_1 + \frac{p_2}{4} \end{array} \right), \ \
\end{equation}
\begin{equation} \label{S05f2B}
\hf_{2B} = \left(\begin{array}{cccc} 0 & 0 & 0 & 0 \\ \frac{p_2}{2} - p_1 & 0 & 0 & 0 \\ - p_1 - \frac{p_2}{2} & 0 & 0 & 0 \\ 0 & \frac{p_2}{2} - p_1 & - p_1 - \frac{p_2}{2} & 0 \end{array} \right),
\end{equation}
\begin{equation} \label{S05f2C}
\hf_{2C} = \left(\begin{array}{cccc} 0 & - p_1 - \frac{p_2}{2} & \frac{p_2}{2} - p_1 & 0 \\ 0 & 0 & 0 & - p_1 - \frac{p_2}{2} \\ 0 & 0 & 0 & \frac{p_2}{2} - p_1 \\ 0 & 0 & 0 & 0 \end{array} \right),
\end{equation}
\begin{equation} \label{S05f2D}
\hf_{2D} = \left(\begin{array}{cccc} 1 + p_1 + \frac{p_2}{4} & 0 & 0 & 0 \\ 0 & 1 - \frac{p_2}{4} & p_2 & 0 \\ 0 & 0 & 1 - \frac{p_2}{4} & 0 \\ 0 & 0 & 0 & 1 - p_1 + \frac{p_2}{4} \end{array} \right). \ \
\end{equation}

\subsection{Coefficients~$\hq^{1\alpha}$ and~$\hq^{2\alpha}$}

Here we list all nonzero elements of eight~$4 \times 4$ matrices~$\hq^{1\alpha}$ and~$\hq^{2\alpha}$
with~$\alpha=A,B,C,D$. They are calculated from Eqs.~(\ref{hbmQ1a})--(\ref{hDel2D}) of the main text
using the matrix forms of operators~$\hf_{1\alpha}$ and~$\hf_{2\alpha}$ given
in Eqs.~(\ref{S05f1A})--(\ref{S05f2D})).
The quantities~$p_1$ and~$p_2$ are defined in Eqs.~(\ref{defp1}) and~(\ref{defp2}) of the main text, respectively.

The structures of~$\hq^{1\alpha}$ and~$\hq^{2\alpha}$ matrices are
\begin{equation} \label{hq_m1}
                \hq^{1A}=\left(\begin{array}{cccc} \hq^{1A}_{11} & 0 & 0 & 0 \\ 0 & \hq^{1A}_{22} & \hq^{1A}_{23} & 0 \\ 0 & \hq^{1A}_{32} & \hq^{1A}_{33} & 0 \\ 0 & 0 & 0 & \hq^{1A}_{44} \end{array}\right)\end{equation}
\begin{equation}\hq^{1D}=\left(\begin{array}{cccc} \hq^{1D}_{11} & 0 & 0 & 0 \\ 0 & \hq^{1D}_{22} & \hq^{1D}_{23} & 0 \\ 0 & \hq^{1D}_{32} & \hq^{1D}_{33} & 0 \\ 0 & 0 & 0 & \hq^{1D}_{44} \end{array}\right)\end{equation}
\begin{equation}\hq^{1B}=\left(\begin{array}{cccc} 0 & 0 & 0 & 0 \\ \hq^{1B}_{21} & 0 & 0 & 0 \\ \hq^{1B}_{31} & 0 & 0 & 0 \\ 0 & \hq^{1B}_{42} & \hq^{1B}_{43} & 0 \end{array}\right)\end{equation}
\begin{equation}\hq^{1C}=\left(\begin{array}{cccc} 0 & \hq^{1C}_{12} & \hq^{1C}_{13} & 0 \\ 0 & 0 & 0 & \hq^{1C}_{24} \\ 0 & 0 & 0 & \hq^{1C}_{34} \\ 0 & 0 & 0 & 0 \end{array}\right)\end{equation}

\begin{equation}\hq^{2A}=\left(\begin{array}{cccc} \hq^{2A}_{11} & 0 & 0 & 0 \\ 0 & \hq^{2A}_{22} & \hq^{2A}_{23} & 0 \\ 0 & \hq^{2A}_{32} & \hq^{2A}_{33} & 0 \\ 0 & 0 & 0 & \hq^{2A}_{44} \end{array}\right)\end{equation}
\begin{equation}\hq^{2D}=\left(\begin{array}{cccc} \hq^{2D}_{11} & 0 & 0 & 0 \\ 0 & \hq^{2D}_{22} & \hq^{2A}_{23} & 0 \\ 0 & \hq^{2D}_{32} & \hq^{2D}_{33} & 0 \\ 0 & 0 & 0 & \hq^{2D}_{44} \end{array}\right)\end{equation}
\begin{equation}\hq^{2B}=\left(\begin{array}{cccc} 0 & 0 & 0 & 0 \\ \hq^{2B}_{21} & 0 & 0 & 0 \\ \hq^{2B}_{31} & 0 & 0 & 0 \\ 0 & \hq^{2B}_{42} & \hq^{2B}_{43} & 0 \end{array}\right)\end{equation}
\begin{equation} \label{hq_m2}
               \hq^{2C}=\left(\begin{array}{cccc} 0 & \hq^{2C}_{12} & \hq^{2C}_{13} & 0 \\ 0 & 0 & 0 & \hq^{2C}_{24} \\ 0 & 0 & 0 & \hq^{2C}_{34} \\ 0 & 0 & 0 & 0 \end{array}\right)\end{equation}

The nonzero elements of the above matrices are:

\begin{widetext}
\begin{eqnarray} \label{q_1A_11}
q^{1A}_{11} &=& -4/(4 p_1-p_2-4), \\
q^{1A}_{22} &=& -4 (16 p_1^2+4 p_1 (p_2-4)-3 p_2^2-16)/(32 p_1^2 (3 p_2-4)-4 p_1 (15 p_2^2+8 p_2-16)+9 p_2^3+28 p_2^2-16 p_2+64), \\
q^{1A}_{23} &=& -64 (p_1^2-p_2)/(32 p_1^2 (3 p_2-4)-4 p_1 (15 p_2^2+8 p_2-16)+9 p_2^3+28 p_2^2-16 p_2+64), \\
q^{1A}_{32} &=& 16 (2 p_1-p_2)^2/(32 p_1^2 (3 p_2-4)-4 p_1 (15 p_2^2+8 p_2-16)+9 p_2^3+28 p_2^2-16 p_2+64), \\
q^{1A}_{33} &=& -4 (16 p_1^2+4p_1 (p_2-4)-3 p_2^2-16)/(32 p_1^2 (3 p_2-4)-4 p_1 (15 p_2^2+8 p_2-16)+9 p_2^3+28 p_2^2-16 p_2+64), \\
q^{1A}_{44} &=& 4 (p_2-4)^2/(32 p_1^2 (3 p_2-4)-4 p_1 (15 p_2^2+8 p_2-16)+(p_2+4) (9 p_2^2-8 p_2+16)),
\end{eqnarray}

\begin{eqnarray}
q^{1D}_{11} &=& 4 (p_2-4)^2/(32 p_1^2 (3 p_2-4)-4 p_1 (15 p_2^2+8 p_2-16)+(p_2+4) (9 p_2^2-8 p_2+16)), \\
q^{1D}_{22} &=& -4 (16 p_1^2+4 p_1 (p_2-4)-3 p_2^2-16)/(32 p_1^2 (3 p_2-4)-4 p_1 (15 p_2^2+8 p_2-16)+9 p_2^3+28 p_2^2-16 p_2+64), \\
q^{1D}_{23} &=& 16 (2 p_1-p_2)^2/(32 p_1^2 (3 p_2-4)-4 p_1 (15 p_2^2+8 p_2-16)+9 p_2^3+28 p_2^2-16 p_2+64), \\
q^{1D}_{32} &=& 64 (p_1^2-p_2)/(32 p_1^2 (3 p_2-4)-4 p_1 (15 p_2^2+8 p_2-16)+9 p_2^3+28 p_2^2-16 p_2+64), \\
q^{1D}_{32} &=& -4 (16 p_1^2+4 p_1 (p_2-4)-3 p_2^2-16)/(32 p_1^2 (3 p_2-4)-4 p_1 (15 p_2^2+8 p_2-16)+9 p_2^3+28 p_2^2-16 p_2+64), \\
q^{1D}_{44} &=& -4/(4 p_1-p_2-4),
\end{eqnarray}

\begin{eqnarray}
q^{1B}_{21} &=& 8 (p_2 (3 p_2+4)-2 p_1 (5 p_2-4))/(32 p_1^2 (3 p_2-4)-4 p_1 (15 p_2^2+8 p_2-16)+9 p_2^3+28 p_2^2-16 p_2+64), \\
q^{1B}_{31} &=& 8 (4-p_2) (2 p_1-p_2)/(32 p_1^2 (3 p_2-4)-4 p_1 (15 p_2^2+8 p_2-16)+9 p_2^3+28 p_2^2-16 p_2+64), \\
q^{1B}_{42} &=& 8 (p_2 (3 p_2+4)-2 p_1 (5 p_2-4))/(32 p_1^2 (3 p_2-4)-4 p_1 (15 p_2^2+8 p_2-16)+(p_2+4) (9 p_2^2-8 p_2+16)), \\
q^{1B}_{43} &=& 8 (4-p_2) (2 p_1-p_2)/(32 p_1^2 (3 p_2-4)-4 p_1 (15 p_2^2+8 p_2-16)+(p_2+4) (9 p_2^2-8 p_2+16)),
\end{eqnarray}

\begin{eqnarray}
q^{1C}_{12} &=& 8 (4-p_2) (2 p_1-p_2)/(32 p_1^2 (3 p_2-4)-4 p_1 (15 p_2^2+8 p_2-16)+(p_2+4) (9 p_2^2-8 p_2+16)), \\
q^{1C}_{13} &=& 8 (p_2 (3 p_2+4)-2 p_1 (5 p_2-4))/(32 p_1^2 (3 p_2-4)-4 p_1 (15 p_2^2+8 p_2-16)+(p_2+4) (9 p_2^2-8 p_2+16)), \\
q^{1C}_{24} &=& 8 (4-p_2) (2 p_1-p_2)/(32 p_1^2 (3 p_2-4)-4 p_1 (15 p_2^2+8 p_2-16)+9 p_2^3+28 p_2^2-16 p_2+64), \\
q^{1C}_{34} &=& 8 (p_2 (3 p_2+4)-2 p_1 (5 p_2-4))/(32 p_1^2 (3 p_2-4)-4 p_1 (15 p_2^2+8 p_2-16)+9 p_2^3+28 p_2^2-16 p_2+64),
\end{eqnarray}

\begin{eqnarray}
q^{2A}_{11} &=& -4/(4 p_1-p_2-4) \\
q^{2A}_{22} &=& -4 (16 p_1^2+4 p_1 (p_2-4)-3 p_2^2-16)/(32 p_1^2 (3 p_2-4)-4 p_1 (15 p_2^2+8 p_2-16)+9 p_2^3+28 p_2^2-16 p_2+64), \\
q^{2A}_{23} &=& 16 (2 p_1-p_2)^2/(32 p_1^2 (3 p_2-4)-4 p_1 (15 p_2^2+8 p_2-16)+9 p_2^3+28 p_2^2-16 p_2+64), \\
q^{2A}_{32} &=& 64 (p_1^2-p_2)/(32 p_1^2 (3 p_2-4)-4 p_1 (15 p_2^2+8 p_2-16)+9 p_2^3+28 p_2^2-16 p_2+64), \\
q^{2A}_{33} &=& -4 (16 p_1^2+4 p_1 (p_2-4)-3 p_2^2-16)/(32 p_1^2 (3 p_2-4)-4 p_1 (15 p_2^2+8 p_2-16)+9 p_2^3+28 p_2^2-16 p_2+64), \\
q^{2A}_{44} &=& 4 (p_2-4)^2/(32 p_1^2 (3 p_2-4)-4 p_1 (15 p_2^2+8 p_2-16)+(p_2+4) (9 p_2^2-8 p_2+16)),
\end{eqnarray}

\begin{eqnarray}
q^{2D}_{11} &=& 4 (p_2-4)^2/(32 p_1^2 (3 p_2-4)-4 p_1 (15 p_2^2+8 p_2-16)+(p_2+4) (9 p_2^2-8 p_2+16)), \\
q^{2D}_{22} &=& -4 (16 p_1^2+4 p_1 (p_2-4)-3 p_2^2-16)/(32 p_1^2 (3 p_2-4)-4 p_1 (15 p_2^2+8 p_2-16)+9 p_2^3+28 p_2^2-16 p_2+64), \\
q^{2D}_{23} &=& 64 (p_1^2-p_2)/(32 p_1^2 (3 p_2-4)-4 p_1 (15 p_2^2+8 p_2-16)+9 p_2^3+28 p_2^2-16 p_2+64), \\
q^{2D}_{32} &=& 16 (2 p_1-p_2)^2/(32 p_1^2 (3 p_2-4)-4 p_1 (15 p_2^2+8 p_2-16)+9 p_2^3+28 p_2^2-16 p_2+64), \\
q^{2D}_{33} &=& -4 (16 p_1^2+4 p_1 (p_2-4)-3 p_2^2-16)/(32 p_1^2 (3 p_2-4)-4 p_1 (15 p_2^2+8 p_2-16)+9 p_2^3+28 p_2^2-16 p_2+64), \\
q^{2D}_{44} &=& -4/(4 p_1-p_2-4),
\end{eqnarray}

\begin{eqnarray}
q^{2B}_{21} &=& 8 (4-p_2) (2 p_1-p_2)/(32 p_1^2 (3 p_2-4)-4 p_1 (15 p_2^2+8 p_2-16)+9 p_2^3+28 p_2^2-16 p_2+64), \\
q^{2B}_{31} &=& 8 (p_2 (3 p_2+4)-2 p_1 (5 p_2-4))/(32 p_1^2 (3 p_2-4)-4 p_1 (15 p_2^2+8 p_2-16)+9 p_2^3+28 p_2^2-16 p_2+64),\\
q^{2B}_{42} &=& 8 (4-p_2) (2 p_1-p_2)/(32 p_1^2 (3 p_2-4)-4 p_1 (15 p_2^2+8 p_2-16)+(p_2+4) (9 p_2^2-8 p_2+16)), \\
q^{2B}_{43} &=& 8 (p_2 (3 p_2+4)-2 p_1 (5 p_2-4))/(32 p_1^2 (3 p_2-4)-4 p_1 (15 p_2^2+8 p_2-16)+(p_2+4) (9 p_2^2-8 p_2+16)),
\end{eqnarray}

\begin{eqnarray}
q^{2C}_{12} &=& 8 (p_2 (3 p_2+4)-2 p_1 (5 p_2-4))/(32 p_1^2 (3 p_2-4)-4 p_1 (15 p_2^2+8 p_2-16)+(p_2+4) (9 p_2^2-8 p_2+16)), \\
q^{2C}_{13} &=& 8 (4-p_2) (2 p_1-p_2)/(32 p_1^2 (3 p_2-4)-4 p_1 (15 p_2^2+8 p_2-16)+(p_2+4) (9 p_2^2-8 p_2+16)), \\
q^{2C}_{24} &=& 8 (p_2 (3 p_2+4)-2 p_1 (5 p_2-4))/(32 p_1^2 (3 p_2-4)-4 p_1 (15 p_2^2+8 p_2-16)+9 p_2^3+28 p_2^2-16 p_2+64), \\
q^{2C}_{34} &=& 8 (4-p_2) (2 p_1-p_2)/(32 p_1^2 (3 p_2-4)-4 p_1 (15 p_2^2+8 p_2-16)+9 p_2^3+28 p_2^2-16 p_2+64).
\label{q_2C_34}
\end{eqnarray}
\end{widetext}
The remaining components of the above matrices are zero. To understand mathematical
structure of~$\hq^{1\alpha}$ and~$\hq^{2\alpha}$ matrices it is convenient to analyze their form
for small~$p_1$ and~$p_2$ values. Then one obtains
\begin{equation} \hq^{1A} \simeq \left(\begin{array}{cccc} 1-p_1+\frac{p_2}{4} & 0 & 0 & 0 \\ 0 & 1-\frac{p_2}{4} & p_2 & 0\\ 0 & 0 & 1-\frac{p_2}{4} & 0 \\ 0 & 0 & 0 & 1+p_1+\frac{p_2}{4} \end{array} \right),\end{equation}
\begin{equation} \hq^{1B} \simeq \left(\begin{array}{cccc} 0 & 0 & 0 & 0 \\ -p_1-\frac{p_2}{2} & 0 & 0 & 0 \\ \frac{p_2}{2}-p_1 & 0 & 0 & 0 \\ 0 & -p_1-\frac{p_2}{2} & \frac{p_2}{2}-p_1 & 0 \end{array} \right),\end{equation}
\begin{equation} \hq^{1C} \simeq \left(\begin{array}{cccc} 0 & \frac{p_2}{2}-p_1 & -p_1-\frac{p_2}{2} & 0 \\ 0 & 0 & 0 & \frac{p_2}{2}-p_1 \\ 0 & 0 & 0 & -p_1-\frac{p_2}{2} \\ 0 & 0 & 0 & 0 \end{array} \right),\end{equation}
\begin{equation} \hq^{1D} \simeq \left(\begin{array}{cccc} 1+p_1+\frac{p_2}{4} & 0 & 0 & 0 \\ 0 & 1-\frac{p_2}{4} & 0 & 0 \\ 0 & p_2 & 1-\frac{p_2}{4} & 0 \\ 0 & 0 & 0 & 1-p_1+\frac{p_2}{4} \end{array} \right),\end{equation}
\begin{equation} \hq^{2A} \simeq \left(\begin{array}{cccc} 1-p_1+\frac{p_2}{4} & 0 & 0 & 0 \\ 0 & 1-\frac{p_2}{4} & 0 & 0 \\ 0 & p_2 & 1-\frac{p_2}{4} & 0 \\ 0 & 0 & 0 & 1+p_1+\frac{p_2}{4} \end{array} \right),\end{equation}
\begin{equation} \hq^{2B} \simeq \left(\begin{array}{cccc} 0 & 0 & 0 & 0 \\ \frac{p_2}{2}-p_1 & 0 & 0 & 0 \\ -p_1-\frac{p_2}{2} & 0 & 0 & 0 \\ 0 & \frac{p_2}{2}-p_1 & -p_1-\frac{p_2}{2} & 0 \end{array} \right),\end{equation}
\begin{equation} \hq^{2C} \simeq \left(\begin{array}{cccc} 0 & -p_1-\frac{p_2}{2} & \frac{p_2}{2}-p_1 & 0 \\ 0 & 0 & 0 & -p_1-\frac{p_2}{2} \\ 0 & 0 & 0 &\frac{p_2}{2}-p_1 \\ 0 & 0 & 0 & 0 \end{array} \right),\end{equation}
\begin{equation} \hq^{2D} \simeq \left(\begin{array}{cccc} 1+ p_1+\frac{p_2}{4} & 0 & 0 & 0 \\ 0 & 1-\frac{p_2}{4} & p_2 & 0 \\ 0 & 0 & 1-\frac{p_2}{4} & 0 \\ 0 & 0 & 0 & 1-p_1+\frac{p_2}{4} \end{array} \right),\end{equation}

From the above equation one notes that for small~$p_1$ and~$p_2$ the matrices~$\hq^{1A},\hq^{1D},\hq^{2A},\hq^{2D}$ are
diagonal, and for~$p_1, p_2 \rightarrow 0$ they tend to the identity matrix~$\hI$. The remaining matrices are nondiagonal and
they vanish in the limit~$p_1, p_2 \rightarrow 0$.

\subsection{Coefficients~$\hk^{a\alpha}$ and~$\hk^{b\alpha}$}

Here we list the coefficients~$x_{ij}$ in Eq.~(\ref{TrK_ab_m}) of the main text and the nonzero
coefficients of matrices~$\hk^{a\alpha}, \hk^{b\alpha}$ in Eqs.~(\ref{DefKc}) and~(\ref{hkaA})
of the main text
\begin{eqnarray} \label{xx11}
 x_{11} = x_{44} &=& \frac{8(6p_1^3+3p_1^2-8p_1-4)}{(3p_1^2-4p_1-4)^2}+2, \\
 x_{22} = x_{33} &=& \frac{2p_1^2(3p_1-2)}{(p_1-2)(3p_1^2-4p_1-4)}, \\
 x_{23} = x_{32} &=& \frac{16p_1^2}{(3p_1^2-4p_1-4)^2}, \label{xx23}
\end{eqnarray}

\begin{eqnarray} \label{kaA11}
 k^{aA}_{11} = k^{aA}_{11} = k^{aD}_{33} = k^{aD}_{44} &=& \frac{2}{2-p_1}, \\
 k^{aA}_{33} = k^{aA}_{44} = k^{aD}_{11} = k^{aD}_{22} &=& \frac{-2(p_1+2)}{3p_1^2-4p_1-4} \\
 k^{aB}_{31} = k^{aB}_{42} = k^{aC}_{13} = k^{aC}_{24} &=& \frac{-4p_1}{3p_1^2-4p_1-4},
\end{eqnarray}

\begin{eqnarray}
 k^{bA}_{11} = k^{bA}_{33} = k^{bD}_{22} = k^{bD}_{44} &=& \frac{2}{2-p_1},\\
 k^{bA}_{22} = k^{bA}_{44} = k^{bD}_{11} = k^{bD}_{33} &=& \frac{-2(p_1+2)}{3p_1^2-4p_1-4}, \\
 k^{bB}_{21} = k^{bB}_{43} = k^{bC}_{12} = k^{bC}_{34} &=& \frac{-4p_1}{3p_1^2-4p_1-4}. \label{kbB21}
\end{eqnarray}
The remaining elements are zero. The coefficients~$k^{a\alpha}_{ij}$ can be obtained from
the corresponding coefficients~$q^{a\alpha}_{ij}$, while the coefficients~$k^{b\alpha}_{ij}$
from the corresponding coefficients~$q^{b\alpha}_{ij}$, see Eqs.~(\ref{q_1A_11})--(\ref{q_2C_34}),
by approximating~$p_2 \rightarrow p_1^2$.

\subsection{Two spinless delta potentials}

Here we calculate the GF of the electron gas in the presence of two
scalar delta like potentials placed in~${\bm r}_a$ and~${\bm r}_b$, respectively.
We use a similar notation and symbols as in Sections~\ref{Sec_Prelim} and~\ref{Sec_GFDyson}
of the main text.
Since the matrix elements of scalar potentials commute,
the GF of the system is much simpler than that for spin operators in Eq.~(\ref{Vr}).
This derivation of GF may help to understand the main steps of
calculating the GF in Sections~\ref{Sec_Prelim} and~\ref{Sec_GFDyson}
of the main text on a simpler example.

Using the same assumptions about the system as in Section~\ref{Sec_Prelim} of
the main text one obtains instead of Eq.~(\ref{Vr}) of the main text
\begin{equation} \label{AVr}
 V(\bm r) = V_a \delta({\bm r}-{\bm r}_a) + V_b\delta({\bm r}-{\bm r}_b).
\end{equation}
From the Dyson equation one obtains, see Eq.~(\ref{hG_Dyson})--~(\ref{hG_12_a}) of the main text
\begin{equation} \label{AhG_12_a}
 \hG_{12} = \hg_{12} + \hg_{1a} V_a \hG_{a2} + \hg_{1b} V_b \hG_{b2}.
\end{equation}
On setting~${\bm r}_1 \rightarrow {\bm r}_a$ and~${\bm r}_1 \rightarrow {\bm r}_b$
one obtains, see Eq.~(\ref{2Eqs_ab}) of the main text
\begin{equation} \label{A2Eqs_ab}
\left(\begin{array}{cc} 1 - \hg_0 V_a & -\hg_{ab} V_b \\ -\hg_{ba} V_a & 1 - \hg_0 V_b \end{array} \right)
 \left(\begin{array}{c} \hG_{a2} \\ \hG_{b2} \end{array} \right) =
 \left(\begin{array}{c} \hg_{a2} \\ \hg_{b2} \end{array} \right).
\end{equation}
The above equation is a set of two linear equations for two unknown
functions~$\hG_{a2}$ and~$\hG_{b2}$. We solve them in the standard way. Using
\begin{equation}
 \hat{Y}^{-1} = \left[\begin{array}{cc} A & B \\ C & D \end{array} \right]^{-1}
 =F^{-1} \left[\begin{array}{cc} D & -B \\ -C & A \end{array} \right],
\end{equation}
with~$F=AD-BC$ we find
\begin{equation}
 \left(\begin{array}{c} \hG_{a2} \\ \hG_{b2} \end{array} \right) = \frac{1}{F}
 \left(\begin{array}{cc} 1 - \hg_0 V_b & \hg_{ab} V_b \\ \hg_{ba} V_a & 1 - \hg_0 V_a \end{array} \right)
 \left(\begin{array}{c} \hg_{a2} \\ \hg_{b2} \end{array} \right),
\end{equation}
and
\begin{equation}
 F = (1 - \hg_0 V_a)(1 - \hg_0 V_b) + \hg_{ab}\hg_{ba} V_a V_b.
\end{equation}
For scalar potentials, the quantities~$A,B,C,D,F$ are also scalars and their order is irrelevant.
Then the GF in Eq.~(\ref{AhG_12_a}) is
\begin{widetext}
\begin{equation} \label{AhbmG_12_c}
\hG_{12} = \hg_{12} + \frac{1}{F} \left[
         \hg_{1a}V_a(1 - \hg_0 V_b)\hg_{a2} + \hg_{1a} \hg_{ab} V_a V_b \hg_{b2} +
         \hg_{1b}\hg_{ba}V_b V_a \hg_{a2} + \hg_{1b} V_b (1 - \hg_0 V_a)\hg_{b2} \right],
\end{equation} \end{widetext}
which is analogues to Eq.~(\ref{hbmG_12_c}) of the main text.
Taking~$F\approx (1 - \hg_0 V_a)(1 - \hg_0 V_b)$ and~$F\approx 1$ one obtains an approximate form of GF
\begin{eqnarray} \label{AhbmG_12_apr}
\hG_{12} = \hg_{12} + \hg_{1a} \hg_{a2} V_a /(1 - \hg_0 V_a) + \hg_{1a} \hg_{ab} \hg_{b2} V_a V_b \nonumber \\
                    + \hg_{1b} \hg_{ba} \hg_{a2} V_b V_a + \hg_{1b} \hg_{b2} V_b/ (1 - \hg_0 V_b). \ \ \ \
\end{eqnarray}
In this approximation the GF separates on three independent parts: the first and the third
terms in Eq.~(\ref{AhbmG_12_apr}) describe two separate one-impurity GFs,
see Eq.~(\ref{GF_Ziman}) of the main text,
while the second term in Eq.~(\ref{AhbmG_12_apr}) is the inter-impurity coupling,
analogous to the RKKY interaction for spin-dependent potentials.


\begin{thebibliography}{99}

\bibitem{Ruderman1954}     M. A. Ruderman, C. Kittel, Phys. Rev. {\bf 96}, 99 (1954).
\bibitem{Kasuya1956}       T. Kasuya, Progr. Theoret. Phys. (Kyoto) {\bf 16}, 450 (1956).
\bibitem{Yosida1957}       K. Yosida, Phys.Rev. {\bf 106}, 893 (1957).
\bibitem{FreemanBook}      A. J. Freeman, {\it Magnetic Properties of Rare-Earth Metals}, edited by R. J. Elliott, (Plenum Press, London, 1972), p.~245.
\bibitem{KittelBook}       C. Kittel, {\it Quantum Theory of Solids} (Wiley, New York, 2nd ed.1987).
\bibitem{Koster1954}       G. F. Koster and J. C. Slater, Phys. Rev. {\bf 96}, 1208 (1954). 
\bibitem{Vertogen1966}     G. Vertogen and W. J. Gaspers, Phys. Rev. Lett. {\bf 16}, 904 (1966). 
\bibitem{Bowen1968}        S. P. Bowen, Phys. Rev. Lett. {\bf 20}, 726 (1968).
\bibitem{Kittel1968}       C. Kittel, in {\it Solid State Physics}, edited by F. Seitz, D. Turnbull, and H. Ehrenreich (Academic, New York, 1968), Vol. {\bf 22}, p.~1.
\bibitem{ZimanBook}        J. M. Ziman, {\it Elements of Advanced Quantum Theory} (Cambridge University Press, Cambridge 1969) p.~131.
\bibitem{Wolff1961}        P. A. Wolff, Phys. Rev. {\bf 124}, 1030 (1961). 
\bibitem{Clogston1962}     A. M. Clogston, Phys. Rev. {\bf 125}, 439 (1962). 
\bibitem{Wiegmann1981}     P. B. Wiegmann, J. Phys. C {\bf 14}, 1463 (1981). 
\bibitem{Andrei1983}       N. Andrei, K. Furuya, and J. H. Lowenstein. Rev. Mod. Phys. {\bf 55}, 331 (1983). 
\bibitem{Woodbury50}       M. A. Woodbury, {\it Inverting modified matrices}, Memorandum Rept. {\bf 42}, Statistical Research Group, Princeton University,
                           Princeton, NJ, 4pp MR38136 (1950); see also https://en.wikipedia.org/wiki/Woodbury\_matrix\_identity, (2019).
\bibitem{Stone2005}        N. J. Stone, Atomic Data and Nuclear Data Tables {\bf 90}, 75 (2005).
\bibitem{EconomouBook}     E. N. Economou {\it Green's Functions in Quantum Physics}, 3rd.ed.(Springer,Berlin,2006).
\bibitem{Wiertz1976}       W. Wiertz and R. R. Gerhardts, Z. Physik B {\bf 25}, 19 (1976).
\bibitem{Wildberger1995}   K. Wildberger, P. Lang, R. Zeller, and P. H. Dederichs, Phys. Rev. B {\bf 52}, 11502 (1995).
\bibitem{Furdyna1998}      J. K. Furdyna, J. Appl. Phys. {\bf 64}, R29 (1988).
\bibitem{Daniel2005}       B. Daniel, K. C. Agarwal, J. Lupaca-Schomber, C. Klingshirn, and M. Hetterich, Appl. Phys. Lett. {\bf 87}, 212103 (2005). 
\bibitem{Ravichandran2011} J. Ravichandran, W. Siemons, M. L. Scullin, S. Mukerjee, M. Huijben, J. E. Moore, A. Majumdar, and R. Ramesh,
                           Phys. Rev. B {\bf 83}, 035101 (2011).
\bibitem{Benthema2001}     K. van Benthema, C. Elsasser, and R. H. French, J. Appl. Phys. {\bf 90}, 6156 (2001). 
\bibitem{Brooks1991}       M. S. S. Brooks, T. Gasche, S. Auluck, L. Nordstrom, L. Severin, J. Trygg, and B. Johansson, J. Appl. Phys. {\bf 70}, 5972 (1991).
\bibitem{Brooks1997}       M. S. S. Brooks, in {\it Magnetism in Metals}, A Symposium in Memory of Allan Mackintosh,
                           ed. Edited by D.F. McMorrow, J. Jensen and H. M. Ronnow, The Royal Danish Academy of Sciences and Letters
                           (Commissioner: Munksgaard, Copenhagen 1997) p.~291; see also: https://www.fys.ku.dk/~jjensen/Book/Allansympc.pdf.
\bibitem{OtherDef}         The factor of two in~$J_{4f-5d} = -J/(2 \Omega_0)$ follows from other definition of
                           the coupling constant used in Eq.~(\ref{Vr}) and Ref.~\cite{Brooks1997} or Eq.~(\ref{HLiu}).
\bibitem{Kim1995}          J. G. Kim, E. K. Lee, and S. Lee, Phys. Rev. B {\bf 51}, 670(R) (1995).
\bibitem{Zawadzki1974}     W. Zawadzki, Adv. Phys. {\bf 23}, 435 (1974).
\bibitem{Case1950}         K. M. Case, Phys. Rev. {bf 80}, 797 (1950). 
\bibitem{Calkin1987}       M. G. Calkin, D. Kiang and, Y. Nogami, Am. J. Phys. {\bf 55}, 737 (1987).
\bibitem{Frisken1986}      S. J. Frisken and D. J. Miller, Phys. Rev. Lett. {\bf 57}, 2971 (1986).
\bibitem{Coleman2015}      P. Coleman in {\it Many-Body Physics: From Kondo to Hubbard}
                           (eds E. Pavarini, E. Koch and P. Coleman), (Publisher: Forschungszentrum Julich),
                           Chapter 1, 1.1-1.34 (2015); see also arXiv:1509.05769v1 (2015).
\bibitem{de Gennes1962}    P. G. de Gennes, J. Phys. Radium {\bf 23}, 630 (1962).
\bibitem{Bulaevski1986}    L. N. Bulaevski and S. V. Panyukov, Pis'ma Zh. Eksp. Teor. Fiz. {\bf 43}, 190 (1986). 
\bibitem{Rusin2018}        T. M. Rusin and W. Zawadzki, Phys. Rev. B {\bf 97} 205410 (2018).
\bibitem{Proetto1982}      C. Proetto and A. Lopez, Phys. Rev. B {\bf 25}, 7037 (1982). 
\bibitem{Yafet1987}        Y. Yafet, Phys. Rev. B {\bf 36}, 3948 (1987).
\bibitem{Litvinov1998}     V. I. Litvinov and V. K. Dugaev, Phys. Rev. B {\bf 58}, 3584 (1998).
\bibitem{Rusin2017}        T. M. Rusin and W. Zawadzki, J.M.M.M. {\bf 441}, 387 (2017).
\bibitem{Liu1961}          S. H. Liu, Phys. Rev. {\bf 121}, 451 (1961).
\bibitem{Gorman2014}       P. D. Gorman, J. M. Duffy, S. R. Power, and M. S. Ferreira, Phys. Rev. B {\bf 90}, 125411 (2014).
\bibitem{LLoyd1967}        P. Lloyd, Proc. Phys. Soc. London {\bf 90}, 207 (1967); {\bf 90}, 217 (1967).

\end{thebibliography}
\end{document}